\documentclass[12pt]{article}
\usepackage {palatino}

\usepackage{amsfonts,amsmath,amsthm,amssymb,verbatim,nicefrac,thmtools}
\usepackage[pdftex,letterpaper,left=.95in,top=.95in,bottom=.95in,right=.95in]{geometry} 
\usepackage{afterpage} 
\usepackage[round,longnamesfirst]{natbib}
\usepackage{graphicx, color}
\usepackage{setspace}
\usepackage[colorlinks,pagebackref=false]{hyperref}
\usepackage{epsf}

\usepackage{latexsym, stmaryrd, bm}
\usepackage[titletoc,title]{appendix}
\usepackage{titlesec}
\usepackage[dotinlabels]{titletoc}
\usepackage{marginnote}
\usepackage{graphicx, subfigure, ftnxtra}
\usepackage{minitoc}
\usepackage[nottoc]{tocbibind} 
\graphicspath{{Figures/}} 

\usepackage{enumitem}
\setlist[enumerate]{itemsep=1pt, topsep=0pt}
\setlist[itemize]{itemsep=1pt, topsep=0pt}

\definecolor{dark-red}{rgb}{0.60,0.15,0.15}
\definecolor{dark-blue}{rgb}{0.15,0.15,0.65}
\definecolor{medium-blue}{rgb}{0,0,0.5}
\definecolor{dark-green}{rgb}{0.1,.6,0.5}

\hypersetup{
    pdftitle={Single-Crossing Differences in Convex Environments},
    pdfauthor={Kartik, Lee, Rappoport},
    citecolor={dark-blue},
    bookmarksnumbered=true,
    urlcolor=dark-red,
    linkcolor=dark-red,
}

\newenvironment{appendixref}[1]{\hyperref[#1]{Appendix \ref{#1}}}

\newtheorem{theorem}{Theorem}
\newtheorem{lemma}{Lemma}
\newtheorem{claim}{Claim}
\newtheorem{proposition}{Proposition}

\newtheorem{corollary}{Corollary}

\theoremstyle{definition}
\newtheorem{definition}{Definition}
\declaretheorem[qed=\textsquare]{example}

\theoremstyle{remark}
\newtheorem{remark}{Remark}
\newtheorem*{example*}{Example}

\newcommand{\citepos}[1]{\citeauthor{#1}'s (\citeyear{#1})}
\newcommand{\citeauthorpos}[1]{\citeauthor{#1}'s}

\makeatletter
  \renewcommand\@seccntformat[1]{\csname the#1\endcsname.{\hskip.7em\relax}} 
\makeatother

\titlespacing*{\paragraph}{0pt}{1.5ex plus 1ex minus .2ex}{1ex plus 1ex minus .2ex}
\titlespacing\section{0pt}{10pt plus 2pt minus 2pt}{4pt plus 2pt minus 2pt} 
\titlespacing\subsection{0pt}{6pt plus 2pt minus 2pt}{2pt plus 2pt minus 2pt} 
\titlespacing\subsubsection{0pt}{6pt plus 2pt minus 2pt}{0pt plus 2pt minus 2pt} 

\newcommand{\ind}{{1}\hspace{-2.5mm}{1}}
\renewcommand{\epsilon}{\varepsilon}
\renewcommand{\phi}{\varphi}
\renewcommand{\bar}{\overline}
\newcommand{\mailto}[1]{\href{mailto:#1}{\texttt{#1}}}

\def\o{\overline}
\def\u{\underline}

\DeclareMathOperator*{\argmax}{arg\,max}

\DeclareMathOperator{\sign}{sign}
\DeclareMathOperator{\rank}{rank}

\def\bi{\begin{itemize}}
\def\ei{\end{itemize}}

\def\E{\mathbb{E}}
\def\Reals{\mathbb{R}}

\def\TP2{\text{TP}_2}
\def\SCDstar{SCD$^\star$}
\def\MDstar{MD$^\star$}
\def\X{X}
\def\x{x}
\def\Z{Z}
\def\z{z}

\let\oldfootnote\footnote
\renewcommand\footnote[1]{\oldfootnote{\hspace{.4mm}#1}}

\makeatletter
\renewenvironment{proof}[1][\proofname] {\par\pushQED{\qed}\normalfont\topsep6\p@\@plus6\p@\relax\trivlist\item[\hskip\labelsep\bfseries#1\@addpunct{.}]\ignorespaces}{\popQED\endtrivlist\@endpefalse}
\makeatother

\let\oldFootnote\footnote
\newcommand\nextToken\relax

\renewcommand\footnote[1]{%
    \oldFootnote{#1}\futurelet\nextToken\isFootnote}

\newcommand\isFootnote{%
    \ifx\footnote\nextToken\textsuperscript{,}\fi}

\begin{document}
\pagestyle{empty}
\onehalfspacing 

\title{Single-Crossing Differences\\[3pt] in Convex Environments\footnote{Previous versions of this paper were circulated under the title ``Single-Crossing Differences on Distributions''. We thank Nageeb Ali, John Duggan, Federico Echenique, Mira Frick, Ben Golub, Ryota Iijima, Ian Jewitt, Alexey Kushnir, Shuo Liu, Daniele Pennesi, Jacopo Perego, Lones Smith, Bruno Strulovici, and various audiences for helpful comments.
}}
\date{\today}
\author{Navin Kartik\thanks{Department of Economics, Columbia University. E-mail: \mailto{nkartik@gmail.com}} \and SangMok Lee\thanks{Department of Economics, Washington University in St. Louis. E-mail: \mailto{sangmoklee@wustl.edu}} \and Daniel Rappoport\thanks{Booth School of Business, University of Chicago. E-mail: \mailto{Daniel.Rappoport@chicagobooth.edu}}}
\maketitle
\thispagestyle{empty}

\begin{abstract}
An agent's preferences depend on an ordered parameter or type. We characterize the set of utility functions with single-crossing differences (SCD) in \emph{convex environments}. These include preferences over lotteries, both in expected utility and rank-dependent utility frameworks, and preferences over bundles of goods and over consumption streams.  Our notion of SCD does not presume an order on the choice space. This unordered SCD is necessary and sufficient for ``interval choice'' comparative statics.  We present applications to cheap talk, observational learning, and collective choice, showing how convex environments arise in these problems and how SCD/interval choice are useful. Methodologically, our main characterization stems from a result on linear aggregations of single-crossing functions.
\end{abstract}

\bigskip

\begin{quote}
\textbf{Keywords:} monotone comparative statics, choice among lotteries, interval equilibria, aggregating single crossing
\end{quote}

\newpage
\setcounter{tocdepth}{2} 
\tableofcontents

\newpage
\setcounter{page}{1}
\pagestyle{plain}
\section{Introduction}
\label{sec:intro}
\onehalfspacing 

\subsection{Overview}

Single-crossing properties and their implications for choices are at the heart of many economic models, as highlighted by \citet{milgrom1994monotone}. Consider a utility function $u(a,\theta)$, where $a$ is the choice object and $\theta$ a preference parameter. In this paper, we completely characterize the structure of $u(\cdot)$ when it has \emph{single-crossing differences (SCD) in a rich---specifically, convex---environment}. 
Before defining this property more precisely and explaining our results, we begin with some background and motivation for our work in a leading context: choice among lotteries.

\paragraph{Motivation.} In various applications of single crossing, choices have been restricted to deterministic outcomes when it would be desirable to accommodate lotteries. Consider \citepos{CS82} canonical cheap-talk model: a sender with private type $\theta \in \Theta \subseteq \Reals$ chooses a costless message to send a receiver, who then takes a decision $a$. A single-crossing property on the sender's utility $u(a,\theta)$ ensures that any equilibrium is an ``interval equilibrium'', i.e., $\Theta$ is partitioned into intervals that induce the same decision. This result is predicated on assumptions ensuring that in equilibrium, the sender can fully anticipate the decision induced by each message. However, in assorted contexts, one would like to model a sender who is uncertain about the receiver's preferences; but then, from the sender's perspective, each message would induce a lottery over decisions.

A challenge with extending single-crossing properties to choice among lotteries is that there is no natural order on the set of all lotteries. Moreover, it is not always apparent a priori what restrictions are reasonable on the set of feasible lotteries, in particular whether some form of stochastic dominance can order every choice set an agent may face. In the cheap-talk problem, the nature of the receiver's preferences may well imply that a higher message (in equilibrium) induces lotteries that have both higher mean and higher variance.  In other applications, the lotteries may be the result of still further interactions---e.g., they may represent continuations in dynamic strategic problems---that are intractable to structure ex ante.

An alternative to restricting the set of lotteries is to require that an agent's utility difference between \emph{any} pair of lotteries---or, more abstractly, any pair of choice objects---is single crossing in the agent's preference parameter or type. (For any ordered set $\Theta$, a function $\Theta \mapsto \Reals$ is single crossing if its sign is monotonic.) It is this single-crossing property that we study, which we refer to as \hyperref[def:SCD]{\emph{single-crossing differences}} (SCD). Our notion is closely related to that of \citet{milgrom1994monotone} for comparative statics, and the ``order restriction'' property \citet{Rothstein90} introduced for collective choice. As we explain in \autoref{sec:compstats}, SCD characterizes \emph{interval choice} (\autoref{lem:scd_interval_choice}), a fundamental property for applications. Loosely, interval choice says that given any choice set and any option in that set, if a low and a high type both find that option optimal, then so do all intermediate types. In the cheap-talk application, the sender's options are messages, and interval choice yields the aforementioned interval equilibrium desideratum.

A key question, then, is: for an expected-utility agent, which (von Neumann Morgenstern) utility functions assure SCD over lotteries? It is not enough that SCD holds over pure outcomes:
\begin{example*}
\label{eg:intro}
Let $\Theta=[-1,1]$, $A=\{0,1,2\}$, and $u(a,\theta)=a$ for $a\neq 1$ while $u(1,\theta)=\theta^2+1/2$. For any $a,a'\in A$, $u(a,\theta)-u(a',\theta)$ is single-crossing in $\theta$ as its sign does not depend on $\theta$. But for $G$ the uniform lottery over outcomes 
$0$ and $2$, the expected utility difference $u(1,\theta)-\E_{a\sim G}[u(a,\theta)]=\theta^2+1/2-(1/2)2=\theta^2-1/2$ is not single crossing in $\theta$.
Hence, $u$ has SCD over pure outcomes but does not have SCD over lotteries. 
\end{example*}

\paragraph{SCD in Convex Environments.} In the foregoing example, SCD holds for ``extreme'' alternatives (i.e., pure outcomes) but fails over ``in-between'' alternatives (i.e., lotteries). We find this broader viewpoint valuable, as various economic settings have such richness in the alternatives. Accordingly, consider a utility function $u(a,\theta)$, where $a\in A$ is an action (i.e., a choice alternative) and $\theta \in \Theta$ the ordered type. Say that the choice environment is \emph{convex} if the set of functions $\{u(a,\cdot):\Theta\to \Reals\}_{a \in A}$ is convex.\footnote{That is, for all $a,a'\in A$ and $\lambda \in (0,1)$, there is $a''\in A$ such that for all $\theta\in \Theta$, $u(a'',\theta)=\lambda u(a,\theta)+(1-\lambda)u(a',\theta)$.} 

Convex choice environments abound. The function $u$ can be expected utility and $A$ the set of lotteries over arbitrary outcomes. But, as explained by \autoref{eg:wEU} in \autoref{sec:main}, $u$ can also be rank-dependent utility \citep{quiggin82}. Or, as detailed in \autoref{eg:MA}, $A$ can be a product set representing different dimensions of the choice object, and $u$ can be a multidimensional utility function with convex range. This class captures examples of deterministic settings in mechanism design, dynamic consumption streams, and choices over bundles of goods or products with multiple characteristics. 

Our paper's main result, \autoref{char_scd_f}, is a characterization of SCD in any convex environment. \autoref{char_scd_f} establishes that in a convex environment, $u$ has SCD if and only if
	\begin{equation}
	\label{e:sced_intro}	u(a,\theta)=g_1(a)f_1(\theta)+g_2(a)f_2(\theta)+h(\theta),
	\end{equation} 
where $f_1$ and $f_2$ are single-crossing functions that satisfy a \hyperref[def:RO]{\textit{ratio-ordering}} property we define in \autoref{sec:main}. Roughly speaking, ratio ordering requires that the relative importance placed on $g_1(\cdot)$ versus $g_2(\cdot)$ changes monotonically with type.\footnote{In particular, if either $f_1(\cdot)$ or $f_2(\cdot)$ is strictly positive, then ratio ordering reduces to saying that the ratio of the two functions is monotonic. 
More generally, \autoref{equiv_cond} establishes that ratio ordering is necessary and sufficient for all linear combinations of two single-crossing functions to be single crossing.} The idea is transparent when $\Theta$ has a minimum $\u \theta$ and a maximum $\o \theta$. Then, $u$ having SCD is equivalent to the existence of a (type-dependent) representation $\tilde u(a,\theta)$ that satisfies 
$$\tilde u(a, \theta) = \lambda(\theta)\tilde u(a,\o \theta)+(1-\lambda(\theta))\tilde u(a,\u \theta),$$
where $\lambda:\Theta \to [0,1]$ is increasing (\autoref{rmk:convex_comb}). In other words, in such a convex environment, SCD is equivalent to being able to represent each type's preferences by a utility function that is a convex combination of those of the extreme types, with higher types putting more weight on the highest type's utility.

In the context of expected utility, there are canonical (von Neumann Morgenstern) functional forms that induce SCD over lotteries: in mechanism design and screening, $u((q,t),\theta)=\theta q - t$, where $q\in \Reals$ is the quantity, $t\in \Reals$ is the transfer, and $\theta \in \Reals$ is the agent's marginal rate of substitution; in optimal delegation without transfers, $u((q, t), \theta)= \theta q +g(q)-t$, where $q\in \Reals$ is the allocation, $t\in \Reals_+$ is money burning, and $\theta\in \Reals$ is the agent's type \citep[cf.~][]{AB2013}; in communication/delegation and voting, $u(a,\theta)=-(a-\theta)^2=2\theta a-a^2-\theta^2$, where $a\in \Reals$ is an outcome and $\theta\in \Reals$ is the agent's bliss point. On the other hand, our characterization also makes clear that SCD is quite stringent.  For example, within the class of power loss functions, only the quadratic loss function generates SCD over lotteries (\autoref{lossfunctions}).
Outside of expected utility, we explain in \autoref{sec:implications} when discounted utility and Cobb-Douglas utility satisfy SCD in a convex environment; in particular, it holds for the simple two-good case of $u((x,y),\theta)=\theta \log x + (1-\theta)\log y$, where $x,y \in \Reals_{++}$ are the quantities and $\theta$ parameterizes the marginal rate of substitution.

\paragraph{Applications.} \autoref{sec:applications} applies SCD in convex environments to three economic problems, highlighting the implications of interval choice. Among other things, the cheap-talk application with uncertain receiver preferences in \autoref{sec:cheaptalk} demonstrates concretely how choices from \emph{all} lotteries emerge naturally. In \autoref{sec:obslearn}'s observational-learning application, a convex environment stems from multidimensional utility. \autoref{sec:voting} considers collective choice over lotteries, showing how our results contribute to a long-standing question of whether equilibrium exists when political candidates can offer lottery platforms \citep{zeckhauser69majority,shepsle72ambiguity}.

\subsection{An Intuition}

A key step towards \autoref{char_scd_f}'s characterization of SCD in convex environments is establishing that every type's utility is an affine combination of two (type-independent) functions: \autoref{e:sced_intro}. We can provide a succinct intuition when $\Theta$ has a minimum $\u \theta$ and a maximum $\o \theta$. It suffices to show that there are three actions, $a_1$, $a_2$, and $a_3$, such that any type $\theta$'s utility from any action $a$ satisfies
\begin{equation}
\label{e:sced_intro_2}
u(a,\theta)=\lambda_1(a) u(a_1,\theta) +\lambda_2(a) u(a_2,\theta) +\lambda_3(a) u(a_3,\theta),	
\end{equation}
for some $\lambda(a)\equiv (\lambda_1(a),\lambda_2(a),\lambda_3(a))$ with $\sum_{i=1}^3 \lambda_i(a)=1$. (\autoref{e:sced_intro} follows by setting, for $i=1,2$, $f_i(\theta)=u(a_i,\theta)-u(a_3,\theta)$, $g_i(a)=\lambda_i(a)$, and $h(\theta)=u(a_3,\theta)$.)
The desired $\lambda(a)$ is the solution to
\begin{align*}
\begin{bmatrix}
u(a,\u \theta)\\
u(a,\o \theta)\\
1
\end{bmatrix}=
\begin{bmatrix}
u(a_1,\u \theta) &u(a_2,\u \theta) &u(a_3,\u \theta) \\
u(a_1,\o \theta) &u(a_2,\o \theta) &u(a_3,\o \theta) \\
1 & 1 &1
\end{bmatrix}
\begin{bmatrix}
\lambda_1(a) \\
\lambda_2(a)\\
\lambda_3(a)
\end{bmatrix},
\end{align*}
which exists at least when there are three actions for which 
the $3 \times 3$ matrix on the right-hand side is invertible. To interpret this matrix equation, note that in a convex environment, any convex combination of utilities from $\{a_1,a_2,a_3,a\}$ is the utility from some action. Hence, the equation says that one can find two distinct actions with utilities equal to convex utility combinations of $\{a_1,a_2,a_3,a\}$ such that the lowest and highest types are both indifferent between those two actions.\footnote{Take one utility combination to be that corresponding to the uniform distribution $P=(1/4,1/4,1/4,1/4)$ on $\{a_1,a_2,a_3,a\}$ and the other corresponding to $P+\epsilon (\lambda_1(a),\lambda_2(a),\lambda_3(a),-1)$ for any sufficiently small $\epsilon>0$.} By SCD, \emph{all} types must be indifferent between these two actions, which amounts to \autoref{e:sced_intro_2}.

A second key step towards \autoref{char_scd_f} is showing that the two type-independent utilities in \autoref{e:sced_intro} must be ratio ordered. The intuition for this step is provided in \autoref{sec:main_result_intuition}.

\subsection{Related Literature}

We now offer a summary of related work, supplying additional details later.

\citet{quah2012aggregating} consider an expected-utility agent choosing under uncertainty about her preferences, which depend on some unknown ``state''. They ask when single-crossing differences in the \citet{milgrom1994monotone} sense is preserved regardless of the state distribution. In our expected-utility application, we consider an agent who knows her preferences but chooses among lotteries. Although these are conceptually different questions, there is a mathematical connection in portions of our analysis. \citepos{quah2012aggregating} question concerns when single crossing is preserved by positive linear combinations. 
On the other hand, our problem turns on \emph{arbitrary} linear combinations preserving single crossing. As elaborated after \autoref{equiv_cond}, this explains the difference between \citeauthorpos{quah2012aggregating} signed-ratio-monotonicity condition 
and our ratio-ordering condition. Moreover, there is no analog in their analysis to the linear dependence we deduce in \autoref{equiv_cond_general}, which is crucial to our main characterization's functional form \eqref{e:funcform}.

When $\Theta \subseteq \Reals$, the utility specification $u(a,\theta)=\theta g_1(a)+g_2(a)$ induces expected utility with SCD over lotteries; indeed, the expected-utility difference between any two lotteries is monotonic in $\theta$. The usefulness of this utility specification (or slight variants) to structure choices among arbitrary lotteries has been highlighted by \citet{duggan2014majority}, \citet{celik15}, and \citet{kushnir2017equivalence}. In \autoref{sec:discussion}, we show how SCD preferences always have such a ``monotonic differences'' representation in convex environments that satisfy a reasonable additional condition. This is a striking consequence of convex environments, as we are not aware of any such result more generally.

In the operations research literature, there has been interest in functional forms for ``multi-attribute utility functions'' \citep[e.g.,][]{fishburn1974neumann}. When there are two attributes and the agent has expected-utility preferences, \citet{abbas2011one} study a ``one-switch condition'' that is akin to SCD over lotteries on one attribute. In that setting, they offer a result related to our \autoref{rmk:convex_comb}. However, they do not identify ratio ordering as the property that characterizes when single crossing is preserved under aggregation, which is a key contribution of our analysis (\autoref{equiv_cond} and \autoref{equiv_cond_general}). Also novel to our paper are the comparative statics characterizations of SCD (\autoref{lem:scd_interval_choice} and \autoref{prop_mcs}) and our economic applications.

For choice among restricted sets of lotteries (which we wish to largely avoid, for reasons mentioned earlier), there are various prior results on the conditions for monotone comparative statics under expected-utility preferences. Standard restricted classes of lotteries include those ordered by first-order stochastic dominance \citep{Topkis78} and likelihood-ratio dominance \citep{karlin1968total,athey2002mcs}.\footnote{For first-order stochastic dominance, the requirement is that the utility function is supermodular; for likelihood-ratio dominance, it is log-supermodularity. \citet{smith2011frictional} considers choice among arbitrary lotteries and their certainty equivalents.}

Outside of expected utility on lottery spaces, we are not aware of any work highlighting SCD in convex environments. In our view, the lens of convex environments is a contribution of our paper.

\section{Single-Crossing Differences and Interval Choice}
\label{sec:compstats}

Our analysis begins by formalizing comparative statics results that justify a notion of single-crossing differences without reference to an order over the choice space.

Let $(\Theta, \leq)$ be a (partially) ordered set containing upper and lower bounds for all pairs.\footnote{A partial order---hereafter, also referred to as just an order---is a binary relation that is reflexive, anti-symmetric, and transitive (but not necessarily complete). An upper (resp., lower) bound of $\Theta_0 \subseteq \Theta$ is $\overline{\theta} \in \Theta$ (resp., $\underline{\theta} \in \Theta$) such that $\theta \leq \overline{\theta}$ (resp., $\underline{\theta} \leq \theta$) for all $\theta \in \Theta_0$.
While none of our results require any assumptions on the cardinality of $\Theta$, the results in \autoref{sec:main} are trivial when $|\Theta|<3$.  See an earlier version of this paper, \citet[Appendix I]{KLR19}, for how our results extend when $(\Theta,\leq)$ is only a pre-ordered set, i.e., when $\leq$ does not satisfy anti-symmetry.
} We often refer to elements of $\Theta$ as \emph{types}.

\begin{definition}
\label{def:SC}
A function $f: \Theta \to \mathbb{R}$ is:
\begin{enumerate}
\item  \textit{single crossing} (resp., from below or from above) if $\sign[f]$ is monotonic (resp., increasing or decreasing);\footnote{For $x\in \Reals$, $\sign[x]=1$ if $x>0$, $\sign[x]=0$ if $x=0$, and $\sign[x]=-1$ if $x<0$. A function $h: \Theta \to \mathbb{R}$ is increasing (resp., decreasing) if $\theta_h>\theta_l \implies h(\theta_h) \geq h(\theta_l)$ (resp., $h(\theta_h) \leq h(\theta_l)$); it is monotonic if it is either increasing or decreasing. 
An equivalent, and perhaps more familiar, definition of $f$ being single crossing from below is 
	$(\forall \theta < \theta') \ f(\theta)  \geq (>) 0 \implies f(\theta') \geq (>) 0$.}
\item \textit{strictly single crossing} if it is single crossing and there are no $\theta' < \theta''$ such that $f(\theta') = f(\theta'') = 0$.
\end{enumerate}
\label{def:str_RO}
\end{definition}

\begin{definition}
\label{def:SCD}
Given any set $A$, a function $u:A \times \Theta \to \Reals$ has:
\begin{enumerate}
\item  \textit{single-crossing differences (SCD)} if $\forall a, a' \in A$, the difference $D_{a,a'}(\theta) \equiv u(a,\theta)-u(a',\theta)$ is single crossing in $\theta$;

\item \textit{strict single-crossing differences (SSCD)} if $\forall a, a' \in A$ such that $a\neq a'$, 
$D_{a, a'}(\theta)$ is strictly single crossing in $\theta$.
\end{enumerate}
\end{definition}

Our definition of (S)SCD is related to but different from \citet{milgrom1994monotone}, who stipulate that $u: A \times \Theta \to \Reals$ has (strict) single-crossing differences given an order $\succeq$ on $A$ if for all $a' \succ a''$, $D_{a', a''}(\theta)$  is (strictly) single crossing from below. (Here, $\succ$ is the strict component of $\succeq$. Note that the (S)SCD terminology is due to \citet{milgrom2004book}.) 
We do not presume that $A$ is ordered, but we consider differences for all pairs of elements of $A$. If $A$ is completely ordered, then our definition is weaker than \citepos{milgrom1994monotone} because ours does not constrain the direction of single crossing. As established below, our notion characterizes related but distinct comparative statics from theirs.

Our notion of SCD is also closely related to \citepos{Rothstein90} notion of order restriction, which he introduced in the context of collective choice. Like SCD, order restriction does not presume an order on the choice space.  Order restriction is more permissive insofar as the order on the type space can be chosen to generate SCD; it is more restrictive in requiring that order to be complete.

\paragraph{Interval Choice.}
\label{strictviolation} We say that $\Theta_0 \subseteq \Theta$ is an \emph{interval} if $\theta_l, \theta_h \in \Theta_0$ and $\theta_l < \theta_m<\theta_h$ imply $\theta_m \in \Theta_0$. 
Let $C: 2^A \times \Theta \rightrightarrows A$ with $C(S, \theta) \subseteq S$ for each $S \subseteq A$ and $\theta \in \Theta$. We say that $C$ has \emph{interval choice} if $\{\theta : a \in C(S ,\theta)\}$ is an interval for each $S \subseteq A$ and $a \in S$. That is, interpreting $C$ as a choice correspondence, the set of types choosing any option given any choice set is an interval. 
We say that $u: A \times \Theta \to \mathbb{R}$ \emph{strictly violates SCD} if there are $a, a' \in A$ and $\theta_l<\theta_m<\theta_h$ such that $\min\{D_{a, a'}(\theta_l),D_{a,a'}(\theta_h)\}>0>D_{a,a'}(\theta_m)$.

\begin{theorem}
\label{lem:scd_interval_choice}
Let $u:A \times \Theta \to \Reals$ and $C_u(S,\theta)\equiv \argmax_{a\in S} u(a,\theta)$ for any $S\subseteq A$ and $\theta$.
\begin{enumerate}
\item \label{interval_choice} If $u$ has SCD, then the choice correspondence $C_u$ has interval choice. 
If $u$ strictly violates SCD, then $C_u$ does not have interval choice.
\item \label{interval_selection} If $|\Theta| \geq 3$, then $u$ has SSCD if and only if every selection from 
$C_u$ has interval choice.
\end{enumerate}
\end{theorem}

The intuition for the sufficiency of (S)SCD in \autoref{lem:scd_interval_choice} is straightforward. Regarding necessity, we note that a violation of SCD---as opposed to a strict violation---is compatible with the choice correspondence having interval choice: e.g., $A=\{a', a''\}$, $\Theta = \{\theta_l, \theta_m, \theta_h\}$ with $\theta_l < \theta_m < \theta_h$, and $\min\{D_{a',a''}(\theta_l) , D_{a',a''}(\theta_h)\} > 0 =D_{a',a''}(\theta_m)$.\footnote{On the other hand, a strict violation of SCD is slightly stronger than needed: one could weaken its requirement to $\min\{D_{a', a''}(\theta_l),D_{a', a''}(\theta_h)\} \geq 0>D_{a',a''}(\theta_m)$. Our formulation with both inequalities being strict amounts to putting aside indifferences, which proves  convenient for the applications in \autoref{sec:applications}.} In Part \ref{interval_selection} of the theorem, if $|\Theta| =2$ then any selection from any choice correspondence trivially has interval choice,  yet $u$ does not have SSCD when $D_{a, a'}(\theta)=0$ for some $a, a'$ and all $\theta$.

\paragraph{Monotone Comparative Statics.}
\label{sec:MCS_summary}
Our choice space $A$ is unordered. Intuitively, interval choice is intimately related to there being monotone comparative statics (MCS)---i.e., in some sense, higher types make higher choices---with respect to \emph{some} complete order on the choice space. We formalize this connection in \autoref{sec:MCS} by tying (S)SCD to MCS. In brief, given any order on $A$, we order subsets of $A$ by the corresponding strong set order, and say that the function $u$ has MCS if, for every choice set $S\subseteq A$ and all types $\theta_h>\theta_l$, the higher type chooses a higher set: $C_{u}(S,\theta_h)\geq C_u(S,\theta_l)$. Roughly speaking, \autoref{prop_mcs} in \autoref{sec:MCS} shows that an order on $A$ induces MCS if and only if $u$ has SCD and the order is a refinement of a natural ``SCD-order'' generated by $u$.

\section{Single-Crossing Differences in Convex Environments}
\label{sec:main}

This section characterizes single-crossing differences in ``rich'' environments. We now assume the existence of a strictly increasing real-valued function on $(\Theta,\leq)$.\footnote{That is, we assume $\exists h: \Theta \to \mathbb{R}$ such that $\underline \theta < \overline \theta \implies h(\underline \theta) < h(\overline \theta)$. This requirement is related to utility representations for possibly incomplete preferences \citep[][Chapter B.4.3]{ok2007real}. A sufficient condition is that $\Theta$ has a countable order dense subset, i.e., there is a countable set $\Theta_0 \subseteq \Theta$ such that $(\forall \underline \theta,\overline \theta \in \Theta\setminus \Theta_0) \quad  \underline \theta < \overline \theta \implies \exists \theta_0 \in \Theta_0 \text{ s.t. } \underline \theta < \theta_0 < \overline \theta$ \citep[][Corollary 1]{Jaffray1975order}. The assumption only plays a technical role in establishing our characterization of strict SCD, i.e., in the second statement of \autoref{char_scd_f}.} This requirement is satisfied, for example, when $\Theta$ is finite, or $\Theta \subseteq \mathbb{R}^n$ is endowed with the usual order. We assume the environment $(A,\Theta,u)$ is \emph{convex} in the following sense:
\begin{equation} 
\tag{$\star$}
\text{ the set of functions } \{u(a,\cdot):\Theta\to \Reals\}_{a \in A} \text{ is convex.}
\label{star}
\end{equation}
That is, a convex environment is rich enough insofar as for any pair of actions and any weighting, there is a third action that replicates the weighted sum of utilities of the original actions.
We stress that convexity is in terms of utilities: we have not assumed any structure on $A$. However, if $A$ is convex, then it is straightforward that \eqref{star} is assured by linearity of $u$ in its first argument.

\begin{example}[Expected Utility]
\label{eg:EU}
Consider an expected-utility agent who chooses among lotteries. There is a set of consequences $\X$ and the agent has utility $v(\x,\theta)$. Letting $A\equiv \Delta \X$ be the set of all finite-support lotteries over $\X$, the agent's utility from lottery $P\in A$ is given by $u(P,\theta)\equiv \int_x v(x, \theta) \mathrm{d}P$.\footnote{We restrict attention to finite-support distributions throughout the paper for ease of exposition, as it guarantees that expected utility is well defined no matter the distribution and utility function. Nevertheless, we write integrals rather than summations when it simplifies notation regarding the domain of integration/summation.} This is a case in which Condition \eqref{star} holds because $u$ is linear in its first argument and $A$ is convex.
\end{example}

\begin{example}[Rank-Dependent Expected Utility]
\label{eg:wEU}
Continuing with lotteries, non-expected utility environments can also be convex. 
By virtually the same logic as above for expected utility, it is sufficient for \eqref{star} that the utility from lottery $P$ be given by $u(P,\theta)\equiv \int_x v(x, \theta)  \mathrm d(w\circ P)$, 
with $w :\Delta \X \to \Delta \X$ an arbitrary reweighting function whose image is convex. For example, a standard formulation of rank-dependent utility \citep{quiggin82,DW01} corresponds to $\X\equiv \{\x_1,\ldots,\x_n\} \subset \Reals $ with $\x_1<\cdots<\x_n$, and a strictly increasing function $\hat w:[0,1]\to [0,1]$ satisfying $\hat w(0)=0$ and $\hat w(1)=1$ such that for any lottery $P$ and consequence $\x_i$, the reweighting is given by $(w\circ  P)(\x_i)\equiv \hat w\left(\sum_{j=1}^i p(\x_j)\right)-\hat w\left(\sum_{j=1}^{i-1} p(\x_j)\right)$, where $p$ is the probability mass function of the lottery $P$. So long as $\hat w$ is continuous, $w$ has a convex image as required; the image is simply $\Delta \X$.
\end{example}

\begin{example}[Multidimensional Utility]
\label{eg:MA}
While lotteries naturally induce a convex environment, they are not necessary. Another example is choice among multidimensional actions. Specifically, $A\equiv A_1 \times \ldots \times A_n\subseteq \Reals^n$ and for any $a\equiv (a_1,\ldots,a_n)$, utility is given by $u(a,\theta)\equiv \sum_{i=1}^n g_i(a_i)f_i(\theta)$ for some pairs of functions $(g_i,f_i)_{i=1}^n$, with each $g_i$ having a convex image. That $A$ is a product set and each $g_i$ has a convex image ensures \eqref{star}. We refer to this specification as \emph{multidimensional utility}.

Here are some economic contexts in which there is multidimensional utility. First, a consumer chooses among products with multiple characteristics or a bundle of multiple goods, denoted $(a_1,\ldots,a_n)$. Each characteristic or good $i$ with quality or quantity $a_i$ has a common value $g_i(a_i)$, but the tradeoff across characteristics/goods varies with the consumer's preference parameter $\theta$, as given by $f_i(\theta)$. Second, a designer uses an incentive-compatible direct mechanism $\phi: T \to A$ that maps an agent's private type $t \in T\equiv \{1,\ldots,n\}$ to an action $\phi(t)\in A$. The designer's payoff is $\sum_{t} v(\phi(t),t) f(t; \theta)$, where $v(a, t)$ is the designer's utility from allocating $a$ to $t$, and $f(\cdot ; \theta) \in \Delta T$ is a type distribution that depends on some parameter $\theta$. Third, an agent chooses consumption $c_t$ in each period $t\in T\equiv\{1,\dots, n\}$. The agent's present discounted value is $\sum_t v(c_t)\rho(t; \theta)$, where $\rho (t ; \theta)$ is discount function parameterized by $\theta$.
\end{example}

\begin{example}[Experiments]
\label{eg:experiment}
Our final example is one in which an expected-utility agent can only choose from a proper subset of lotteries.
Let $\Omega$ be a finite set of ``states''. We refer to $\Delta \Omega$ as the set of beliefs or posteriors and $\Delta \Delta \Omega$ as the set of experiments (with finite support).  In Bayesian persuasion \citep{kamenica11} or, more broadly, information design \citep{bergemann17}, an expected-utility agent has preferences represented by $v(p,\theta)$, where $p\in\Delta \Omega$ and $\theta$ is a preference parameter. The expected utility from experiment $Q\in \Delta \Delta \Omega$ is given by $u(Q, \theta) \equiv \int_p v(p, \theta) \mathrm{d}Q$. If we consider all experiments, then this setting is a special case of \autoref{eg:EU}. But given a prior $p^*\in \Delta \Omega$, any experiment must in fact be Bayes-plausible, i.e., its distribution of posteriors must average to the prior $p^*$. So, given $p^*$, the agent can only choose an experiment in $A \equiv \{ Q \in \Delta\Delta \Omega : \int_{p\in \Delta \Omega} p \mathrm d Q= p^*\}$. Nevertheless, the linearity of $u$ in its first argument and the convexity of $A$ again imply \eqref{star}. Indeed, based on \autoref{eg:wEU}, Condition \eqref{star} will hold even for rank-dependent expected-utility agents choosing among Bayes-plausible experiments.
\end{example}

\subsection{The Characterization}\label{sec:char}

Our characterization of SCD in convex environments (\autoref{char_scd_f} below) requires the following definition.

\begin{definition}
\label{def:RO}
Let $f_1,f_2 : \Theta \to \Reals$ each be single crossing.
\begin{enumerate}
\item  $f_1$ \textit{ratio dominates} $f_2$ if
\begin{align}
\hspace{-.3in}(\forall \theta_l < \theta_h) & \quad f_1(\theta_l) f_2(\theta_h) \leq f_1(\theta_h) f_2(\theta_l), \quad\text{and} \label{e:RD}\\
\hspace{-.3in}(\forall \theta_l < \theta_m < \theta_h) & \quad  f_1(\theta_l)f_2(\theta_h) = f_1(\theta_h)f_2(\theta_l) \iff  
\left\{\begin{array}{c}
f_1(\theta_l)f_2(\theta_m) = f_1(\theta_m)f_2(\theta_l),\\
f_1(\theta_m)f_2(\theta_h) = f_1(\theta_h)f_2(\theta_m).
\end{array}
\right.
\label{e:RD_add}
\end{align}
\item $f_1$ \textit{strictly ratio dominates} $f_2$ if Condition \eqref{e:RD} holds with strict inequality.
\item $f_1$ and $f_2$ are \textit{(strictly) ratio ordered} if either $f_1$ (strictly) ratio dominates $f_2$ or vice-versa.
\end{enumerate}
\end{definition}

Condition \eqref{e:RD} contains the essential idea of ratio dominance and is what we focus on in the main text; Condition \eqref{e:RD_add} only deals with some knife-edged cases that are discussed in \appendixref{app:RD_add}.  The definition of strict ratio dominance does not make reference to Condition \eqref{e:RD_add} because that condition is vacuous when Condition \eqref{e:RD} holds with strict inequality.

We use the terminology ``ratio dominance'' because when $f_2$ is a strictly positive function, Condition \eqref{e:RD} is the requirement that the ratio $f_1(\theta)/f_2(\theta)$ must be increasing in $\theta$. Indeed, if both $f_1$ and $f_2$ are probability densities of random variables $Y_1$ and $Y_2$, then \eqref{e:RD} says that $Y_1$ stochastically dominates $Y_2$ in the sense of likelihood ratios.\footnote{From the viewpoint of information economics, think of $\theta$ as a signal of a state $s\in \{1,2\}$, drawn from the density $f(\theta|s)\equiv f_s(\theta)$. Condition \eqref{e:RD} is \citepos{milgrom1981news} monotone likelihood-ratio property for $f(\theta|s)$.}

Condition \eqref{e:RD} is a natural generalization of the increasing ratio property to functions that may change sign. To get a geometric intuition, suppose $f_1$ strictly ratio dominates $f_2$. Let $f(\theta)\equiv (f_1(\theta),f_2(\theta))$.
For every $\theta_l < \theta_h$, $f_1(\theta_l)f_2(\theta_h) - f_1(\theta_h)f_2(\theta_l) < 0$ implies that the vector $f(\theta_l)$ moves to $f(\theta_h)$ through a rescaling of magnitude and a clockwise---rather than counterclockwise---rotation (throughout our paper, a ``rotation'' must be no more than $180$ degrees).\footnote{To elaborate on the direction of rotation, recall that from the definition of cross product,
\begin{align*}
(f_1(\theta_l), f_2(\theta_l),0) \times (f_1(\theta_h), f_2(\theta_h),0) = \lVert f(\theta_l)\rVert \lVert f(\theta_h) \rVert \sin(r) e_3 = \left(f_1(\theta_l)f_2(\theta_h)-f_1(\theta_h)f_2(\theta_l)\right) e_3,
\end{align*} 
where $r$ is the counterclockwise angle from $f(\theta_l)$ to $f(\theta_h)$, $e_3 \equiv (0,0,1)$, $\times$ is the cross product, and $\lVert \cdot \rVert$ is the Euclidean norm. If $\sin(r)< 0$ (resp., $\sin(r) > 0$), then $f(\theta_l)$ moves to $f(\theta_h)$ through a clockwise (resp., counterclockwise) rotation.}

\begin{figure}[h!]
\begin{center}
    \subfigure[Condition \eqref{e:RD} holds for $\theta_l<\theta_m<\theta_h$.]{\includegraphics[width=2.5in]{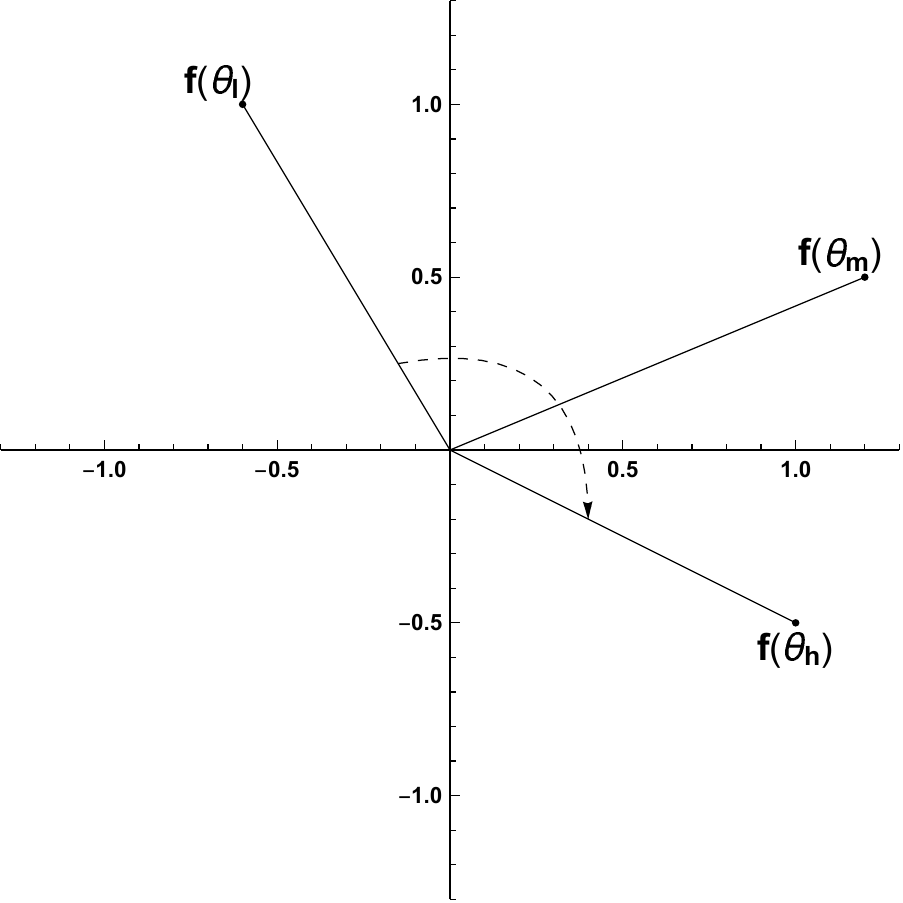}
    \label{fig:not_ordered1}}
    \qquad
    \subfigure[Condition \eqref{e:RD} fails both when $\Theta=\{\theta_l,\theta_m,\theta_h\}$ with $\theta_l<\theta_m<\theta_h$, and when  $\Theta=\{\theta_l,\theta_m,\hat \theta_h\}$ with $\theta_l<\theta_m<\hat \theta_h$.]{\includegraphics[width=2.5in]{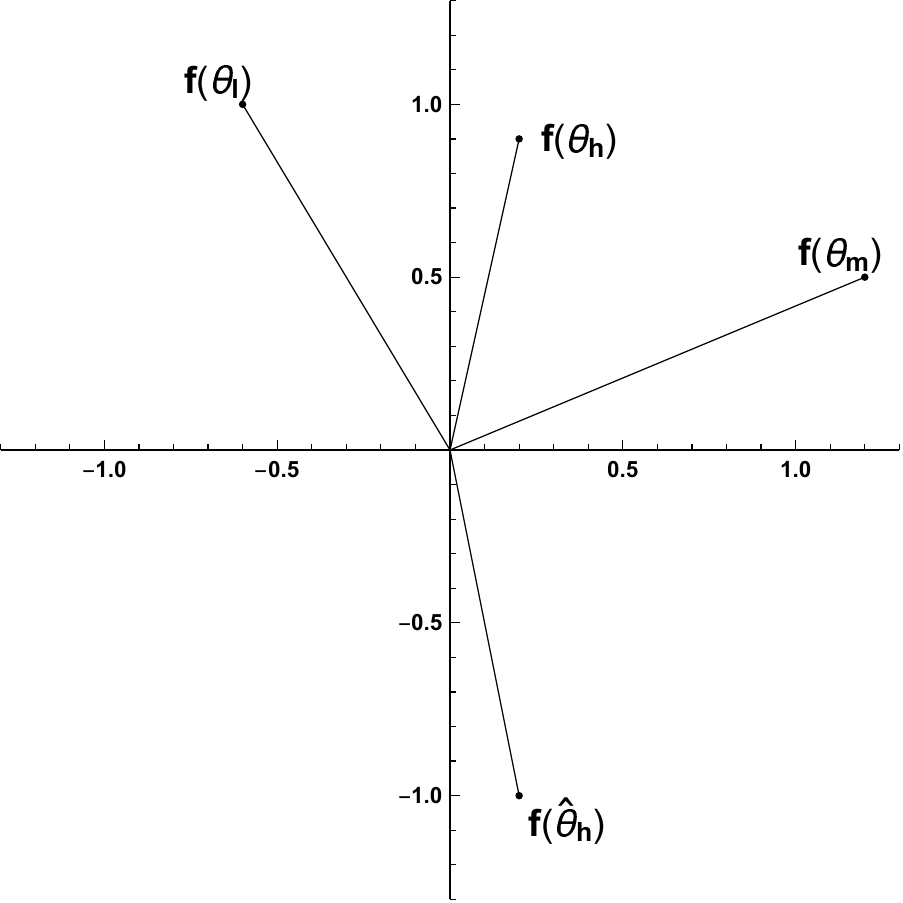}
    \label{fig:not_ordered2}}
	    \caption{\small Geometric representation of Condition \eqref{e:RD}.}
           \label{fig:ordering_intro}
\end{center}
\end{figure}

Hence, $f_1$ and $f_2$ are ratio ordered only if $f(\theta)$ \emph{rotates monotonically} as $\theta$ increases, either always clockwise or always counterclockwise.\footnote{The preceding discussion establishes this point under the presumption of strict ratio ordering;
however, because of the hypothesis in \autoref{def:RO} that $f_1$ and $f_2$ are single crossing and because of Condition \eqref{e:RD_add}, the conclusion holds for ratio ordering too.
Furthermore, it can be confirmed that a monotonic rotation of $f(\cdot)$ implies ratio ordering if there are no $\theta'$ and $\theta''$ such that $f(\theta')$ and $f(\theta'')$ are collinear.}
It follows that the set \mbox{$\{f(\theta): \theta \in \Theta\}$} must be contained in a closed half-space of $\Reals^2$ defined by a hyperplane that passes through the origin: otherwise, there will be two pairs of vectors such that an increase in $\theta$ corresponds to a clockwise rotation in one pair and a counterclockwise rotation in the other. See \autoref{fig:ordering_intro}. In its panel \subref{fig:not_ordered2}, the non-monotonic rotation of $f(\theta)$ is clear when $\Theta=\{\theta_l,\theta_m,\theta_h\}$ with $\theta_l<\theta_m<\theta_h$, but it is also present when $\Theta=\{\theta_l,\theta_m,\hat \theta_h\}$ with $\theta_l<\theta_m<\hat \theta_h$; in the latter case, the aforementioned half-space requirement is violated. 

We can now state our main characterization. For brevity, we say that \emph{$u$ has (S)\SCDstar} as shorthand for ``In a convex environment, $u$ has (S)SCD.''
Furthermore, we say that $A$ is \emph{minimal} if there is no pair of utility-indistinguishable actions: $\forall a, a' \in A$, $\exists \theta$ such that $D_{a,a'}(\theta)\equiv u(a,\theta)-u(a',\theta)\neq 0$.

\begin{samepage}
\begin{theorem}
\label{char_scd_f}
The function $u:A\times \Theta \to \Reals$ has \SCDstar \ if and only if it takes the form 
\begin{equation}
\label{e:funcform}
u(a, \theta) = g_1(a)f_1(\theta) + g_2(a)f_2(\theta) + h(\theta),
\end{equation}
with $f_1$ and $f_2$ each single crossing and ratio ordered. 
In addition, if $A$ is minimal, then $u$ has S\SCDstar \ if and only if $f_1$ and $f_2$ are strictly ratio ordered.\footnote{If $A$ is not minimal, then so long as $|\Theta|>1$, $u$ violates S\SCDstar \ because for some $a$ and $a$', $D_{a,a'}$ is a zero function and hence not strictly single crossing.  Nonetheless, the characterization applies when we consider utility-indistinguishable actions as equivalence classes and the corresponding utility function on those equivalence classes.}
\end{theorem}
\end{samepage}

We make a number of observations to help interpret \autoref{char_scd_f}.

The theorem says that for $u$ to have \SCDstar, it must be possible to write it in the form \eqref{e:funcform}. 
Notice that given \eqref{e:funcform}, for any $a_0,a_1\in A$, the function $u(a_1,\cdot)-u(a_0,\cdot)$ is a linear combination of $f_1(\cdot)$ and $f_2(\cdot)$. Therefore, to rule out the possibility of the form \eqref{e:funcform}, it is sufficient to find $a_0,a_1, a_2, a_3 \in A$ and $\theta_l < \theta_m < \theta_h$ such that the $3 \times 3$ matrix $M \equiv [u(a_i,\theta_j)-u(a_0,\theta_j)]_{i \in \{1,2, 3\}, j \in \{l,m,h\}}$ is invertible. This procedure is often useful to reject \SCDstar, as we illustrate subsequently in \autoref{lossfunctions}. 

Given the functional form \eqref{e:funcform}, not only is \SCDstar \ assured by $f_1$ and $f_2$ each being single crossing and ratio ordered, but these properties are almost necessary.\footnote{\label{footnote:affine_independent}``Almost'' excludes the case in which $g_1$ and $g_2$ are affinely dependent, i.e., $g_1=\lambda g_2 + \gamma$ for some $\lambda,\gamma \in \Reals$. Intuitively, affine independence ensures that neither $g_1$ nor $g_2$ is dispensable in \eqref{e:funcform}.}

An asymmetry between $a$ and $\theta$ in \autoref{e:funcform} bears noting: the function $h:\Theta\to \Reals$ does not have a counterpart function $A \mapsto \Reals$. The reason is that whether the utility difference between two actions is single crossing 
could be altered by adding a function of $a$ alone to 
$u(a,\theta)$. On the other hand, adding a function of $\theta$ alone has no such effect because SCD is an ordinal property that is invariant to any (type-dependent) increasing transformation, or \emph{representation}, of $u$, i.e., $\tilde{u}(a, \theta) \equiv m(u(a, \theta) , \theta)$, where each $m(\cdot, \theta): \Reals \to \Reals$ is strictly increasing.

In general, whether the convexity condition \eqref{star} holds can depend on which representation one chooses. But since SCD is ordinal, the scope of our analysis is in fact broader than it may seem: if a utility function satisfies SCD and some representation satisfies \eqref{star}, then \autoref{char_scd_f} applies to that representation.\footnote{For example, given a product set $A\subseteq \Reals^2_+$ and $\theta \in [0,1]$, the Cobb-Douglas utility $a_1^\theta a_2^{1-\theta}$ does not satisfy \eqref{star}, but the representation $\theta \log{a_1}+(1-\theta)\log{a_2}$ is a multidimensional utility (\autoref{eg:MA}) and therefore satisfies \eqref{star}, indeed \SCDstar.} We illustrate how this is useful in \autoref{sec: multi}.

If $u(a,\theta)$ has the form \eqref{e:funcform} with strictly positive functions $f_1$ and $f_2$, then up to a positive affine transformation (viz., subtracting $h(\theta)$ and dividing by $f_1(\theta)+f_2(\theta)$),\footnote{In general, a positive affine transformation of $u(a,\theta)$ is any $b(\theta)u(a,\theta)+c(\theta)$ where $b(\cdot)>0$. Unlike arbitrary representations, positive affine transformations preserve condition \eqref{star}.} any type's utility becomes a convex combination of two type-independent utility functions over actions, $g_1$ and $g_2$. \autoref{char_scd_f}'s ratio ordering requirement then simply says that the relative weight on $g_1$ and $g_2$ changes monotonically with the agent's type. This idea underlies the following proposition.

\begin{proposition}
\label{rmk:convex_comb}
Let $\Theta$ have both a minimum and a maximum (i.e., $\exists \ \underline{\theta}, \overline{\theta} \in \Theta$ such that $(\forall \theta)$ $\underline{\theta} \leq \theta  \leq \overline{\theta}$), and the environment $(A, \Theta, u)$ be convex. Then, $u$ has \SCDstar \ if and only if $u$ has a positive affine transformation $\tilde u$ satisfying \begin{equation}\tilde{u}(a, \theta) = \lambda(\theta)\tilde u(a,\o \theta) + (1-\lambda(\theta))\tilde u(a,\u \theta),\label{eqn:convex_med_rep}\end{equation} with $\lambda: \Theta \to [0,1]$ increasing. In addition, if $A$ is minimal, then $u$ has S\SCDstar \ if and only if $\lambda$ is strictly increasing.
\end{proposition}

\autoref{rmk:convex_comb} provides an economic interpretation of \SCDstar \ as capturing preferences that monotonically shift weight from one extreme type's to the other's. We note, though, that even when $\Theta$ has extreme types, it could be easier to verify whether a given function has \SCDstar \ using \autoref{char_scd_f} because one does not have to search among the affine transformations allowed by \autoref{rmk:convex_comb}.

We offer one additional interpretational comment. On the one hand, \autoref{char_scd_f} and \autoref{rmk:convex_comb} indicate that SCD---while desirable for tractability (including \hyperref[lem:scd_interval_choice]{interval choice}), economic intuition, etc.---is a demanding property in a convex environment. On the other hand, because the functions $g_1$ and $g_2$ in \autoref{char_scd_f} are arbitrary, \SCDstar \ nevertheless allows for a broad economic landscape. Specifically, when one assumes SCD in any environment (convex or not), one generally has in mind that there is an underlying tradeoff---e.g., risk vs.~expected return among lotteries, delay vs.~total amount in payment streams, or price vs.~product quality in markets---whose balance shifts monotonically with type. Our characterizations show that \SCDstar \ is broad enough to capture such desiderata because the $g_1$ and $g_2$ functions in \autoref{char_scd_f} can evaluate the tradeoff differently: e.g., compared to $g_2$, the function $g_1$ can be more risk averse, discount the future more, or be more price sensitive; furthermore, as highlighted by \autoref{rmk:convex_comb}, higher types put more weight on one criterion.

\subsection{Implications for Leading Examples}
\label{sec:implications}

This subsection develops the implications of \autoref{char_scd_f} for our leading examples. For brevity, we focus on the implications of \SCDstar, stating the S\SCDstar \ counterpart only in \autoref{cor:sced_char}. 

\subsubsection{Single-Crossing Expectational Differences}
\label{sec:SCED}

Suppose, following \autoref{eg:EU}, that $A \equiv \Delta X$ is a set of lotteries and $u$ is the expected utility induced by $v(x,\theta)$. We say that $v$ has \textit{(strict) single-crossing expectational differences}, or \textit{(S)SCED,} if the expected utility function $u$ has (S)\SCDstar.\footnote{In general, an expected utility function $u$ may not have S\SCDstar \ simply because $A\equiv\Delta X$ is not minimal. This arises, for example, when $u$ is a mean-variance utility function and multiple lotteries have the same mean and variance. In such cases we consider utility-indistinguishable lotteries as equivalence classes and the corresponding utility function $\tilde{u}$ defined on these equivalence classes of lotteries. We say $v$ has SSCED if $\tilde{u}$ has S\SCDstar.} 
SCED is not implied by SCD or even supermodularity of $v$.\footnote{For instance, given $x,\theta\in \Reals$, any power loss function $v(x,\theta)=-|x-\theta|^z$ is supermodular when $z>1$, but \autoref{lossfunctions} below establishes that SCED fails for $z\neq 2$.} Rather:

\begin{corollary}
\label{cor:sced_char}
The von Neumann Morgenstern utility function $v$ has (S)SCED if and only if it has the same form as $u$ in \autoref{char_scd_f}.
\end{corollary}

The corollary's proof is straightforward: if $v$ satisfies the characterization, then $u(P,\theta) = (\int_x g_1(x) \mathrm{d} P) f_1(\theta) + (\int_x g_2(x) \mathrm{d} P) f_2(\theta)+h(\theta)$ with $f_1$ and $f_2$ satisfying the conditions given in \autoref{char_scd_f}, so $u$ has  \SCDstar. Conversely, if $u(P, \theta)\equiv \int_x v(x, \theta) \mathrm{d} P$ has \SCDstar, and hence has the form in \autoref{char_scd_f}, then so does $v$ because $v(\x, \theta) \equiv u(\delta_{\x}, \theta)=g_1(x)f_1(\theta) + g_2(x)f_2(\theta) + h(\theta)$, where $\delta_x$ denotes the degenerate lottery on $x$.

\autoref{cor:sced_char} is related to \citet[Theorem 1]{abbas2011one}. They study expected utility over lotteries with some additional restrictions on the environment (e.g., $\Theta$ is finite and completely ordered, and preferences satisfy some substantive economic conditions). For that setting, their Theorem 1 states a similar result to the version of \autoref{cor:sced_char} that would obtain using \autoref{rmk:convex_comb} instead of \autoref{char_scd_f}.\footnote{\citet{abbas2011one} use the terminology of ``one-switch independence'', which appears equivalent to SSCED up to one minor detail that can be set aside here (concerning the treatment of distinct lotteries that all types are indifferent over). Translated into our notation, they assert that a utility function $v(x,\theta)$ has SSCED if and only if $v(\x,\theta)=f_1(\theta)g_1(\x)+f_2(\theta)g_2(\x)+h(\theta)$, with $f_1$ strictly positive and $f_2/f_1$ strictly increasing. This can be seen from our \autoref{rmk:convex_comb} because, given its hypothesis of extreme types, it implies that $v$ has SSCED if and only if $v(x,\theta)=b(\theta)\left[g_1(x)+\lambda(\theta)(g_2(x)-g_1(x)\right]+c(\theta)$, where $b$ is strictly positive, $\lambda$ is strictly increasing, and $g_1$ and $g_2$ are, respectively, monotonic transformations of $v(\cdot,\bar{\theta})$ and $v(\cdot,\underline \theta)$.}

Even aside from the ratio-ordering and single-crossing requirements in \autoref{char_scd_f}, the functional form \eqref{e:funcform} deserves emphasis: there are only two ``interaction terms'', each of which is multiplicatively separable in $a$ and $\theta$. This means that preferences over lotteries must be summarized by two linear statistics: for any lottery $P\in \Delta X$, the statistics are $\int_x g_1(x)\mathrm{d}P$ and $\int_x g_2(x)\mathrm{d}P$.\footnote{The utility function $v(x,\theta)=\exp{(x \theta)}$ cannot be written in the form of \autoref{e:funcform} and hence does not have SCED. Indeed, its expected utility from an arbitrary lottery cannot be summarized by any finite number of linear statistics, let alone two.} This point underlies the following corollary, which identifies quadratic loss as the unique power loss function that has SCED. 

\begin{corollary}
\label{lossfunctions}
Let $X = \Reals$ and $\Theta \subseteq \Reals$ with $|\Theta| \geq 3$. A loss function \mbox{$v(x,\theta)=-|x-\theta|^{z}$} with $z>0$ has SCED if and only if $z=2$.
\end{corollary}

Under quadratic loss, preferences over lotteries are summarized by the lotteries' first and second moments. We note that some non-power-loss generalizations of quadratic loss, such as $v(x,\theta)=x\theta+g(x)+h(\theta)$ with $g:\Reals \to \Reals$, also satisfy SCED; these functional forms---or variants that also augment a quasilinear money burning component, which continues to preserve SCED---are used in the study of delegation with stochastic mechanisms or money burning \citep{AB2013,Kleiner22}.

SCED is useful in dynamic problems, as seen in \citet{BanksDuggan06}, \citet{duggan2014majority}, \citet{celik15}, and \citet{AKK22}. It is thus noteworthy that:

\begin{corollary}
\label{cor:discountedutility}
Suppose $v(x,\theta)$ has SCED. Then, denoting $x^\infty\equiv (x_t)_{t=0}^\infty$, so does the discounted utility function $\tilde v
(x^\infty,\theta) = \sum_{t=0}^\infty \rho(t) v(x_t,\theta)$ for any $\rho:\{0,1,\ldots\}\to \Reals$.
\end{corollary}
We omit a proof as the result follows straightforwardly from \autoref{cor:sced_char}. Note that \autoref{cor:discountedutility} does not require exponential discounting. (Of course, implicitly $\rho$ must ensure that $\tilde v(\cdot)$ is finite.)

\subsubsection{Single-Crossing Rank-Dependent Expected Utility}

Suppose an agent's preferences over lotteries $A\equiv  \Delta \X$ have a rank-dependent expected utility (RDEU) representation as described in \autoref{eg:wEU}, with underlying utility $v(x,\theta)$. 

\begin{corollary}\label{RDEU}
An RDEU function has \SCDstar \ if and only if the underlying utility $v$ 
has the same form as $u$ in \autoref{char_scd_f}.

\end{corollary}

We omit a proof because it is analogous to that for \autoref{cor:sced_char}. An RDEU agent thus evaluates a lottery $P$ according to two summary statistics: $ \int_x g_1(x) \mathrm{d}(w\circ P)$ and $\int_x g_2(x) \mathrm{d}(w\circ P)$. The difference with expected utility is that these summary statistics are no longer linear in $P$. Instead the statistics reweight probabilities according to the original reweighting function.

\subsubsection{Multidimensional Utility}\label{sec: multi}

Suppose, following \autoref{eg:MA}, that an agent has a multidimensional utility function $u(a,\theta)\equiv \sum_{i=1}^n g_i(a_i)f_i(\theta)$.

\begin{corollary}
\label{cor:multidim}
A multidimensional utility function $u$ has \SCDstar \  if and only if
\begin{align}
    u(a,\theta)=\left(\sum_{i=1}^n \lambda^{I}_i g_i(a_i) \right) f^{I}(\theta)+\left(\sum_{i=1}^n \lambda^{II}_i g_i(a_i) \right) f^{II}(\theta)+h(\theta),\label{multiform}
\end{align}
with $f^{I}$ and $f^{II}$ each single crossing and ratio ordered, and $\lambda^{I},\lambda^{II}\in\mathbb{R}^n$.
\end{corollary}

The interpretation is that a multidimensional utility has \SCDstar \ when at most two ``summary" dimensions matter. The values on these summary dimensions are weighted sums of the original values $g_i(a_i)$ over the primitive dimensions $i=1, \dots, n$. Higher types place relatively more weight on products/bundles that are more valuable on one of the two summary dimensions; recall \autoref{rmk:convex_comb}. 

As mentioned in \autoref{eg:MA},  multidimensional utility can capture a consumer choosing among consumption bundles. A canonical specification is the Cobb-Douglas utility $u(a,\theta)=\prod_{i=1}^n a_i^{f_i(\theta)}$, where $a\in \Reals^n_+$ is the consumption bundle and the vector of $f_i(\theta)\geq 0$ parameterizes the consumer's marginal rates of substitution (MRS). Note that an alternative representation is $\sum_{i=1}^n f_i(\theta) \log(a_i)$, which has the multidimensional form. \autoref{cor:multidim} implies that for Cobb-Douglas utility to have SCD, it must be representable as $$u(a,\theta)=\left(\prod_{i=1}^n a_i^{\lambda^{I}_i}\right)^{f(\theta)}\left(\prod_{i=1}^n a_i^{\lambda^{II}_i}\right)^{1-f(\theta)},$$
with $f:\Theta\to [0,1]$ monotonic and $\lambda^I,\lambda^{II}\geq 0$.\footnote{More precisely, \autoref{cor:multidim} yields the representation 
$\left(\prod_{i=1}^n a_i^{\lambda^{I}_i}\right)^{f^I(\theta)}\left(\prod_{i=1}^n a_i^{\lambda^{II}_i}\right)^{f^{II}(\theta)}$,
where $f^I$ and $f^{II}$ are each single crossing and ratio ordered, and \autoref{rmk:convex_comb} yields the further simplification.} The interpretation is that to satisfy SCD, the Cobb-Douglas utility must have ``two layers": the consumer first evaluates the $\theta$-independent Cobb-Douglas value of two composite goods, and then trades off these composite goods according to a Cobb-Douglas utility function with MRS that is monotonic in $\theta$. SCD guarantees that for any choice set (e.g., a budget set), the value of each composite good in the chosen consumption bundle changes monotonically (necessarily in opposite directions) in $\theta$. A special case is the textbook example of two goods, say $1$ and $2$, and $u(a,\theta)=a_1^\theta a_2^{1-\theta}$ with $\theta\in [0,1]$. In that case, we recover the  textbook observation that given any budget set, the consumption of each good is monotonic in the MRS.

As another example, consider discounted utility over consumption streams. For a consumption stream $(c_t)_{t=1}^T$, the discounted utility $\sum_{t=1}^T c_t \rho (t, \theta)$  is of the multidimensional form. \autoref{cor:multidim} implies that the discounted utility has \SCDstar \ only if, for any consumption stream $(c_t)$, $\sum_t c_t \rho(t, \theta)$ is linearly generated by two functions of $\theta$. By considering consumption streams that are positive only in a single period, we see that each $\rho(t, \cdot)$ must in fact be generated by the same two functions of $\theta$, i.e., $\rho(t, \theta) = \lambda^I_t f^I(\theta) + \lambda^{II}_t f^{II}(\theta)$. When $T\geq 3$, such linear dependency does not hold for exponential discounting, i.e., when each type $\theta$ has a discount factor $\delta_\theta$ such that $\rho(t,\theta)=(\delta_\theta)^t$. Consequently, exponential discounting is incompatible with interval choice: given any three discount factors, there are consumption streams $(c_t)$ and $(c'_t)$ such that an agent with either low or high patience strictly prefers $(c_t)$ while an agent with intermediate patience strictly prefers $(c'_t)$.\footnote{For example, let $(c_t)_{t=1}^3= (1,0,6)$ and $(c'_t)_{t=1}^3 =(0,5,0)$. The difference $u(c', \theta) - u(c, \theta) = \delta_\theta (1- 5\delta_\theta + 6\delta_\theta^2)$ is strictly positive if and only if $\delta_\theta \in (1/3, 1/2)$.} In fact, our analysis reveals that in general choice between an arbitrary pair of consumption streams will be monotonic in the time preference parameter $\theta$ only if $\rho(t,\theta)$ has the aforementioned linear dependence. An example is linear time cost, say $\theta \in [\u\theta,\overline \theta]\subset \Reals$ and $\rho(t,\theta)=\alpha-\theta t$ for some $\alpha>\overline \theta T$.\footnote{The fact that SCD imposes strong restrictions on the nature of discounting can be related to difficulties in aggregating time preferences \citep{JacksonYariv15}. Specifically, with three types (or agents), $\underline \theta<P<\overline \theta$, where ``$P$'' stands for Planner, SCD is very similar to a Pareto principle: whenever types $\u \theta$ and $\o \theta$ have the same preference over a pair of consumption streams, so should type $P$. Our discussion indicates that if types $\u \theta$ and $\o \theta$ are exponential discounters, then $P$ cannot be, as was highlighted by \citet{JacksonYariv15}. In fact, our result says that $P$'s preference must linearly depend on the other two types, echoing \citet{harsanyi55}.}

\subsubsection{Single-Crossing Expectational Differences over Experiments}
\label{subsec:info_design}

Suppose, following \autoref{eg:experiment}, that $\Omega$, $\Delta \Omega$, and $\Delta\Delta\Omega$ are a set of states, posteriors, and experiments respectively. Given a prior $p^*\in\Delta \Omega$, an agent can only choose a Bayes-plausible experiment: $A \equiv \{ Q \in \Delta\Delta \Omega : \int_{p\in \Delta \Omega} p \mathrm d Q= p^*\}$. We say that $v(p,\theta)$ has \textit{single-crossing expectational differences over experiments (SCED-X)} if the expected utility function $u(Q,\theta)\equiv \int_p v(p, \theta) \mathrm{d} Q$ over $A$ has \SCDstar. Since $A \subsetneq \Delta\Delta \Omega$, SCED is sufficient for SCED-X but not necessary. SCED-X is, instead, characterized as follows.

\begin{corollary}\label{p:SCED-X}
For any full-support prior $p^*$, the function $v$ 
has SCED-X if and only if
\begin{equation}
\label{eqn:SCED-X}
v(p,\theta) = g_1(p)f_1(\theta) + g_2(p)f_2(\theta) + \sum_\omega v(\delta_\omega, \theta) p(\omega),
\end{equation}
with $f_1$ and $f_2$ each single crossing and ratio ordered. 
\end{corollary}

As the environment is convex, \autoref{char_scd_f}'s characterization applies to the expected utility function $u$. Similar to the SCED characterization in \autoref{cor:sced_char}, the form of the expected utility function $u$ passes to the von Neumann Morgenstern utility function $v$.\footnote{This is not as straightforward as in the case SCED, where one can appeal to degenerate lotteries on any posterior. Under SCED-X, the only posterior on which there can be a degenerate lottery is the prior.} The difference is that the last term in \autoref{eqn:SCED-X} can depend on the posterior $p$ as well as the type $\theta$, unlike the $h(\theta)$ term in \autoref{e:funcform}. This is because the expected utility having the form in \autoref{char_scd_f} only imposes, for SCED-X, that the expectation of that term is constant over lotteries that average to the prior, rather than over all lotteries. This means that the term can admit a linear dependence in $p$, as seen in \autoref{eqn:SCED-X}.

\subsection{Aggregating Single-Crossing Functions}
\label{sec:main_result_intuition}

This subsection explains the logic behind \autoref{char_scd_f}; we focus on single crossing here, deferring strict single crossing to Supplementary \appendixref{proof_str_sc}.
Owing to the convex environment, the central issue is when linear aggregations of functions are single crossing.
\autoref{equiv_cond} below shows that ratio ordering is the characterizing property when aggregating two functions; \autoref{equiv_cond_general} then establishes that when aggregating more than two functions, no more than two can be linearly independent, which leads to \autoref{char_scd_f}.\footnote{Real-valued functions $f_1,\ldots,f_n$ are linearly independent if $(\forall \lambda \in \Reals^n \backslash \{0\})$ $\sum_{i=1}^n\lambda_i f_i$ is not a zero function, i.e., is not everywhere zero.}

\label{sec:aggregation}
\begin{lemma}
\label{equiv_cond}
Let $f_1, f_2: \Theta \to \mathbb{R}$. The linear combination \mbox{$\alpha_1 f_1(\theta) + \alpha_2 f_2(\theta)$} is single crossing $\forall \alpha \in \mathbb{R}^2 $ if and only if $f_1$ and $f_2$ are each single crossing and ratio ordered.
\end{lemma}

\begin{figure}[h!]
\begin{center}
    \subfigure[Sufficiency of ratio ordering.]{{\includegraphics[width=3in]{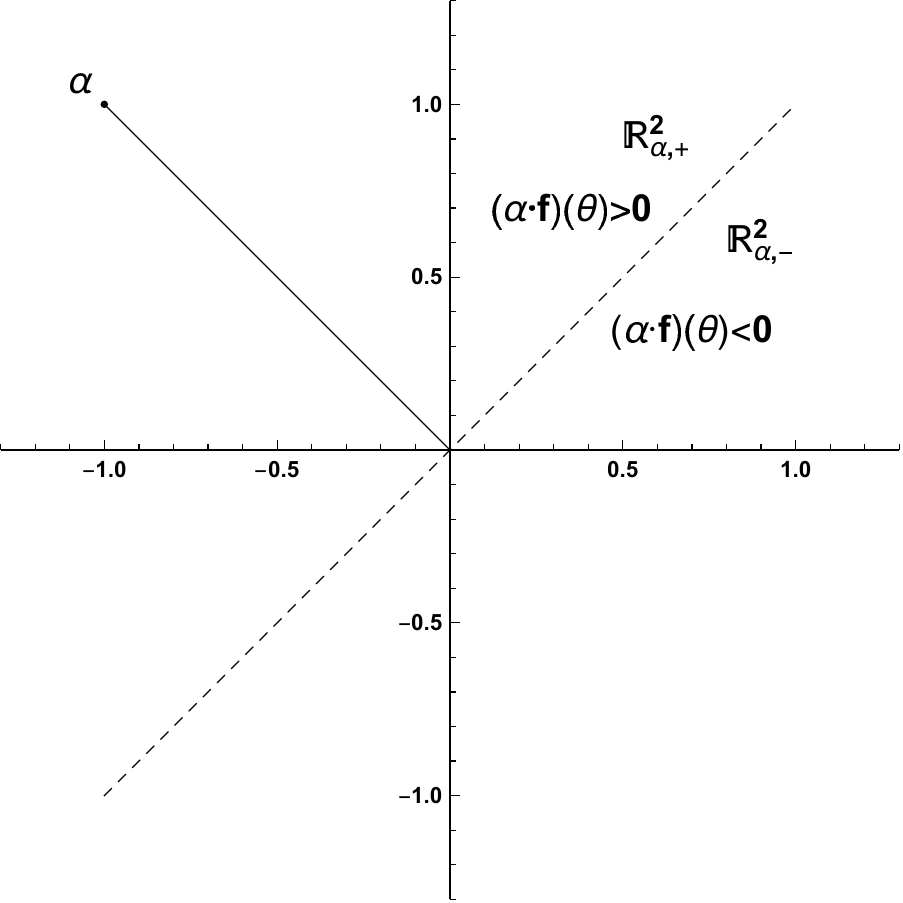}}
    \label{fig:ordering_suff}}
    \qquad
    \subfigure[Necessity of ratio ordering, with $\theta_l<\theta_m<\theta_h$.]{{\includegraphics[width=3in]{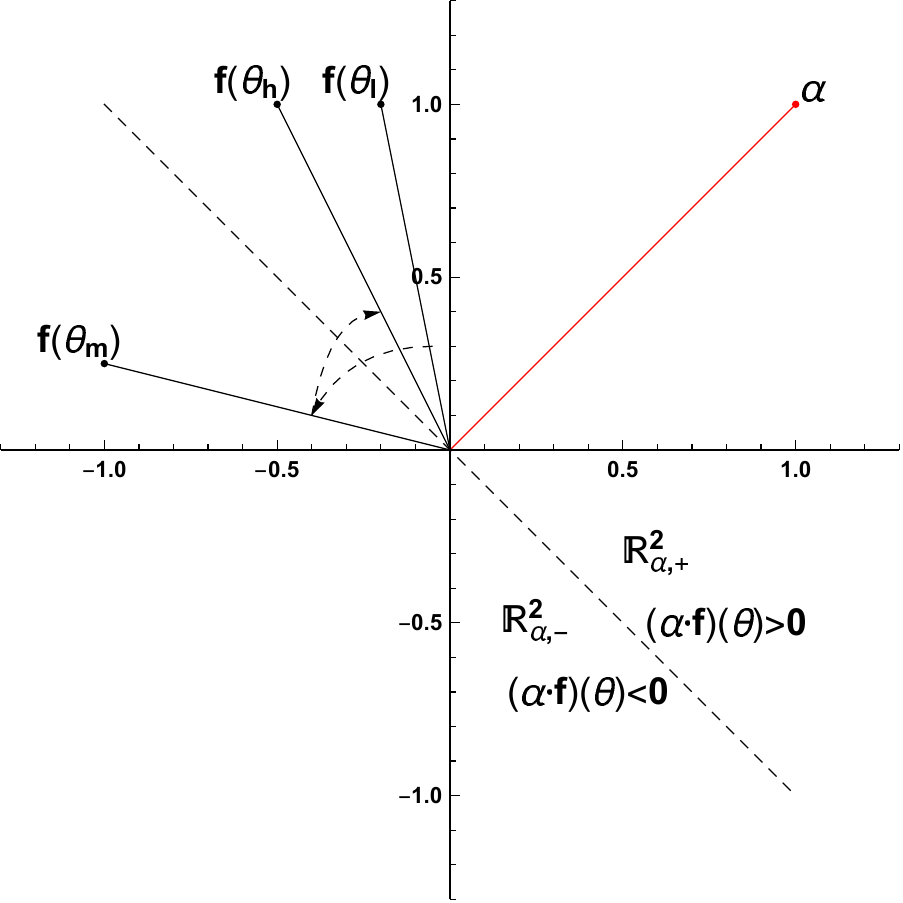}}
    \label{fig:ordering_nec}}
\caption{Ratio ordering and single crossing of all linear combinations.}
\label{fig:ordering_suff_nec}
\end{center}
\end{figure}

Here is the lemma's intuition. For sufficiency, consider any linear combination $\alpha_1 f_1 + \alpha_2 f_2$. Assume $\alpha \in\mathbb{R}^2\backslash \{0\}$, to avoid triviality. The vector $\alpha$ defines  two open half spaces \mbox{$\Reals^2_{\alpha, -} \equiv \{ x \in \Reals^2 : \alpha \cdot x < 0\}$} and $\Reals^2_{\alpha, +} \equiv \{ x \in \Reals^2 : \alpha \cdot x > 0\}$, where $\cdot$ is the dot product; see \autoref{fig:ordering_suff_nec}\subref{fig:ordering_suff}. As explained earlier after \autoref{def:RO}, ratio ordering of $f_1$ and $f_2$ implies that the vector $f(\theta) \equiv (f_1(\theta), f_2(\theta))$ rotates monotonically as $\theta$ increases. If the rotation is from $\Reals^2_{\alpha, -}$ to $\Reals^2_{\alpha, +}$ (resp., from $\Reals^2_{\alpha, +}$ to $\Reals^2_{\alpha, -})$, then $\alpha \cdot f\equiv \alpha_1f_1 + \alpha_2f_2$ is single crossing only from below (resp., only from above). If $\bigcup_{\theta\in \Theta} f(\theta) \subseteq \Reals^2_{\alpha,-}$ or $\bigcup_{\theta\in \Theta} f(\theta) \subseteq \Reals^2_{\alpha,+}$, then $\alpha \cdot f$ is single crossing both from below and above. Other cases are similar.\footnote{\label{fn:AB-error}Notice that this argument does not require either $f_1$ or $f_2$ to be single signed. By contrast, \citet[p.~769, in their last paragraph on ``Necessity'']{abbas2011one} incorrectly claim that all linear combinations of two (strictly) single crossing functions are (strictly) single crossing only if one function is single signed.}

To see why Condition \eqref{e:RD} of ratio ordering is necessary, suppose the vector $f(\theta)$ does not rotate monotonically. \autoref{fig:ordering_suff_nec}\subref{fig:ordering_nec} illustrates a case in which, for $\theta_l < \theta_m < \theta_h$, $f(\theta_l)$ rotates counterclockwise to $f(\theta_m)$, but $f(\theta_m)$ rotates clockwise to $f(\theta_h)$. As shown in the figure, one can find $\alpha \in \Reals^2$ such that $f(\theta_m) \in \Reals^2_{\alpha, -}$ while both $f(\theta_l), f(\theta_h) \in \Reals^2_{\alpha,+}$, which implies that $\alpha \cdot f$ is not single crossing. See \appendixref{app:RD_add} for why Condition \eqref{e:RD_add} is necessary.

\autoref{equiv_cond} relates to \citet[][Proposition 1]{quah2012aggregating}. 
They establish that for any two functions $f_1$ and $f_2$ that are single crossing from below, $\alpha_1f_1 + \alpha_2f_2$ is single crossing from below for all $\alpha \in \mathbb{R}^2_+$ if and only if $f_1$ and $f_2$ satisfy a condition they call signed-ratio monotonicity (see Supplementary \appendixref{sec:qs_karlin} for the definition). In general, ratio ordering is not comparable with signed-ratio monotonicity because we consider a different aggregation problem from \citeauthor{quah2012aggregating}: (i) the input functions may be single crossing in either direction; (ii) the linear combinations involve coefficients of arbitrary sign; and (iii) the resulting combination can be single crossing in either direction.  The \hyperref[eg:intro]{example} in the introduction highlights the importance of point (ii). There, $\theta\in [-1,1]$ and $u(a,\theta)=(\theta^2+1/2)\ind_{\{a=1\}}+2\ind_{\{a=2\}}$. Both $f_1(\theta)=\theta^2+1/2$ and $f_2(\theta)=2$ are positive functions (hence, single crossing from below), and so all positive linear combinations are also positive functions, but $f_1(\theta)-(1/2)f_2(\theta)=\theta^2-1/2$ is not single crossing because $f_1$ and $f_2$ are not ratio ordered. If the input functions in \autoref{equiv_cond} are restricted to be single crossing from below, then ratio ordering implies signed-ratio monotonicity.

\autoref{equiv_cond} implies a characterization of likelihood-ratio ordering for random variables with strictly positive densities.  While this likelihood-ratio ordering characterization does not appear to be well-known among economists, it is a special case of \citepos{karlin1968total} results on the variation diminishing property of totally positive functions. More generally, however, \autoref{equiv_cond} differs from \citet{karlin1968total} because it considers aggregations of functions that can change sign, whereas he only studies non-negative functions.\footnote{Under some conditions, it is possible to transform the lemma's aggregation problem into one of aggregating non-negative functions, and then apply \citepos{karlin1968total} results. However, the precise conditions are opaque \citep*[Appendix H.2]{KLR19}; moreover, the procedure is somewhat involved and, in our view, less instructive than our direct proof of \autoref{equiv_cond}.}

\autoref{char_scd_f} requires an extension of \autoref{equiv_cond} to more than two functions.  
Consider any set $\Z$ and $f:\Z \times \Theta \to \Reals$. 
We say that $f$ is \textit{linear combinations SC-preserving} 
if $\int_z f(\z, \theta) \mathrm{d} \mu$ is single crossing in $\theta$ for every function $\mu:\Z \to \Reals$ with finite support.\footnote{The notation $\int_z f(\z,\theta) \mathrm{d} \mu$ should be read as $\sum_{\{z: \mu(z) \neq 0\}} f(z,\theta) \mu(z)$.}

\begin{proposition}
\label{equiv_cond_general}
Let $f:\Z \times \Theta \to \Reals$ for some set $\Z$. The function $f$ is linear combinations SC-preserving if and only if there exist $\z_1, \z_2\in \Z$ and $\lambda_1,\lambda_2:\Z \to \Reals$ such that 
	\begin{enumerate}
	\item  \label{equiv_cond_general_1} $f(\z_1,\cdot):\Theta \to \Reals$ and $f(\z_2,\cdot):\Theta \to \Reals$ are (i) each single crossing and (ii) ratio ordered, and
	\item \label{equiv_cond_general_2} $(\forall \z) \ f(\z,\cdot)= \lambda_{1}(\z) f(\z_1,\cdot) + \lambda_2(\z) f(\z_2,\cdot)$.
	\end{enumerate}
\end{proposition}

\autoref{equiv_cond_general} says that a family of single-crossing functions $\{f(\z,\cdot)\}_{\z\in \Z}$ preserves single crossing of all finite linear combinations if and only if the family is ``linearly generated'' by two single-crossing functions that are ratio ordered. In particular, given any three single-crossing functions, $f_1$, $f_2$, and $f_3$, all their linear combinations will be single crossing if and only if there is a linear dependence in the triple, say $\lambda_1 f_1 + \lambda_2 f_2=f_3$ for some $\lambda\in \Reals^2$, and $f_1$ and $f_2$ are ratio ordered.

\begin{figure}[h!]
\begin{center}
            \includegraphics[width=4in]
            {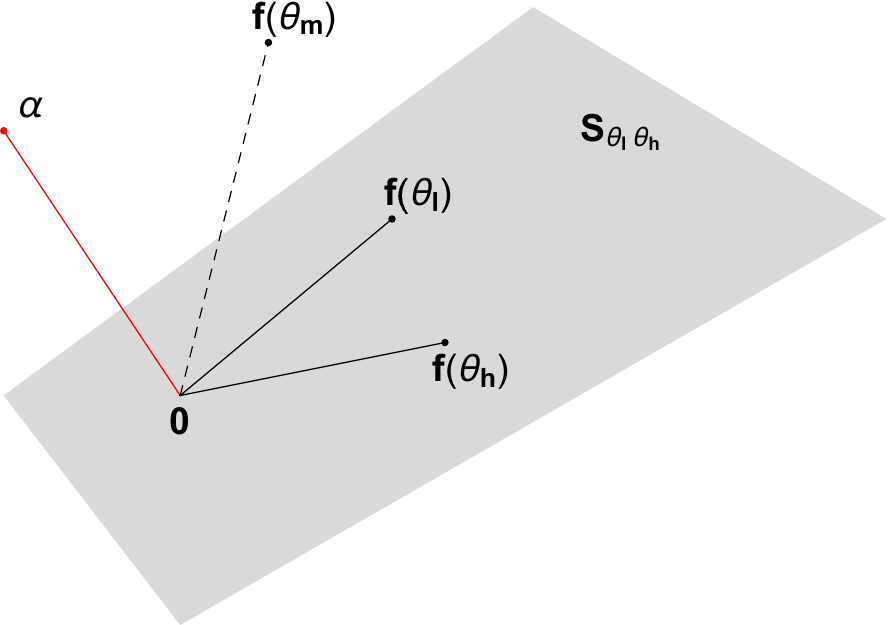}
    \caption{\small The necessity of linear dependence in \autoref{equiv_cond_general}.}
            \label{fig:prop1_intuition}
\end{center}
\end{figure}

The sufficiency direction of \autoref{equiv_cond_general} follows from \autoref{equiv_cond}, as does necessity of the ``generating functions'' being ratio ordered. The intuition for the necessity of linear dependence is as follows. Assume $\Theta$ is completely ordered. For any $\theta$, let $f(\theta) \equiv (f_1(\theta), f_2(\theta), f_3(\theta))$. If $\{f_1, f_2, f_3\}$ is linearly independent, then there exist $\theta_l < \theta_m < \theta_h$ such that $\{f(\theta_l), f(\theta_m), f(\theta_h)\}$ spans $\Reals^3$. Take any $\alpha \in \Reals^3\setminus \{0\}$ that is orthogonal to the plane $S_{\theta_l, \theta_h}$ that is spanned by $f(\theta_l)$ and $f(\theta_h)$, as illustrated in \autoref{fig:prop1_intuition}. The linear combination $\alpha \cdot f$ is not single crossing because $(\alpha \cdot f)(\theta_l) = (\alpha \cdot f)(\theta_h) = 0$ while \mbox{$(\alpha \cdot f)(\theta_m) \neq 0$}.

While the necessity portion of \autoref{equiv_cond_general} only asserts ratio ordering of the ``generating functions'', \autoref{equiv_cond} implies that if $f:\Z \times \Theta \to \Reals$ is linear combinations SC-preserving, then for all $\z,\z'\in \Z$, the pair $f(\z,\cdot):\Theta\to \Reals$ and $f(\z',\cdot):\Theta\to \Reals$ 
must be ratio ordered.

\paragraph{Proof Sketch of \autoref{char_scd_f}.}
\label{sec:SCEDsketch}

We can now sketch the argument for \autoref{char_scd_f}. That its characterization is sufficient for \SCDstar \ is straightforward from \autoref{equiv_cond}. For necessity, suppose as a simplification that, for some $a_0 \in A$, $(\forall \theta)$ $u(a_0, \theta) = 0$.\footnote{This is just a normalization, since $u(a,\theta)$ has \SCDstar \ if and only if $\tilde{u}(a, \theta) \equiv u(a,\theta)-u(a_0,\theta)$ has \SCDstar.} For any $\{a_1,\ldots,a_{n}\} \subset A$ and $(\lambda_1,\ldots,\lambda_n)$, we build on the Hahn-Jordan decomposition of 
$(\lambda_1,\ldots,\lambda_n)$ to write the linear combination $\sum_{i=1}^n \lambda_i v(a_i,\theta)$ as $M \sum_{i=0}^n (p(a_i)-q(a_i)) u(a_i,\theta)$, where $p$ and $q$ are probability mass functions on $\{a_0, a_1, \dots, a_n\}$, and $M$ is a scalar.\footnote{\label{fn:excess}Let $L \equiv \sum_{i=1}^n \lambda_i$. For $i>0$, set $p'(a_i)\equiv \max\{\lambda_i,0\}$ and $q'(a_i)\equiv -\min\{\lambda_i,0\}$. If $L \geq 0$, set $p'(a_0)=0$ and $q'(a_0) \equiv L$; if $L<0$, set $p'(a_0) \equiv -L$ and $q'(a_0) \equiv 0$. Let $M\equiv \sum_{i=0}^n p'(a_i) = \sum_{i=0}^n q'(a_i)$. Finally, for all $a\in \{a_0, \dots, a_n\}$, set $p(a)\equiv p'(a)/M$ and $q(a) \equiv q'(a)/M$.} (Unless $\sum_{i=1}^n \lambda_i =0$, we have $\sum_{i=1}^n p(a_i) \neq \sum_{i=1}^n q(a_i)$; the assumption that $u(a_0,\cdot)=0$ permits us to assign all the ``excess difference'' to $a_0$, as detailed in \autoref{fn:excess}.) Since the environment is convex there exist $a_p, a_q \in A$ such that the linear combination is equal to $M(u(a_p, \theta) - u(a_q, \theta))$, which is single crossing because $u$ has \SCDstar\ . Since every such linear combination is single crossing, \autoref{equiv_cond_general} guarantees $a'$ and $a''$ such that for all $a$, \mbox{$u(a,\cdot)=g_1(a) u(a',\cdot) + g_2(a) u(a'',\cdot)$}, with $u(a',\cdot)$ and $u(a'',\cdot)$ each single crossing and ratio ordered.

\medskip

\begin{remark}
    The proof of \autoref{char_scd_f} is materially no different from that for an expected-utility environment. Indeed, any convex environment $(A,\Theta,u)$ can be transformed into another convex environment $(\tilde A,\Theta,\tilde u)$ where $\tilde A\equiv \{u(a,\cdot)\}_{a\in A}$ is a convex subset of the vector space of functions $\Theta \mapsto \Reals$, and, under mild conditions, Choquet's Theorem implies that each $\tilde u(\cdot,\theta):\tilde A \to \Reals$ is an expected utility function. We develop the lens of convex environments for two reasons: (i) to highlight that our characterization is driven only by Condition \eqref{star}, not other properties of expected utility such as linearity or continuity; and (ii) to cover additional economic settings, as in Examples \ref{eg:wEU}--\ref{eg:experiment}. 
\end{remark}

\section{Applications}
\label{sec:applications}

This section illustrates the usefulness of our results in three applications. Interested readers may consult our earlier working paper, \citet*[Section 4.3]{KLR19}, for another application to costly signaling.

\subsection{Cheap Talk with Uncertain Receiver Preferences}
\label{sec:cheaptalk}

There are two expected-utility players, a sender ($S$) and a receiver ($R$). The sender's type is $\theta \in \Theta$, where $\Theta$ is  ordered by $\leq$. After learning his type, $S$ chooses a payoff-irrelevant message $m \in M$, where $|M|>1$. After observing $m$ but not $\theta$, $R$ takes a decision $x \in X$. The sender's von Neumann Morgenstern utility function is $v^S(x,\theta)$; the receiver's is $v^R(x,\theta,\psi)$, where $\psi \in \Psi$ is a preference parameter that is unknown to $S$ when choosing $m$, and known to $R$ when choosing $x$. Note that $\psi$ does not affect the sender's preferences. The variables $\theta$ and $\psi$ are independently drawn from commonly-known probability distributions.

An example is $\Theta=[0,1]$, $X=\Reals$, $\psi \in \Psi\subseteq \Reals^2$, $v^S(x,\theta)=-(x-\theta)^2$ and $v^R(x,\theta,\psi)=-(x-\psi_1-\psi_2 \theta)^2$. Here the variable $\psi_1$ captures the receiver's ``type-independent bias'' and $\psi_2$ captures the relative ``sensitivity'' to the sender's type. If $\psi$ were commonly known and $\theta$ uniformly distributed, this would be the model of \citet{melumad1991communication}, which itself generalizes the canonical example from \citet{CS82} that obtains when $\psi_1\neq 0$ and $\psi_2=1$.

We focus on (weak Perfect Bayesian) equilibria in which $S$ uses a pure strategy, $\mu: \Theta \to M$, and $R$ plays a possibly-mixed strategy, $\alpha:M \times \Psi \to \Delta X$.\footnote{Our notion of equilibrium requires optimal play for every (not just almost every) type of sender. The restriction to pure strategies for the sender is for expositional simplicity.} Given any $\alpha$, every message $m$ induces some lottery $P_{\alpha}(m) \in \Delta X$ from the sender's viewpoint. An equilibrium $(\mu,\alpha)$ is: (i) an \emph{interval equilibrium} if every message is used by an interval of sender types, i.e., if $(\forall m)$ $\{\theta : \mu(\theta) = m\}$ is an interval; and (ii) \emph{sender minimal} if for all on-the-equilibrium-path $m \neq m'$, there is some $\theta$ such that $\mathbb{E}_{P_\alpha(m)}[v^S(\cdot,\theta)]\neq \mathbb{E}_{P_\alpha(m')}[v^S(\cdot,\theta)]$. In other words, sender minimality rules out all sender types being indifferent between two distinct on-path messages.\footnote{In \citet{CS82} and \citet{melumad1991communication}, all equilibria are outcome equivalent to sender-minimal equilibria. More generally, all equilibria are sender minimal when there is a complete order over messages under which higher messages are infinitesimally more costly for all sender types.}

\begin{claim}
\label{cheaptalk}
If $v^S$ has SSCED (\autoref{cor:sced_char})
then every sender-minimal equilibrium is an interval equilibrium.
\end{claim}
\begin{proof}
From the sender's viewpoint, the lottery over the receiver's decisions that is induced by any message (given any receiver strategy) is independent of $\theta$ because $\psi$ and $\theta$ are independent. The result follows from \autoref{lem:scd_interval_choice}, as sender-minimality implies that one can restrict attention to the sender choosing among 
lotteries that are utility-distinguishable (i.e., if $P$ and $Q$ are equilibrium 
lotteries, then $D_{P,Q}$ is not a zero function).
\end{proof}

\autoref{cheaptalk} relates to \citet{seidmann1990effective}, who first considered an extension of \citet{CS82} to sender uncertainty about the receiver's preferences.  His goal was to illustrate how such uncertainty could facilitate informative communication even when the sender always strictly prefers higher receiver decisions. Example 2 in \citet{seidmann1990effective} constructs a non-interval and sender-minimal equilibrium that is informative. \autoref{cheaptalk} clarifies that the key is a failure of SSCED.

The strict single-crossing property in standard cheap-talk models (e.g., \citet{CS82} and \citet{melumad1991communication}) not only yields interval equilibria, but it also implies that local incentive compatibility is sufficient for global incentive compatibility. This additional tractability also holds under SSCED. Let $\Theta=\{ \theta_i : i \in \mathbb{Z}\}$ such that $\theta_i < \theta_j$ for $i < j$, and $P:\Theta \rightarrow\Delta X$ be a candidate equilibrium outcome (i.e., the distribution of receiver's choices that each sender type induces in equilibrium). Under SSCED, it is sufficient for sender incentive compatibility that $(\forall i\in\mathbb{Z})$ $u(P(\theta_i),\theta_i)\geq \max\{u(P(\theta_{i-1}),\theta_i),u(P(\theta_{i+1}),\theta_i)\}$.

Besides being sufficient, (S)SCED is also necessary to guarantee interval cheap-talk equilibria. Say that $v^S$ \emph{strictly violates} SCED if the expected utility function \hyperref[strictviolation]{strictly violates SCD}, i.e., if there are $P,Q\in \Delta X$ and $\theta_l<\theta_m<\theta_h$ such that $\min\{D_{P,Q}(\theta_l),D_{P,Q}(\theta_h)\}>0>D_{P,Q}(\theta_m)$.

\begin{claim}
\label{flexibility}
Let $\Theta \subseteq \Reals$, $X \equiv \Reals$, $\Psi\subseteq \Reals^2$, and $v^R(x,\theta,\psi)\equiv -(x-\psi_1-\psi_2\theta)^2$. If $v^S$ strictly violates SCED, then for some pair of distributions of $\theta$ and $\psi$ there is a non-interval equilibrium in which each sender type plays its unique best response.
\end{claim}

\begin{proof}
Assume $v^S$ strictly violates SCED and let $P$ and $Q$ be the distributions and $\theta_l<\theta_m<\theta_h$ the types in that definition. In what follows, we can without loss assume $|M|=2$. So let $M\equiv \{m', m''\}$ and consider the sender's strategy
$$\mu(\theta) = 
\begin{cases}
m' & \text{if $\theta \in \{\theta_l, \theta_h\}$}\\
m'' & \text{if $\theta = \theta_m$}. 
\end{cases}
$$

Let $F_{\theta}$ be any distribution with support $\{
\theta_l,\theta_m,\theta_h\}$  and $\theta' \equiv \E_{F_{\theta}}[\theta | \theta\in\{\theta_l, \theta_h\}] \neq \theta_m$. 
Then, the unique best response against $\mu$ for a receiver of type $\psi=(\psi_1, \psi_2)$ is
$$\alpha(m', \psi) = \psi_1 + \psi_2 \theta' \quad \text{and} \quad \alpha(m'', \psi) = \psi_1 + \psi_2 \theta_m.$$
It can be verified that there is a distribution $F_\psi$ such that, from the sender's viewpoint, the message $m'$ leads to the distribution $P$ and the message $m''$ leads to the distribution $Q$, and so $\mu$ is the sender's unique best response.\footnote{Let $x$ and $y$ be random variables with distributions $P$ and $Q$, respectively. Let $F_{\psi}$ be the distribution of a random vector $\psi=(\psi_1, \psi_2)$ defined by
$$
\begin{bmatrix}
\psi_1\\ \psi_2 
\end{bmatrix}
\equiv
\begin{bmatrix}
1 & \theta' \\
1 & \theta_m
\end{bmatrix}^{-1}
\begin{bmatrix}
x \\ y
\end{bmatrix}.$$

As $\psi \sim F_{\psi}$ (i.e., $\psi$ has distribution $F_\psi$),
$$
\begin{bmatrix}
1 & \theta' \\
1 & \theta_m
\end{bmatrix}
\begin{bmatrix}
\psi_1\\ \psi_2 
\end{bmatrix}
=
\begin{bmatrix}
\psi_1 + \psi_2 \theta'\\
\psi_1 + \psi_2 \theta_m
\end{bmatrix}$$
is stochastically equivalent to $(x, y)$. Thus, $\alpha(m', \psi) = \psi_1 + \psi_2 \theta' \sim P$ and $\alpha(m'', \psi) = \psi_1 + \psi_2 \theta_m \sim Q$.}
\end{proof}

The particular specification of $v^R$ in \autoref{flexibility} is not critical; what is important  
is that there be enough flexibility to generate appropriate lotteries from the sender's viewpoint using best responses for the receiver.\footnote{In particular, the result in \autoref{flexibility} holds under the following more general assumptions: $\Theta,A \subseteq \Reals$, the receiver's preferences are represented by $u^R(a,\theta,\psi)=g_1(\psi, a)\theta+g_2(\psi, a)$, and for every $\theta_l<\theta_h$ and $a', a''\in A$, there exists $\psi \in \Psi$ such that $a' \in \argmax_a u^R(a,\psi,\theta_l)$ and $a'' \in \argmax_a u^R(a,\psi,\theta_h)$. The proof is very similar to that of \autoref{flexibility}.}  For example, the result would also hold---more straightforwardly, but less interestingly---if the receiver were totally indifferent over all actions for some preference realization. On the other hand, if $\psi \in \Reals$ and $v^R(x,\theta,\psi)\equiv -(x-\theta-\psi)^2$, then (S)SCED is not necessary, because any pair of lotteries that the sender may face are ranked by first-order stochastic dominance. Strict supermodularity of $v^S(x,\theta)$ then guarantees that all sender-minimal equilibria are interval equilibria; however, strict supermodularity does not imply (S)SCED, as noted in \autoref{sec:SCED}.

In our cheap-talk application it is uncertainty about the receiver's preferences that leads to the sender effectively choosing among lotteries over the receiver's decisions. Similar results could also be obtained when the receiver's preferences are known but communication is noisy, \`{a} la \citet{blume2007noisy}.

\subsection{Observational Learning with Multidimensional Utility}
\label{sec:obslearn}

The classic sequential observational learning  model \citep{Banerjee92,BHW92} considers agents sequentially choosing products with uncertainty about product values but learning from predecessors' choices.

For concreteness, suppose that students $t=1, 2, \dots$ sequentially purchase laptops before entering an engineering school. Each laptop is identified by its attribute vector $a\equiv(a_1,\ldots,a_n)\in [0,1]^n$, which consists of processing power, memory, screen size, price, etc. The students are uncertain about the nature of the work required to complete the degree, which is captured by a state $\theta \in \Theta \subset \Reals$, where $\Theta$ is countable. The state affects how students weigh each attribute of a laptop. Specifically, following \autoref{eg:MA}, students have a common multidimensional utility function $u(a, \theta) \equiv \sum_{i=1}^n g_i(a_i) f_i(\theta)$. Each $g_i$ has a convex range, which ensures a convex environment, i.e., Condition \eqref{star}. Students are expected-utility maximizers.

Given a finite set of available laptops $\mathcal A\subseteq [0,1]^n$ and a prior $\mu_0$ over the state, each student chooses a laptop using two information sources. Conditional on the state, student $t$ obtains an independent private signal $s_t$ about $\theta$ (e.g., her own impression from reading syllabi and talking to alumni). A canonical example is \textit{normal information}: $s_t \sim\mathcal{N}(\theta,\sigma^2)$, where $\sigma>0$ is a known constant. Each student also observes all predecessors' choices.

The question is whether there is \emph{adequate learning}, as defined by \citet{KLLR22}: no matter the prior $\mu_0$ and the choice set $\mathcal A$, do students' sequential choices eventually settle on the laptop they would choose if the state were known?\footnote{Here is a precise definition. The full-information expected utility given a belief $\mu \in \Delta \Theta$ is the expected utility under that belief if the state will be revealed before an action is chosen: 
$$
U^*(\mu) \equiv \sum_{\theta \in \Theta} \max_{a\in \mathcal A} u(a, \theta) \mu(\theta). 
$$
Given a prior $\mu_0$ and a strategy profile $\sigma$, student $t$'s utility $U_t$ is a random variable. Let $\E_{\sigma, \mu_0} [U_t]$ be student $t$'s ex-ante expected utility. There is adequate learning at a choice set $\mathcal A$ if for every prior $\mu_0$ and every equilibrium strategy profile $\sigma$, 
$\E_{\sigma, \mu_0} [U_t] \to U^*(\mu_0)$ as $t\to \infty$. There is adequate learning if there is adequate learning at all choice sets.
} Those authors identify SCD as the necessary and sufficient condition on the utility function $u(a, \theta)$ for learning under normal information, and more generally, for information structures that have ``directionally unbounded beliefs''.\footnote{Roughly, an information structure has directionally unbounded beliefs if for every state there exist signals that provide arbitrarily-close-to relative certainty about that state vs.~all lower states, and analogously (potentially different) signals for that state vs.~all higher states. Under normal information, for any state, arbitrarily high signals deliver the former while arbitrarily low signals deliver the latter.} In the current multidimensional utility environment, the following characterization obtains, which we state without proof.

\begin{claim}
\label{claim:obslearning}
There is adequate learning under normal information if and only if $u(a,\theta)$ has \SCDstar, i.e., it has the form 
stated in \autoref{cor:multidim}. 
\end{claim}

The claim implies that adequate learning under normal information obtains if only if students' preferences depend on at most two ``summary attributes'', which are each linear combinations over the original value over attributes, i.e., linear combinations of the functions $g_i(a_i)$. For example, students may trade off ``convenience'', which is an aggregate of price, weight, and size, with ``performance'', which aggregates processing speed, memory, and storage. Learning requires that the convenience and performance aggregators do not vary with the state. A second implication derives from $f^I$ and $f^{II}$ in \autoref{multiform} being ratio ordered. The interpretation is that for learning, students must value one summary attribute more as the state increases, e.g., a higher state indicates more computationally intensive work and hence a higher value of performance relative to convenience.

\subsection{Collective Choice over Lotteries}
\label{sec:voting}

Collective choice over lotteries manifests in many contexts: elections entail uncertainty about what policies elected politicians will implement; and a board of directors may view each candidate for CEO as a probability distribution over firm earnings. \citet{zeckhauser69majority} first pointed out that pairwise-majority comparisons in these settings can be cyclical, even when comparisons over deterministic outcomes are not. Our results shed light on when such difficulties do not arise. 

Consider a finite group of individuals indexed by $i \in \{1,2,\ldots,N\}$, where for simplicity $N$ is odd. The group must choose from a set of lotteries over $X,$ where $X$ is the space of outcomes (political policies, earnings, etc.) with generic element $x$.  
Each individual $i$ has von Neumann Morgenstern utility function $v(x,\theta_i)$, where $\theta_i \in \Theta$ is a preference parameter or $i$'s type. We assume $\Theta$ is completely ordered; without further loss of generality, let $\Theta\subset \Reals$ and $\theta_1\leq \cdots \leq \theta_N$. The expected utility for an individual of type $\theta$ from lottery $P$ is $u(P,\theta) \equiv \int_x v(x,\theta) \mathrm{d} P$. We denote individual $i$'s preference relation over lotteries by $\succeq_i$, with strict component $\succ_i$. 

Define the group's preference relation, $\succeq_{maj}$, over lotteries $P$ and $Q$ by majority rule: $$P \succeq_{maj}Q \text{ if } \left|\{i:P \succeq_i Q\}\right|\geq N/2.$$

\begin{claim}
\label{collective}
If $v$ has SCED (\autoref{cor:sced_char}), 
then the group's preference relation is transitive and coincides with that of individual $(N+1)/2$.
\end{claim}

The claim follows from the arguments of \citet{Rothstein90} and \citet{gans96majority}, but we include a direct proof given how short it is.

\begin{proof}
Let $M\equiv (N+1)/2$. By SCD, (i) if $P \succeq_M Q$ , then $P \succeq_{maj} Q$ because there cannot exist $i<M<j$ such that $Q \succ_i P$ and $Q\succ_j P$; analogously, (ii) if $P \succ_M Q$, then $P \succ_{maj} Q$. Hence, $\succeq_{maj}$ coincides with $\succeq_{M}$.
\end{proof}

\autoref{collective} can be applied to a well-known problem in political economy \citep{shepsle72ambiguity}. For simplicity, we now further assume the policy space is a finite set $X \subset \mathbb{R}$ and 
there is a group of $N$ voters whose types
are their ideal points, i.e., $\{\theta_i\}=\argmax_{x \in \Reals} v(x,\theta_i)$. There are two office-motivated candidates, $L$ and $R$; each $j\in\{L,R\}$ can commit to any lottery from some given set $\mathcal{A}_j \subseteq \Delta X$.  A restricted set $\mathcal{A}_j$ may capture various kinds of constraints; for example, \citet{shepsle72ambiguity} assumed the incumbent candidate could only choose degenerate lotteries. In our setting, what ensures the existence of an equilibrium, and which policy lotteries are offered in an equilibrium?\footnote{More precisely: the two candidates simultaneously choose their lotteries, and each voter then votes for his preferred candidate (assuming, for concreteness, that a voter randomizes between the candidates with equal probability if indifferent). A candidate wins if he receives a majority of the votes. Candidates maximize the probability of winning.  We seek a Nash equilibrium of the game between the two candidates.}

\autoref{collective} implies that if voters' utility functions $v$ have SCED and if voter \mbox{$M\equiv (N+1)/2$} is indifferent between her most-preferred lottery in $\mathcal{A}_L$ and in $\mathcal{A}_R$, 
then there is a unique equilibrium: each candidate offers the best lottery for voter $M$; in particular, both candidates converge to $\delta_{\theta_M}$, the degenerate lottery on $\theta_M$, if that is feasible for both. It bears emphasis that in this case policy convergence at the median ideal point is not driven by all voters being globally ``risk averse'' (e.g., $v(x,\theta)=-(x-\theta)^2$); rather, it is because SCED ensures the existence of a decisive voter whose most-preferred lottery is degenerate.\footnote{An example may be helpful. Let $X=[-1,1]$, $\Theta=\{-1,0,1\}$, and $v(x,\theta)=x \theta + 1/(|x|+1)+1$. The corresponding functions $f_1(\theta)=\theta$ and $f_2(\theta)=1$ are each strictly single crossing from below and strictly ratio ordered. For all $\theta$, $v(\cdot,\theta)$ is maximized at $x=\theta$ but is convex on some subinterval of $X$.}

There is a sense in which SCED is necessary to guarantee that each candidate $j$ will offer the median ideal-point voter's most-preferred lottery from the feasible set $\mathcal{A}_j$.  Suppose $v(x,\theta)$ strictly violates SCED, i.e., there are $P,Q\in \Delta X$ and $\theta_l<\theta_m<\theta_h$ such that $\min\{D_{P,Q}(\theta_l),D_{P,Q}(\theta_h)\}>0>D_{P,Q}(\theta_m)$.  Then, if the population of voters is just $\{l,m,h\}$ and $\mathcal{A}_L= \mathcal{A}_R=\{P,Q\}$, the unique equilibrium is for both candidates to offer lottery $P$, which is voter $m$'s less-preferred lottery.

\section{Discussion}
\subsection{Single Crossing vs.~Monotonicity}
\label{sec:discussion}
\label{sec:KL}

We have characterized when $u:A \times \Theta \to \Reals$ has SCD in a convex environment. SCD is an {ordinal} property. Analogous to the interest in monotonic functions rather than just single-crossing functions, one may also be interested in the following {cardinal} property, which we term \textit{monotonic differences} (MD):
$$
\left(\forall a,a' \in  A\right) \ u(a,\theta)-u(a',\theta) 
\text{ is monotonic  in } \theta.
$$
In a convex environment, we write \MDstar \ analogously to \SCDstar.
\begin{theorem}
\label{char_mdstar}
The function $u: A \times \Theta \to \Reals$ has \MDstar \ if and only if it takes the form 
\begin{equation}
\label{ASCP}
u(a,\theta) = g_1(a)f_1(\theta) + g_2(a) + h(\theta),
\end{equation}
where $f_1$ is monotonic. 
\end{theorem}

Suppose that an expected-utility agent with von Neumann Morgenstern utility $v(x,\theta)$ is choosing among lotteries, so $A \equiv \Delta X$. We say that $v$ has \textit{monotonic expectational differences} (MED) if the expected utility function $u$ has \MDstar. Analogous to the SCED characterization in \autoref{cor:sced_char}, it is straightforward that
the function $v$ has MED if and only if it has the same characterization as given for $u$ in \autoref{char_mdstar}. This MED characterization has largely been obtained by \citet{kushnir2017equivalence} in their study of the equivalence between Bayesian and dominant-strategy implementation. 

A function $u$ with \SCDstar \ has \MDstar \ when the function $f_2$ in the form \eqref{e:funcform} is identically equal to one; $f_1$ and $f_2$ being ratio ordered is then equivalent to $f_1$ being monotonic. 
In general, \SCDstar \ functions need not have \MDstar; however, by \autoref{rmk:convex_comb}, \SCDstar \ functions have \MDstar \  \emph{representations} if $\Theta$ has both a minimum and a maximum.\footnote{Absent this condition, an \SCDstar \ function may not even have an \MDstar \ representation. In the context of expected utility, an earlier version of our paper characterized precisely when such representations do not exist and provided an example \citep*[Appendix G]{KLR19}. \citet[Section 4]{duggan2014majority} also discusses the difficulty of finding SCED functions that do not have MED representations.}
We view this finding as unexpected; outside of convex environments, we are not aware of any general result on when SCD functions have MD representations.

\subsection{Interval Choice and Monotone Comparative Statics}
\label{sec:MCS}

In \autoref{sec:compstats}, we showed that (S)SCD characterizes interval choice.
There is a sense in which interval choice is intimately related to monotone comparative statics holding with respect to \emph{some} order over the choice space. In light of \autoref{lem:scd_interval_choice}, the connection is elucidated below by tying (S)SCD to monotone comparative statics.

Throughout this subsection, we consider an ordered set of alternatives, $(A, \succeq)$. To simplify exposition and avoid dealing with equivalence classes, we will focus on comparative statics for a function $u:A\times \Theta \to \Reals$ such that $A$ is {minimal} (with respect to $u$), i.e., $(\forall a \neq a') (\exists \theta) \, u(a, \theta) \neq u(a', \theta)$.
For any $a, a' \in A$, let $a \vee a'$ and $a \wedge a'$ denote the usual join and meet respectively.\footnote{Alternative $\overline{a} \in A$ is the join (or supremum) of $\{a, a'\}$ if (i) $\overline{a}\succeq a$ and $\overline{a} \succeq a'$, and (ii) if $b\succeq a$ and $b\succeq a'$, then $b \succeq \overline{a}$. The meet (or infimum) of $\{a, a'\}$ is defined analogously.} Neither need exist.  Given any $S, S' \subseteq A$, we say that $S$ dominates $S'$ in the \textit{strong set order}, denoted $S \succeq_{SSO} S'$, if for every $a \in S$ and $a'\in S'$, (i) $a \vee a'$ and $a \wedge a'$ exist, and (ii) $a \vee a' \in S$ and $a \wedge a' \in S'$. It can be verified that $\succeq_{SSO}$ is transitive on non-empty subsets of $(A, \succeq)$.

\begin{definition}
\label{def:MCS}
$u:A \times \Theta \to \Reals$ has \textit{monotone comparative statics (MCS)} on $(A,\succeq)$ if
$$(\forall S \subseteq A) \ (\forall \theta_l \leq \theta_h) \quad \argmax_{a \in S} u(a,\theta_h) \succeq_{SSO} \argmax_{a \in S} u(a,\theta_l).$$
\end{definition}

Our definition of MCS is closely related to but not the same as \citet{milgrom1994monotone}. We take $(A, \succeq)$ to be any ordered set while they require a lattice. We focus only on monotonicity of choice in $\theta$ but require the monotonicity to hold for every subset \mbox{$S\subseteq A$}; \citeauthor{milgrom1994monotone} require monotonicity of choice jointly in the pair $(\theta, S)$, but this implicitly only requires choice monotonicity in $\theta$ to hold for every sublattice $S\subseteq A$.

\label{def:SCDorder}Define binary relations $\succ_{SCD}$ and $\succeq_{SCD}$ on $A$ as follows: $a \succ_{SCD} a'$ if $D_{a,a'}(\theta) \equiv u(a,\theta)-u(a',\theta)$ is single crossing \emph{only} from below; $a \succeq_{SCD} a'$ if either $a \succ_{SCD} a'$ or $a=a'$. It is clear that $\succeq_{SCD}$ is reflexive and anti-symmetric. If $u:A \times \Theta \to \Reals$ has SCD, then $\succeq_{SCD}$ is also transitive. Moreover, given SCD (and minimality), the $\succeq_{SCD}$ order is incomplete only over pairs with ``dominance'': if $a \nsucceq_{SCD} a'$ and vice-versa, then either $(\forall \theta) \ D_{a,a'}(\theta)>0$, or $(\forall \theta) \ D_{a,a'}(\theta)<0$.

\begin{theorem}
\label{prop_mcs}
$u:A \times \Theta \to \Reals$ has monotone comparative statics on $(A,\succeq)$, where $A$ is minimal, if and only if $u$ has SCD and $\succeq$ is a refinement of $\succeq_{SCD}$.
\end{theorem}

Our \hyperref[def:SCD]{definition of SCD} does not require an order on the set of alternatives, whereas MCS does. \autoref{prop_mcs} says that SCD is necessary and sufficient for there to exist an order that yields MCS. Moreover, the theorem justifies viewing $\succeq_{SCD}$ as the prominent order for MCS: MCS does not hold with any order that either coarsens $\succeq_{SCD}$ or reverses a ranking by $\succ_{SCD}$. The argument for necessity in \autoref{prop_mcs} only makes use of binary choice sets. If SCD fails, then there is no order $\succeq$ for which there is choice monotonicity for all binary choice sets. If SCD holds, then choice monotonicity on all binary choice sets requires $\succeq$ to refine $\succeq_{SCD}$. For sufficiency in \autoref{prop_mcs}, we show that for any choice set, $\succeq_{SCD}$ is a complete order among the set of alternatives that are chosen by some type; thereafter, we appeal to \citet[Theorem 4]{milgrom1994monotone}.

In Supplementary \autoref{sec:monotone_selection}, we state and prove an analog of \autoref{prop_mcs} for monotone selection, i.e., that every selection from the choice correspondence is increasing; similarly to \citet[][Theorem 4']{milgrom1994monotone}, this result uses SSCD rather than SCD.

\section{Conclusion}
\label{sec:conclusion}
This paper has characterized the class of utility functions that have \SCDstar, i.e., single-crossing differences in a convex environment (\autoref{char_scd_f}). Our notion of SCD does not presume an order on the choice space and is the appropriate notion for interval-choice comparative statics (\autoref{lem:scd_interval_choice}). We have given a number of examples of convex environments, most notably expected utility, rank-dependent expected utility, and multidimensional utility, and discussed the implications of our characterization in these contexts. The applications in \autoref{sec:applications} illustrate how \SCDstar \ is useful in economic problems.

While we have emphasized interval choice, it bears noting that there are other familiar implications of single-crossing properties that in fact rely only on our order-independent notion of (S)SCD; for example, SSCD is the key to guaranteeing that local incentive compatibility implies global incentive compatibility.\footnote{We mean in a sense analogous to \citet[Proposition 4]{Carroll2012}. While \citeauthor{Carroll2012} establishes his result by defining a single-crossing property with respect to some given order over alternatives, essentially the same logic applies with our order-independent notion of strict SCD (\autoref{def:SCD}).} 

We close by mentioning some avenues for future research. First, it may be of interest to characterize exactly when preferences have a utility representation that satisfies our convexity condition \eqref{star}. Second, and relatedly, one may explore characterizations of SCD outside convex environments. In particular, we are intrigued by the question of when SCD preferences do \emph{not} possess a representation that takes a similar form to that we have characterized.

Third, our results have direct bearing on problems in which all types of an agent face the same choice set. Such situations are natural. But consider a variation of the cheap-talk application (\autoref{sec:cheaptalk}) in which the sender's type is correlated with the receiver's type.  Even though the receiver's type does not affect the sender's payoff, different sender types will generally have different beliefs about the distribution of the receiver's action that any message induces in equilibrium. Effectively, different sender types will be choosing from different choice sets. An approach that synthesizes the current paper's with that of, for example, \citepos{athey2002mcs} may be useful for such problems.

\newpage
\appendix
\addappheadtotoc 
\section*{Appendices}


\addtocontents{toc}{\protect\setcounter{tocdepth}{-1}} 

\paragraph{Organization.} 

\appendixref{compstats_proofs} contains proofs for our comparative statics results (\autoref{lem:scd_interval_choice} and  \autoref{prop_mcs}); \appendixref{main_proofs} for aggregation of single crossing functions (\autoref{equiv_cond} and \autoref{equiv_cond_general}) and our characterizations of \SCDstar \ (\autoref{char_scd_f} and \autoref{rmk:convex_comb}); and \appendixref{sec:proof_med} for our \MDstar \ characterization (\autoref{char_mdstar}). Some additional material is in the \hyperref[supp_app]{Supplementary Appendices}.

\paragraph{A preliminary result.} Before turning to the proofs, we state a useful equivalence with the violation of single crossing; the result is obvious when $(\Theta, \leq)$ is a completely ordered set but also applies when it is not.
\begin{claim}
\label{claim_reversal}
A function $f:\Theta \to \Reals$ is not single crossing if and only if for some $\theta_l < \theta_m < \theta_h$, either
\begin{align}
& \sign[f(\theta_l)] < \sign[f(\theta_m)] \text{ and } \sign[f(\theta_m)] > \sign[f(\theta_h)], \quad \text{or}
\label{no_scd_cond1}\\
& \sign[f(\theta_l)] >\sign[f(\theta_m)] \text{ and } \sign[f(\theta_m)] < \sign[f(\theta_h)].
\label{no_scd_cond2}
\end{align}
\end{claim}

\begin{proof}[Proof of \autoref{claim_reversal}]
The ``if'' direction of the claim is immediate. For the ``only if'' direction, suppose $f: \Theta \to \Reals$ is single crossing neither from below nor from above:
\[\begin{split}
&(\exists \theta_1 < \theta_2 ) \quad \sign[f(\theta_1)] <\sign[f(\theta_2)],\text{ and }\\
&(\exists \theta_3 < \theta_4 ) \quad \sign[f(\theta_3)] > \sign[f(\theta_4)].
\end{split}\]
Let $\Theta_0 \equiv \{\theta_1, \theta_2, \theta_3, \theta_4\}$ and $\bar{\theta}$ and $\underline{\theta}$ be upper and lower bounds of $\Theta_0$. If $f(\underline{\theta})=f(\overline{\theta})=0$, then $(\theta_l, \theta_m, \theta_h) = (\underline{\theta}, \theta_0, \overline{\theta})$ for some $\theta_0 \in \Theta_0$ with $f(\theta_0) \neq 0$ satisfies either \eqref{no_scd_cond1} or \eqref{no_scd_cond2}. So assume $f(\bar{\theta}) \neq 0$; an similar argument applies if $f(\underline{\theta}) \neq 0$. If $f(\bar{\theta}) < 0$, then $(\theta_l, \theta_m, \theta_h) = (\theta_1, \theta_2, \bar{\theta})$ satisfies \eqref{no_scd_cond1}. If $f(\bar{\theta}) > 0$, then $(\theta_l, \theta_m, \theta_h) = (\theta_3, \theta_4, \bar{\theta})$ satisfies \eqref{no_scd_cond2}. 
\end{proof}

\section{Proofs for Comparative Statics}
\label{compstats_proofs}

\subsection{Proof of \autoref{lem:scd_interval_choice}}
\label{interval_choice_proofs}
\textbf{Part 1.}\ Suppose $u$ has SCD, and consider any $S\subseteq A$, $a' \in S$, and $\theta_l, \theta_h \in \{ \theta : a' \in \argmax_{a \in S} u(a, \theta)\}$ with $\theta_l < \theta_h$. For any $a'' \in S$, $D_{a',a''}(\theta_l) \geq 0$ and $D_{a', a''}(\theta_h) \geq 0$, which imply that $D_{a', a''} (\theta_m) \geq 0$ for all $\theta_m$ with $\theta_l < \theta_m < \theta_h$. It follows that $\{ \theta : a' \in \argmax_{a \in S} u(a, \theta)\}$ is an interval. 

If $u$ strictly violates SCD, $\exists a', a'' \in A$ and $\theta_l < \theta_m < \theta_h$ such that 
$\min\{D_{a', a''}(\theta_l), D_{a', a''}(\theta_h)\}$ $> 0 > D_{a', a''}(\theta_m)$. Clearly,  $\{\theta : a' \in \argmax_{a \in \{a', a''\}} u(a, \theta)\}$ is not an interval. 

\textbf{Part 2.}\ 
Suppose $|\Theta| \geq 3$. A function $u: A \times \Theta \to \mathbb{R}$ does not have SSCD when there exist $a', a'' \in A$ such that $D_{a', a''}$ (and $D_{a'', a'}$) are not strictly single crossing. Alternatively, using \autoref{claim_reversal}, $u$ does not have SSCD if and only if $(\exists a', a'')$ $(\exists \theta_l < \theta_m < \theta_h)$ $D_{a', a''}(\theta_l) \geq 0$, $D_{a', a''}(\theta_m) \leq 0$, and $D_{a', a''}(\theta_h) \geq 0$. This condition is equivalent to $(\exists S \subseteq A \, \text{with}\, a', a'' \in S)$ $(\exists \theta_l < \theta_m < \theta_h) $ $a' \in \bigcap_{\theta \in \{\theta_l,\theta\}}\argmax_{a \in S} u(a, \theta)$ and $D_{a', a''}(\theta_m) \leq 0$, which holds if and only if some selection from the choice correspondence $C_u$ does not have interval choice.

\subsection{Proof of \autoref{prop_mcs}}
\label{Proof_prop_mcs}
\paragraph{$(\implies)$}
Suppose $u:A \times \Theta \to \Reals$ has MCS on $A$ with some order $\succeq$. We first prove the following claim. 
\begin{claim}\label{size2_subset}
For every $a', a'' \in A$, 
if $\exists \theta_l < \theta_h$ such that $\sign[D_{a',a''}(\theta_l)] < \sign[D_{a', a''}(\theta_h)]$, then $a' \succ a''$.
\end{claim}

\begin{proof}
Consider $S = \{a', a''\}$. Since $\sign[D_{a',a''}(\theta_l)] \neq  \sign[D_{a', a''}(\theta_h)]$, we have 
$$\argmax_{a \in S} u(a, \theta_l) \neq \argmax_{a \in S} u(a, \theta_h).$$
Thus, either (i) $a' \in \argmax_{a \in S} u(a, \theta_l)$ and $a'' \in \argmax_{a \in S} u(a, \theta_h)$, or (ii) $a'' \in \argmax_{a \in S} u(a, \theta_l)$ and $a' \in \argmax_{a \in S} u(a, \theta_h)$. Since $u$ has MCS on $(A, \succeq)$, we have $\argmax_{a \in S } u(a, \theta_h) \succeq_{SSO} \argmax_{a \in S} u(a, \theta_l)$. 
Therefore, $a' \wedge a'' \in \argmax_{a \in S} u(a, \theta_l)$ and $a' \vee a'' \in \argmax_{a \in S} u(a, \theta_h)$, which implies that either $a' \succeq a''$ or $a'' \succeq a'$. Since $a' \neq a''$, we have either $a' \succ a''$ or $a'' \succ a'$. If $a'' \succ a'$, then $a'' = a' \vee a'' \in \argmax_{a \in S} u(a, \theta_h)$, contradicting $\sign[D_{a',a''}(\theta_l)] <  \sign[D_{a', a''}(\theta_h)]$. Thus, $a' \succ a''$.
\end{proof}

To show that $u$ has SCD on $A$, suppose not, towards contradiction. \autoref{claim_reversal} implies there exist $a', a'' \in A$ and $\theta_l<\theta_m<\theta_h$ such that either 
\begin{align}
&\sign [D_{a',a''}(\theta_l)]<\sign [D_{a',a''}(\theta_m)] \quad \text{and} \quad \sign [D_{a',a''}(\theta_m)] > \sign [D_{a',a''}(\theta_h)], \quad \text{or} \label{scd_violation1}\\
&\sign [D_{a', a''}(\theta_l)] > \sign [D_{a', a''}(\theta_m)] \quad \text{and} \quad \sign [D_{a', a''}(\theta_m)] < \sign [D_{a',a''}(\theta_h)].\label{scd_violation2}
\end{align}
Given either \eqref{scd_violation1} or \eqref{scd_violation2}, \autoref{size2_subset} implies $a' \succ a''$ and $a'' \succ a'$, a contradiction.

To show that $\succeq$ is a refinement of $\succeq_{SCD}$, it suffices to show that
\begin{equation}
\label{order_refinement_strict}
(\forall a', a'' \in A) \quad a' \succ_{SCD} a'' \implies a' \succ a'',
\end{equation}
because both $\succeq$ and $\succeq_{SCD}$ are anti-symmetric.  Take any $a', a'' \in A$ such that $a' \succ_{SCD} a''$. As $D_{a', a''}$ is single crossing only from below, $\exists \theta_l<\theta_h$ such that $\sign [D_{a',a''}(\theta_l)]<\sign [D_{a',a''}(\theta_h)]$. \autoref{size2_subset} implies $a' \succ a''$, which proves \eqref{order_refinement_strict}.

\paragraph{$(\impliedby)$}
Suppose that $u: A \times \Theta \to \Reals$ has SCD, and $\succeq$ is a refinement of $\succeq_{SCD}$. For any $S \subseteq A$, define $C_u(S) \equiv \bigcup_{\theta \in \Theta} \argmax_{a \in S} u(a, \theta)$. It is clear that 
$$(\forall \theta) \quad \argmax_{a \in S} u(a, \theta) = \argmax_{a \in C_u(S)} u(a, \theta).$$

We claim that $C_u(S)$ is completely ordered by $\succeq_{SCD}$. To see why, take any pair $a', a'' \in C_u(S)$ with $a' \neq a''$. As $u$ has SCD, $D_{a', a''}$ is single crossing in $\theta$. As $A$ is minimal, $D_{a', a''}$ is not a zero function. Also, as $a', a'' \in C_u(S)$, $\sign[D_{a', a''}]$ is not a constant function with value either 1 or -1. Thus, $D_{a', a''}$ is single crossing either only from below, or only from above. It follows that $a' \succ_{SCD} a''$ or $a'' \succ_{SCD} a'$. 

Since $\succeq$ is a refinement of $\succeq_{SCD}$, $\succeq$ coincides with $\succeq_{SCD}$ on $C_u(S)$, and the strong set orders generated by $\succeq$ and $\succeq_{SCD}$ on the collection of all subsets of $C_u(S)$ also coincide. By definition of $\succeq_{SCD}$, when restricted to $C_u(S)\times \Theta$, $u$ satisfies 
\citepos{milgrom1994monotone} single-crossing property in $(a, \theta)$ with respect to $\succeq_{SCD}$ and $\leq$.\footnote{\citeauthor{milgrom1994monotone}'s single-crossing property 
of $u$ on $C_u(S)\times \Theta$
is equivalent to $$(\forall a' \succ_{SCD} a'') (\forall \theta_l < \theta_h) \quad u(a', \theta_l) \geq (>) u(a'', \theta_l) \implies u(a', \theta_h) \geq (>) u(a'', \theta_h).$$} As $C_u(S)$ is completely ordered by $\succeq_{SCD}$, it follows from \citet[Theorem 4]{milgrom1994monotone} that $\forall \theta_l < \theta_h$, 
$$\argmax_{a \in S} u(a, \theta_h) = \argmax_{a \in C_u(S)} u(a, \theta_h)  \succeq_{SSO} \argmax_{a \in C_u(S)} u(a, \theta_l) = \argmax_{a \in S} u(a, \theta_l).
$$

\section{Proofs for Aggregating Single-Crossing Functions and Main Characterizations}
\label{main_proofs}

Before providing the proofs in this appendix, we first clarify Condition \eqref{e:RD_add} in the definition of ratio dominance.

\subsection{On the Definition of Ratio Dominance}
\label{app:RD_add}

\begin{figure}[h!]
\begin{center}
    \subfigure[Failure of $\implies$.]{\includegraphics[width=2.5in]{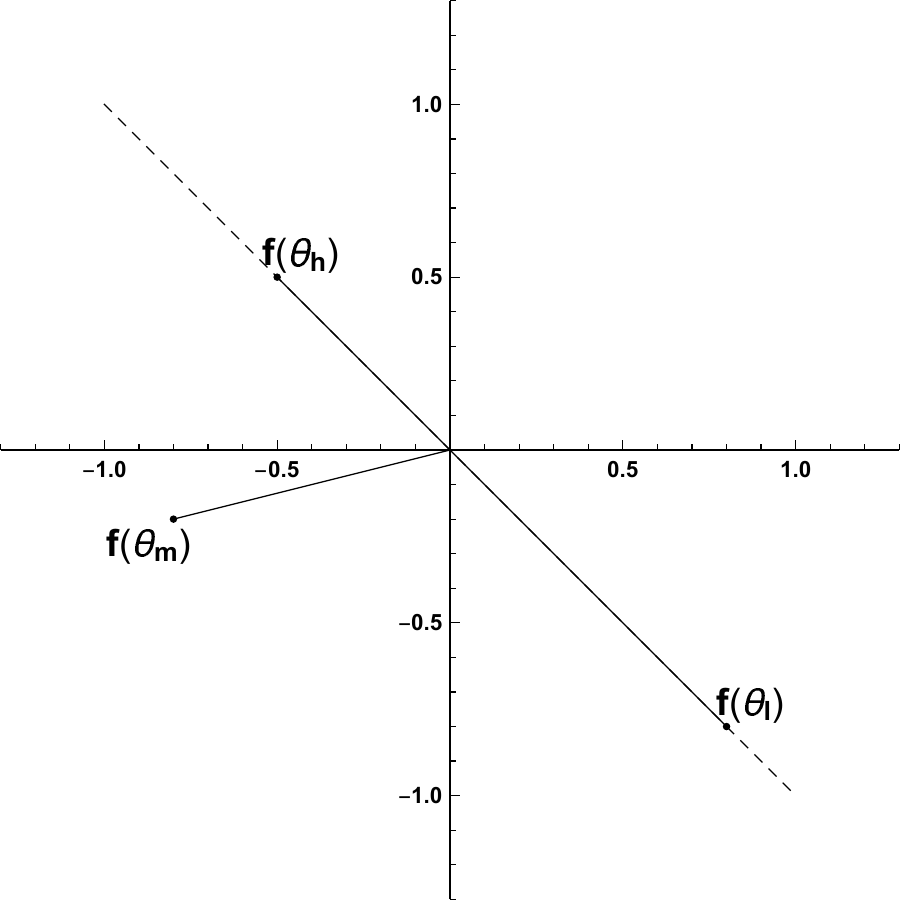}
    \label{fig:not_ordered3}}
    \qquad
    \subfigure[Failure of $\impliedby$.]{\includegraphics[width=2.5in]{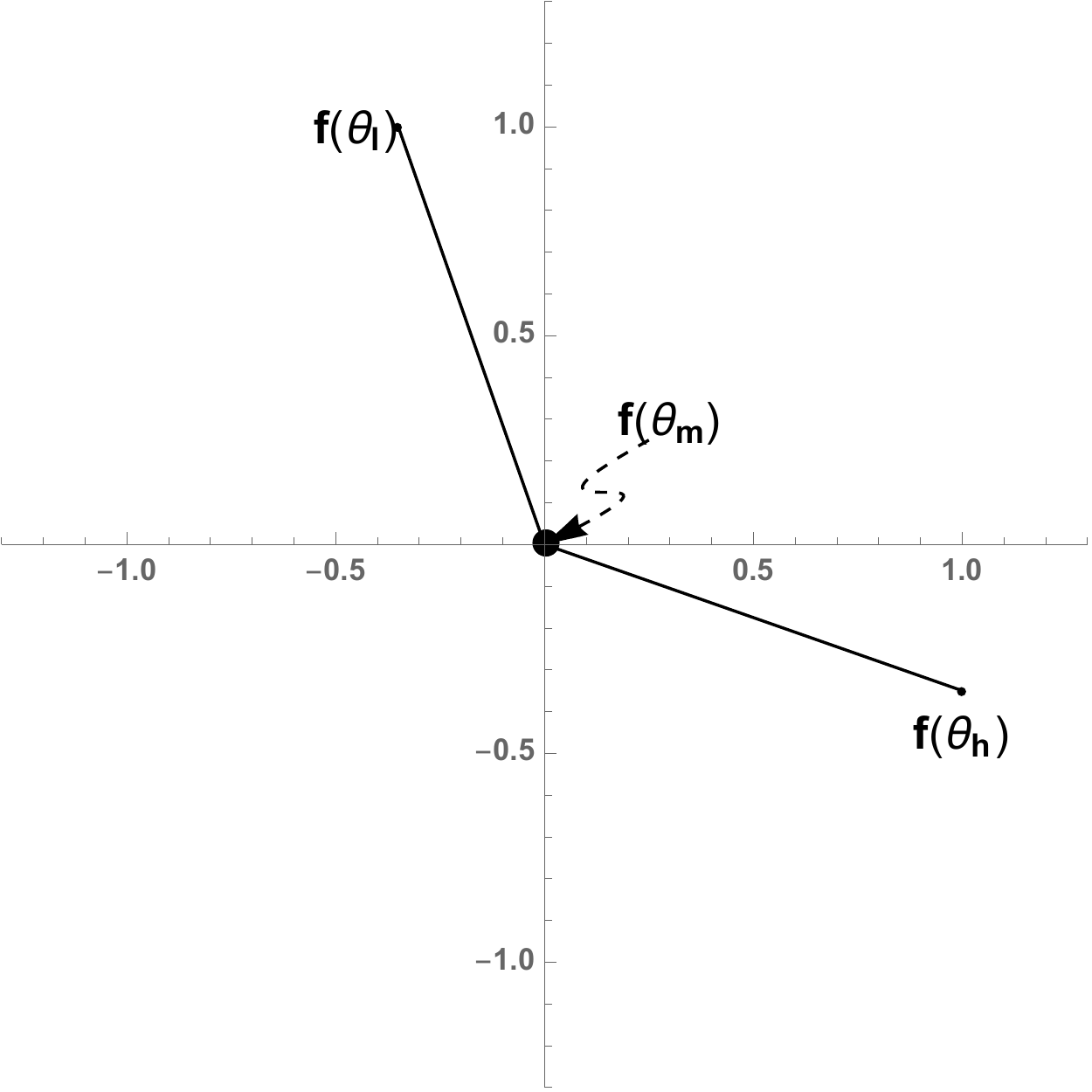}
    \label{fig:not_ordered4}}
\caption{$f_1$ and $f_2$ are not ratio ordered because Condition \eqref{e:RD_add} fails for $\theta_l<\theta_m<\theta_h$.}
\label{fig:not_ordered}
\end{center}
\end{figure}

\autoref{sec:main} in the main text explained Condition \eqref{e:RD} in the definition of ratio dominance. We impose Condition \eqref{e:RD_add} to rule out cases in which, for some $\theta_l < \theta_m < \theta_h$, either (i) $f(\theta_l)$ and $f(\theta_h)$ are collinear in opposite directions while $f(\theta_m)$ is not, or (ii) $f(\theta_l)$ and $f(\theta_h)$ are non-zero vectors while $f(\theta_m)$ is not. See \autoref{fig:not_ordered}, wherein panel \subref{fig:not_ordered3} depicts case (i) and panel \subref{fig:not_ordered4} depicts case (ii). Note that Condition \eqref{e:RD} is satisfied in both panels.

It turns out that for all linear combinations of two single-crossing functions $f_1$ and $f_2$ to be single crossing, Condition \eqref{e:RD_add} of ratio ordering is required; see the \hyperref[sec:proof_aggregation_lemma]{proof} of \autoref{equiv_cond}. \autoref{fig:not_ordered} illustrates: 
in panel \subref{fig:not_ordered1}, $(f_1 + f_2)(\theta_l) = (f_1 + f_2) (\theta_h) = 0$ while $(f_1 + f_2)(\theta_m)<0$; in panel \subref{fig:not_ordered2}, $(f_1 + f_2)(\theta_l)>0$ and $(f_1 + f_2) (\theta_h) > 0$ while $(f_1 + f_2)(\theta_m)=0$. So in both cases, $f_1$ and $f_2$ are each single crossing but $f_1+f_2$ is not. 

\subsection{Proof of \autoref{equiv_cond}}

\label{sec:proof_aggregation_lemma}
When $|\Theta| \leq 2$, the proof is trivial as all functions are single crossing and every pair of functions are ratio ordered. Hereafter, we assume $|\Theta| \geq 3$. 

\paragraph{$(\implies)$}
It is clear that each function $f_1$ and $f_2$ is single crossing. We must show that $f_1$ and $f_2$ are ratio ordered.

To prove \eqref{e:RD}, we suppose towards contradiction that 
\begin{equation}\begin{split}
& (\exists \theta_l < \theta_h) \quad f_1(\theta_l)f_2(\theta_h) < f_1(\theta_h)f_2(\theta_l), \text{ and }\\
& (\exists \theta' < \theta'') \quad f_1(\theta')f_2(\theta'') > f_1(\theta'')f_2(\theta').
\end{split}
\label{e:prof_prop1}
\end{equation}
Take any upper bound $\overline{\theta}$ of $\{\theta_l, \theta_h, \theta', \theta''\}$.

First, let $\alpha_l \equiv (f_2(\theta_l), -f_1(\theta_l))$. Then $(\alpha_l \cdot f)(\theta_l) = (f_2(\theta_l), -f_1(\theta_l)) \cdot (f_1(\theta_l), f_2(\theta_l)) = 0$, and $(\alpha_l \cdot f) (\theta_h) > 0$. Thus, $\alpha_l \cdot f$ is single crossing from below and $(\alpha_l \cdot f) (\overline{\theta}) > 0$. 

Second, let $\alpha' \equiv (f_2(\theta'), -f_1(\theta'))$. Then $(\alpha' \cdot f)(\theta') =0$ and $(\alpha' \cdot f) (\theta'') < 0$. Thus, $\alpha' \cdot f$ is single crossing from above and $(\alpha' \cdot f) (\overline{\theta}) < 0$.

Let $\overline{\alpha} = (f_2(\overline{\theta}), -f_1(\overline{\theta}))$. It follows that
\[\begin{split}
& (\overline{\alpha}\cdot f)(\theta_l) = (f_2(\overline{\theta}), -f_1(\overline{\theta}))\cdot (f_1(\theta_l), f_2(\theta_l)) = - (\alpha_l \cdot f) (\overline{\theta}) < 0,\\
& (\overline{\alpha}\cdot f)(\theta') = - (\alpha' \cdot f) (\overline{\theta}) > 0, \text{ and }\\
& (\overline{\alpha}\cdot f)(\overline{\theta}) =0.\\
\end{split}\]
Therefore, $\overline{\alpha} \cdot f$ is not single crossing, a contradiction.

To prove \eqref{e:RD_add}, take any $\theta_l < \theta_m <\theta_h$. 

First, we show that $f_1(\theta_l)f_2(\theta_h)= f_1(\theta_h)f_2(\theta_l)$ implies $f_1(\theta_m)f_2(\theta_h) = f_1(\theta_h)f_2(\theta_m)$ and $f_1(\theta_m)f_2(\theta_l) = f_1(\theta_l)f_2(\theta_m)$. Assume $f_1$ is  not a zero function on $\{\theta_l, \theta_m, \theta_h\}$, as otherwise the proof is trivial. Since $f_1$ is single crossing, either $f_1(\theta_l) \neq 0$ or $f_1(\theta_h) \neq 0$. We consider the case of $f_1(\theta_h) \neq 0$ (and omit the proof for the other case, as it is analogous). Let $\alpha_h \equiv (f_2(\theta_h), -f_1(\theta_h))$. Since $\alpha_h \cdot f$ is single crossing and $(\alpha_h \cdot f)(\theta) =0$ for $\theta = \theta_l, \theta_h$, it holds that $(\alpha_h \cdot f)(\theta_m) = f_2(\theta_h)f_1(\theta_m) - f_1(\theta_h)f_2(\theta_m) =0$. It follows immediately that $f_1(\theta_m)f_2(\theta_h) = f_1(\theta_h)f_2(\theta_m)$.
As $(f_1(\theta_m), f_2(\theta_m))$ and $(f_1(\theta_h), f_2(\theta_h))$ are linearly dependent and $(f_1(\theta_h), f_2(\theta_h))$ is a non-zero vector, there exists $\lambda\in\mathbb{R}$ such that $f_i(\theta_m)=\lambda f_i(\theta_h)$ for $i=1,2$. Thus, 
$$f_1(\theta_l)f_2(\theta_m) =  \lambda f_1(\theta_l)  f_2(\theta_h) = \lambda f_2(\theta_l)f_1(\theta_h) = f_2(\theta_l)f_1(\theta_m).$$

Next, we show that if $f_1(\theta_l)f_2(\theta_m) = f_1(\theta_m)f_2(\theta_l)$ and $f_1(\theta_m)f_2(\theta_h) = f_1(\theta_h)f_2(\theta_m)$, then $f_1(\theta_l)f_2(\theta_h)= f_1(\theta_h)f_2(\theta_l)$. Let $\alpha \equiv (f_2(\theta_l)-f_2(\theta_h),-f_1(\theta_l)+f_1(\theta_h))$. It follows that
\begin{align*}
(\alpha \cdot f)(\theta_l) & =
\left(f_2(\theta_l) - f_2(\theta_h)\right)f_1(\theta_l) - \left(f_1(\theta_l) - f_1(\theta_h)\right)f_2(\theta_l)=f_1(\theta_h) f_2(\theta_l)-f_1(\theta_l)f_2(\theta_h),\\ 
(\alpha \cdot f)(\theta_h)& =\left(f_2(\theta_l) - f_2(\theta_h)\right)f_1(\theta_h) - \left(f_1(\theta_l) - f_1(\theta_h)\right)f_2(\theta_h)=f_1(\theta_h) f_2(\theta_l)-f_1(\theta_l)f_2(\theta_h), \  \text{and}\\
(\alpha \cdot f)(\theta_m) & = \left(f_2(\theta_l) - f_2(\theta_h)\right)f_1(\theta_m) - \left(f_1(\theta_l) - f_1(\theta_h)\right)f_2(\theta_m)=0.
\end{align*}
As $\alpha \cdot f$ is single crossing, it follows that $(\alpha \cdot f)(\theta_l) = (\alpha \cdot f)(\theta_h) =0$, as we wanted to show.

\paragraph{$(\impliedby)$}
\label{subsection:sc_violation}
Assume that $f_1$ and $f_2$ are each single crossing. 
We prove the result for the case in which $f_1$ ratio dominates $f_2$;
the other case is analogous. For any $\alpha \in \Reals^2$, we prove that $\alpha \cdot f$ is single crossing. We may assume that $\alpha \neq 0$, as the result is trivial otherwise.

Suppose, towards contradiction, that $\alpha \cdot f$ is not single crossing. \autoref{claim_reversal} implies there exist $\theta_l < \theta_m < \theta_h$ such that either
\begin{align}
& \sign[(\alpha \cdot f)(\theta_l)] < \sign[(\alpha \cdot f)(\theta_m)] \text{ and } \sign[(\alpha \cdot f)(\theta_m)] > \sign[(\alpha \cdot f)(\theta_h)], \quad \text{or}
\label{assumption1_in_claim2}\\
& \sign[(\alpha \cdot f)(\theta_l)] >\sign[(\alpha \cdot f)(\theta_m)] \text{ and } \sign[(\alpha \cdot f)(\theta_m)] < \sign[(\alpha \cdot f)(\theta_h)].
\label{assumption2_in_claim2}
\end{align}

First, we consider the case in which $f(\theta)\equiv (f_1(\theta), f_2(\theta))$ for all $\theta \in \{\theta_l, \theta_m, \theta_h\}$ are non-zero vectors. Take any $\theta_1, \theta_2 \in \{\theta_l, \theta_m, \theta_h\}$ such that $\theta_1 < \theta_2$. As $f_1$ ratio dominates $f_2$, by Condition \eqref{e:RD}, $f(\theta_1)$ moves to $f(\theta_2)$ in a clockwise rotation with an angle less than or equal to 180 degrees. Let $r_{12}$ be the clockwise angle from $f(\theta_1)$ to $f(\theta_2)$. 
The vector $\alpha \neq 0$ defines a partition of $\Reals^2$ into $\Reals^2_{\alpha, +} \equiv \{ x \in \Reals^2 \,:\, \alpha \cdot x > 0\}$, $\Reals^2_{\alpha, 0} \equiv \{ x \in \Reals^2 \,:\, \alpha \cdot x = 0\}$, and $\Reals^2_{\alpha, -} \equiv \{ x \in \Reals^2 \,:\, \alpha \cdot x < 0\}$. In both cases \eqref{assumption1_in_claim2} and \eqref{assumption2_in_claim2}, both $f(\theta_l)$ and $f(\theta_h)$ are not in the same part of the partition that $f(\theta_m)$ belongs to. Thus, $r_{lm} > 0$ and $r_{mh} > 0$. On the other hand, both $f(\theta_l)$ and $f(\theta_h)$ are in the same closed half-space, either $\Reals^2_{\alpha, +} \cup \Reals^2_{\alpha, 0}$ or $\Reals^2_{\alpha, -} \cup \Reals^2_{\alpha, 0}$, and $f(\theta_m)$ is in the other closed half-space, either $\Reals^2_{\alpha, -} \cup \Reals^2_{\alpha, 0}$ or $\Reals^2_{\alpha, +} \cup \Reals^2_{\alpha, 0}$, respectively. Thus, $r_{lh} \geq 180$. Since Condition \eqref{e:RD} implies $r_{lh}\leq 180$,  it follows that $r_{lh}=180$. Hence, $f(\theta_l)$ and $f(\theta_m)$ are linearly independent ($0< r_{lm} <180$), and similarly for $f(\theta_m)$ and $f(\theta_h)$. However, $f(\theta_l)$ and $f(\theta_h)$ are linearly dependent ($r_{lh}=180$). This contradicts \eqref{e:RD_add}.

Second, suppose either $f(\theta_l) = 0$ or $f(\theta_h) = 0$. We provide the argument assuming $f(\theta_l)=0$; it is analogous if $f(\theta_h) = 0$. Under either \eqref{assumption1_in_claim2} or \eqref{assumption2_in_claim2}, $f(\theta_m) \neq 0$. By Condition \eqref{e:RD_add}, $f(\theta_m)$ and $f(\theta_h)$ are linearly dependent. In particular, because $f(\theta_m) \neq 0$, there exists a unique $\lambda \in \Reals$ such that $f(\theta_h) = \lambda f(\theta_m)$. Under either \eqref{assumption1_in_claim2} or \eqref{assumption2_in_claim2}, $\lambda \leq 0$, which contradicts the hypothesis that $f_1$ and $f_2$ are single crossing. 

Last, suppose $f(\theta_l) \neq 0$, $f(\theta_m)=0$, and $f(\theta_h) \neq 0$. By Condition \eqref{e:RD_add}, $f(\theta_l)$ and $f(\theta_h)$ are linearly dependent. Hence, there exists a unique $\lambda \in \Reals$ such that $f(\theta_l) = \lambda f(\theta_h)$. Under either \eqref{assumption1_in_claim2} or \eqref{assumption2_in_claim2}, $\lambda > 0$, which contradicts the hypothesis that $f_1$ and $f_2$ are single crossing. 

\subsection{Proof of \autoref{equiv_cond_general}}
\label{Proof_prop1}

The result is trivial if $|\Z|=1$ and it is equivalent to \autoref{equiv_cond} if $|\Z|=2$, so we may assume $|\Z| \geq 3$. The proof is also straightforward if all functions $f(\z,\cdot): \Theta \to \mathbb{R}$ are multiples of one function $f(\z_1,\cdot)$, i.e., if there is $\z_1$ such that $(\exists \lambda: \Z \to \Reals) (\forall \z) f(\z,\cdot) = \lambda(\z) f(\z_1,\cdot)$. Thus, we further assume there exist $\z', \z''$ such that $f(\z', \cdot):\Theta\to \Reals$ and $f(\z'', \cdot):\Theta\to \Reals$ are linearly independent.

\paragraph{$(\impliedby)$}
Assume $f(\z_1,\cdot)$ and $f(\z_2,\cdot)$ are (i) each single crossing and (ii) ratio ordered, and that there are functions $\lambda_1,\lambda_2:\Z \to \Reals$ 
such that $(\forall \z)$ $f(\z,\cdot) = \lambda_1(\z) f(\z_1,\cdot) + \lambda_{2}(\z)f(\z_2,\cdot)$.  Then, for any function $\mu: \Z \to \mathbb{R}$ with finite support,
\begin{align*}
\int_z f(\z,\theta) \mathrm{d} \mu & = \int_z \left(
\lambda_{1}(\z)f(\z_1,\theta) + \lambda_{2}(\z)f(\z_2,\theta) \right) \mathrm{d} \mu = \sum_{i=1,2} \left( \int_z \lambda_{i}(\z) \mathrm{d} \mu\right) f(\z_i,\theta),
\end{align*}
which is single crossing in $\theta$ by \autoref{equiv_cond}.

\paragraph{$(\implies)$}
Take any $\z_1, \z_2 \in \Z$ such that $f_1(\cdot)\equiv f(\z_1, \cdot)$ and $f_2(\cdot) \equiv f(\z_2, \cdot)$ are linearly independent. Then, by \autoref{equiv_cond}, $f_1$ and $f_2$ are each single crossing and ratio ordered, as their linear combinations are all single crossing.

For every $\theta', \theta''$, let 
$$
M_{\theta', \theta''} \equiv 
\begin{bmatrix}
f_{1}(\theta') & f_{2}(\theta') \\
f_{1}(\theta'') & f_{2}(\theta'')
\end{bmatrix}.
$$
We first prove the following claim:
\begin{claim}
\label{claim_full_rank}
There exists $\theta_l < \theta_h$ such that $\rank[M_{\theta_l,\theta_h}]=2$.
\end{claim}

\begin{proof}[Proof of \autoref{claim_full_rank}]
As $f_1$ and $f_2$ are linearly independent, there exists $\theta_0$ such that $f_2(\theta_0) \neq 0$. Let $\lambda \equiv -\frac{f_1(\theta_0)}{f_2(\theta_0)}$. Then, for some $\theta_{\lambda}$, $f_1(\theta_{\lambda}) + \lambda f_2(\theta_{\lambda}) \neq 0$ and $\rank[M_{\theta_0, \theta_{\lambda}}]=2$.

The proof is complete if $\theta_0 > \theta_{\lambda}$ or $\theta_0 < \theta_{\lambda}$. If not, take a lower and upper bound, $\underline{\theta}$ and $\overline{\theta}$, of $\{\theta_0, \theta_{\lambda}\}$. Then $\rank[M_{\underline{\theta},\overline{\theta}}]=2$.
For otherwise, there exists $\alpha \in \mathbb{R}^2\backslash \{0\}$ such that $M_{\underline{\theta},\overline{\theta}} \alpha = 0$. As $\theta_0$ and $\theta_{\lambda}$ are between $\underline{\theta}$ and $\overline{\theta}$, and $\alpha_1f_1 + \alpha_2 f_2$ is single crossing, we have $M_{\theta_0, \theta_{\lambda}} \alpha = 0$, which contradicts $\rank[M_{\theta_0, \theta_{\lambda}}]=2$.
\end{proof}

Now take any $\z\in \Z$, the function $f_{\z}(\cdot) \equiv f(\z,\cdot)$, and $\theta_l, \theta_h$ in \autoref{claim_full_rank}. As \mbox{$\rank[M_{\theta_l, \theta_h}]=2$}, the system
\begin{equation}
\label{e:extending_lin_depen}
\begin{bmatrix}
f_{\z}(\theta_l)\\
f_{\z}(\theta_h)
\end{bmatrix}
=
\begin{bmatrix}
f_1(\theta_l) & f_2(\theta_l) \\
f_1(\theta_h) & f_2(\theta_h) \\
\end{bmatrix}
\begin{bmatrix}
\lambda_1\\
\lambda_2
\end{bmatrix}
\end{equation}
has a unique solution $\lambda \in \mathbb{R}^2$.   We will show that $f_{\z} = \lambda_1 f_1 + \lambda_2 f_2$. 

Suppose, towards contradiction, there exists $\theta_\lambda$ such that
\begin{equation}
\label{e:extending_lin_depend_contra}
f_{\z}(\theta_{\lambda}) \neq \lambda_1 f_1(\theta_{\lambda}) + \lambda_2f_2(\theta_{\lambda}).
\end{equation}
Let $\underline{\theta}$ and $\overline{\theta}$ respectively be a lower and an upper bound of $\{\theta_l, \theta_h, \theta_{\lambda}\}$. 
If $\rank[M_{\underline{\theta}, \overline{\theta}}] < 2$, there is $\lambda' \in \mathbb{R}^2 \backslash \{0\}$ such that $\lambda'_1 f_1(\theta) + \lambda'_2f_2(\theta) = 0$ for $\theta = \underline{\theta}, \overline{\theta}$. As $\lambda'_1 f_1 + \lambda'_2f_2$ is single crossing, we have $\lambda'_1 f_1(\theta) + \lambda'_2f_2(\theta) = 0$ for $\theta = \theta_l, \theta_h$, which contradicts $\rank[M_{\theta_l, \theta_h}]=2$.\footnote{The function $\lambda'_1f_1 + \lambda'_2f_2$ must be single crossing because we can consider $\mu: \Z \to \mathbb{R}$ such that $\mu(\z_1)=\lambda'_1$, $\mu(\z_2)=\lambda'_2$, and $\mu(\z)=0$ for any $\z \neq \z_1, \z_2$. We use similar reasoning subsequently.}

If, on the other hand, $\rank[M_{\underline{\theta}, \overline{\theta}}]=2$, the system
$$
\begin{bmatrix}
f_{\z}(\underline{\theta})\\
f_{\z}(\overline{\theta})
\end{bmatrix}
=
\begin{bmatrix}
f_1(\underline{\theta}) & f_2(\underline{\theta}) \\
f_1(\overline{\theta}) & f_2(\overline{\theta}) \\
\end{bmatrix}
\begin{bmatrix}
\lambda'_1\\
\lambda'_2
\end{bmatrix}
$$
has a unique solution $\lambda' \in \mathbb{R}^2$. As $f_{\z} - \lambda'_1 f_1 - \lambda'_2f_2$ is single crossing,
\begin{align}
f_x(\theta_l) &= \lambda'_1 f_1(\theta_l) + \lambda'_2f_2(\theta_l) \quad \text{and} \quad 
f_x(\theta_h) = \lambda'_1 f_1(\theta_h) + \lambda'_2f_2(\theta_h), \text{ and }
\label{e:extending_lin_depen1}\\
f_x(\theta_{\lambda}) &= \lambda'_1 f_1(\theta_{\lambda}) + \lambda'_2 f_2(\theta_{\lambda}).\label{e:extending_lin_depen2}
\end{align}
\eqref{e:extending_lin_depen1} implies that $\lambda'$ solves \eqref{e:extending_lin_depen}. As the unique solution to \eqref{e:extending_lin_depen} was $\lambda$, it follows that $\lambda'=\lambda$.  But then \eqref{e:extending_lin_depend_contra} and \eqref{e:extending_lin_depen2} are in contradiction. Therefore, there exist $\lambda_1, \lambda_2: \Z \to \Reals$ such that 
\begin{align*}
(\forall \z, \theta) \quad f(\z, \theta) = \lambda_1(\z) f(\z_1, \theta) + \lambda_2(\z) f(\z_2, \theta).
\end{align*}

\subsection{Proof of \autoref{char_scd_f}'s \SCDstar \ Characterization}
\label{Proof_char_scd_f}

Here we only prove \autoref{char_scd_f}'s characterization of \SCDstar. The proof for its S\SCDstar \ characterization is deferred to Supplementary \autoref{proof_str_sc}.

\paragraph{$(\impliedby)$}
Suppose that $u(a, \theta) = g_1(a)f_1(\theta) + g_2(a)f_2(\theta) + h(\theta),$
with $f_1, f_2:\Theta \to \Reals$ each single crossing and ratio ordered. Then, for any $a, a' \in A$, $D_{a, a'}(\theta) = (g_1(a) - g_1(a')) f_1(\theta) + (g_2(a) - g_2(a'))f_2(\theta)$, which is single crossing by \autoref{equiv_cond}.

\paragraph{$(\implies)$}
Assume, without loss of generality, that $|A| \geq 2$. Take any $a_0 \in A$, and define $A' \equiv A \setminus \{a_0\}$. Define $f: A \times \Theta \to \mathbb{R}$ as $f(a, \theta) \equiv u(a, \theta) - u(a_0, \theta)$, which, for every $a$, is single crossing in $\theta$.

We will show that, for every function $\mu': A' \to \mathbb{R}$ with finite support, $\int_{a\in A'} f(a, \theta) \mathrm d \mu'$ can be represented as a multiple of the difference between two convex utility combinations $\int_a u(a, \theta) \mathrm{d}P$ and $\int_a u(a, \theta) \mathrm{d}Q$. Since the environment is convex, there exist $a_P, a_Q \in A$ such that $u(a_P, \theta)=\int_a u(a, \theta) \mathrm{d}P$ and $u(a_Q, \theta)=\int_a u (a, \theta) \mathrm{d}Q$ for all $\theta$. Since the utility difference $u(a_P, \theta) - u(a_Q, \theta)$ is single crossing, so is $\int_{a\in A'} f(a, \theta) \mathrm d \mu'$. The result then follows from \autoref{equiv_cond_general}.

For any function $\mu': A' \to \mathbb{R}$ with finite support, we define a function $\mu: A \to \mathbb{R}$ as an extension of $\mu'$: 
\begin{align*} 
\mu(a_0) \equiv - \sum_{\{a : \mu'(a) \neq 0\}} \mu'(a), \text{ \ and \ } (\forall a \in A') \ \mu(a) \equiv \mu'(a).
\end{align*} 
In a sense, we let $a_0$ absorb the function values on $A'$. In particular, note that $$\sum_{\{a : \mu(a) \neq 0\}} \mu(a) = \mu(a_0) + \sum_{\{a : \mu'(a) \neq 0\}} \mu(a) =0.$$

We construct the Hahn-Jordan decomposition $(\mu_+, \mu_-)$ of $\mu$. That is, we define functions $\mu_+, \mu_-:A \to \mathbb{R}_+$ by $(\forall a \in A)$ $\mu_+(a) \equiv \max\{ \mu(a), 0\}$ and $\mu_-(a) \equiv -\min\{ \mu(a), 0\}$. These are both functions with finite support, and $\mu = \mu_+ - \mu_-$. Let $M \equiv \sum_{\{a: \mu(a) \neq 0\}} \mu_+(a) = \sum_{\{a: \mu(a) \neq 0\}} \mu_-(a)$.
If $M=0$, pick an arbitrary $P\in \Delta A$ with finite support and let $Q=P$. If $M > 0$, define $P, Q \in \Delta A$ with probability mass functions $p, q$ such that for any $a \in A$, 
$$p(a) = \frac{\mu_+(a)}{M} \quad \text{and} \quad q(a) = \frac{\mu_-(a)}{M}.$$ 
Note that $P$ and $Q$ have finite support. Since the environment is convex, there exist $a_P, a_Q \in A$ such that $u(a_P, \theta) = \int_a u(a, \theta) \mathrm{d} P$ and $u(a_Q, \theta) = \int_a u(a, \theta) \mathrm{d} Q$ for all $\theta$. It follows that
\begin{eqnarray*}
\int_{a\in A'} f(a, \theta) \mathrm d \mu'
&=& \int_{a\in A} f(a, \theta) \mathrm d \mu \quad \small{\text{(because $f(a_0, \theta) =0$)}}\\
&=& \int_{a\in A} u(a, \theta) \mathrm d \mu - u(a_0, \theta) \mu(A) \\
&=& \int_{a \in A} u(a, \theta) \mathrm d \mu_+ - \int_{a \in A} u(a, \theta) \mathrm d \mu_-  \quad \small{\text{ (as $\mu(A) = 0$)}}\\
&=& M \cdot \left(u(a_P, \theta) - u(a_Q, \theta) \right),
\end{eqnarray*}
which is single crossing. 

Thus, if $u$ has \SCDstar, then $f: A' \times \Theta \to \Reals$ is linear combinations SC-preserving. By \autoref{equiv_cond_general}, 
there exist $a_1, a_2 \in A'$ and $\lambda_1,\lambda_2: A' \to \mathbb{R}$ such that (i) $f(a_1, \theta)$ and $f(a_2, \theta)$ are each single crossing and ratio ordered, and (ii) $(\forall a \in A')$ $f(a, \cdot) = \lambda_1(a) f(a_1, \cdot) + \lambda_2(a) f(a_2, \cdot)$. Hence, there exist functions $g_1, g_2: A \to \mathbb{R}$ with $g_1(a_0)=g_2(a_0)=0$ such that $(\forall a \in A)$ $f(a, \cdot) = g_1(a) f(a_1, \cdot) + g_2(a)f(a_2, \cdot)$, or equivalently, 
$$(\forall a, \theta) \quad u(a, \theta) = g_1(a) f(a_1, \theta) + g_2(a)f(a_2, \theta) + u(a_0, \theta).$$

\subsection{Proof of \autoref{rmk:convex_comb}}
\label{sec:proof_convex_comb}

We prove \autoref{rmk:convex_comb}'s result about \SCDstar \ and omit an analogous proof for the result about S\SCDstar. If $|\Theta| \leq 2$, then the proof is trivial, so assume $|\Theta| \geq 3$.
The ``if'' direction of the result follows directly from \autoref{char_scd_f}: if $u$ has a positive affine transformation $\tilde{u}$ of the form in \autoref{rmk:convex_comb}, then $u$, as a positive affine transformation of $\tilde{u}$, has \SCDstar.

For the ``only if'' direction, take any $u$ that has \SCDstar. Following the form given in \autoref{char_scd_f}, a positive affine transformation of $u$ is $$\tilde{u}(a, \theta) = g_1(a)f_1(\theta) + g_2(a)f_2(\theta),$$
where $f_1, f_2: \Theta \to \mathbb{R}$ are each single crossing and ratio ordered.

First, we consider the case in which $f(\underline{\theta})$ and $f(\overline{\theta})$ are linearly dependent.\footnote{This case can be ignored in the proof for S\SCDstar, because if $f_1$ and $f_2$ are strictly ratio ordered, then $f(\underline{\theta})$ and $f(\overline{\theta})$ must be linearly independent.} Assume, with a positive affine transformation of $\tilde{u}$, that the length of the vector $f(\theta)\equiv (f_1(\theta), f_2(\theta))$ in $\mathbb{R}^2$ is either $0$ or $1$ for every $\theta$. If $f(\underline{\theta}) = f(\overline{\theta}) = 0$, then because $f_1$ and $f_2$ are each single crossing, we have $(\forall \theta)$ $f_1 (\theta) = f_2 (\theta) = 0$ and $(\forall a, \theta)$ $\tilde{u}(a, \theta) = 0$. We can easily now rewrite $\tilde{u}$ in the form \eqref{eqn:convex_med_rep}, with $\lambda: \Theta \to [0,1]$ increasing. 

Suppose $f(\overline{\theta}) \neq 0$; we omit the analogous proof for the case of $f(\underline{\theta}) \neq 0$. By Condition \eqref{e:RD_add} of ratio ordering, for every $\theta$, the vector $f(\theta) \in \mathbb{R}^2$ is linearly dependent on $f(\overline{\theta})$. As $(\forall \theta)$ $\lVert f(\theta) \rVert \in \{0,1\}$, there exists $\lambda: \Theta \to \{-1, 0, 1\}$ such that $(\forall \theta)$ $f(\theta)=\lambda(\theta)f(\overline{\theta})$. Note that $\lambda$ is increasing because $f_1$ and $f_2$ are each single crossing. If either $\lambda(\underline{\theta}) =0$ (and so $(\forall a) \ \tilde{u}(a, \underline{\theta}) = 0$) or $\lambda(\underline{\theta})=1$ (and so $(\forall \theta) \ \lambda(\theta) = 1$), then
$$\tilde{u}(a, \theta) = \lambda(\theta) \tilde{u}(a, \overline{\theta}) + (1-\lambda(\theta)) \tilde{u}(a, \underline{\theta}),$$
with the last term equal to zero. If, on the other hand, $\lambda(\u \theta) = -1$, then 
\begin{align*}
\tilde{u}(a, \theta) &= \lambda(\theta)\tilde{u}(a, \overline{\theta}) = \frac{\lambda(\theta)+1}{2}\tilde{u}(a, \overline{\theta}) + \frac{\lambda(\theta)-1}{2} \left(-\tilde{u}(a, \underline{\theta})\right)\\
&= \frac{\lambda(\theta)+1}{2}\tilde{u}(a, \overline{\theta}) + \left(1- \frac{\lambda(\theta)+1}{2}\right)\tilde{u}(a, \underline{\theta}).
\end{align*}

Next, suppose that the vectors $f(\underline{\theta}), f(\overline{\theta}) \in \mathbb{R}^2$ are linearly independent, so the angle between the vectors is strictly less than 180 degrees. As $f_1$ and $f_2$ are ratio ordered, for each $\theta$ there exists $\alpha(\theta), \beta(\theta) \in \mathbb{R}_+$ such that
$$f(\theta) = \alpha(\theta) f(\overline{\theta}) + \beta(\theta) f(\underline{\theta}),$$
or equivalently,
$$
\tilde{u}(a, \theta) = \alpha(\theta) \tilde{u}(a, \overline{\theta}) + \beta(\theta) \tilde{u}(a, \underline{\theta}). 
$$
By Condition \eqref{e:RD_add}, $f(\theta) \neq 0$, which implies that $\alpha(\theta) + \beta(\theta) > 0$. A positive affine transformation of dividing $\tilde{u}(\cdot, \theta)$ by $\alpha(\theta)+\beta(\theta)$ results in the form \eqref{eqn:convex_med_rep}, where $\lambda(\theta) \equiv \frac{\alpha(\theta)}{\alpha(\theta) + \beta(\theta)} \in [0,1]$. 

To prove that the function $\lambda$ is increasing, take $\theta_1, \theta_2$ such that $\underline{\theta} \leq \theta_1 \leq \theta_2 \leq \overline{\theta}$. 
To reduce notation below, let $\alpha_i \equiv \alpha(\theta_i)$ and $\beta_i \equiv \beta(\theta_i)$ for $i=1, 2$. We must show that $\frac{\alpha_1}{\alpha_1 + \beta_1} \leq \frac{\alpha_2}{\alpha_2 + \beta_2}$, or equivalently that $\alpha_1\beta_2 \leq \alpha_2 \beta_1$. Suppose $f_1$ ratio dominates $f_2$; the other case is analogous.  Then $f_1(\theta_1) f_2(\theta_2) \leq f_1(\theta_2) f_2(\theta_1)$, and hence
$$\left(\alpha_1 f_1(\overline{\theta}) + \beta_1 f_1(\underline{\theta})\right)\left(\alpha_2 f_2(\overline{\theta}) + \beta_2 f_2(\underline{\theta})\right) \leq \left(\alpha_2 f_1(\overline{\theta}) + \beta_2 f_1(\underline{\theta})\right)\left(\alpha_1 f_2(\overline{\theta}) + \beta_1 f_2(\underline{\theta})\right),
$$
or equivalently,
$$
(\alpha_1 \beta_2 - \alpha_2\beta_1) \left(f_1(\overline{\theta})f_2(\underline{\theta}) -f_1(\underline{\theta})f_2(\overline{\theta}) \right) \leq 0. 
$$
Note that $f_1(\overline{\theta})f_2(\underline{\theta}) - f_1(\underline{\theta})f_2(\overline{\theta}) >0 $ because $f_1$ ratio dominates $f_2$, and $f(\underline{\theta})$ and $f(\overline{\theta})$ are linearly independent. Hence, $\alpha_1\beta_2 \leq \alpha_2 \beta_1$.

\section{Proofs for \MDstar}
\label{sec:proof_med}

The proof of \autoref{char_mdstar} requires 
analogs of \autoref{equiv_cond} and \autoref{equiv_cond_general} for monotonic functions; these results are stated next but their proofs are deferred to Supplementary \autoref{sec:MDproofs}.

\subsection{Aggregating Monotonic Functions}
\label{sec:agg_mon_func}

\begin{lemma}
\label{equiv_cond_mon}
Let $f_1, f_2: \Theta \to \Reals$ be monotonic functions. The linear combination $\alpha_1f_1(\theta) + \alpha_2f_2(\theta)$ is monotonic $\forall \alpha \in \Reals^2$ if and only if either $f_1$ or $f_2$ is an affine transformation of the other, i.e., there exists $\lambda \in \Reals^2$ such that either $f_2 = \lambda_1 f_1+\lambda_2$ or 
$f_1 = \lambda_1 f_2 + \lambda_2$.
\end{lemma}

We say that $f:\Z\times \Theta \to \Reals$ is \textit{linear combinations monotonicity-preserving} if $\int_z f(\z, \theta) \mathrm{d} \mu$ is a monotonic function of $\theta$ for every function $\mu:\Z \to \Reals$ with finite support.

\begin{proposition}
\label{equiv_cond_mon_general}
Let $f:\Z \times \Theta \to \Reals$ for some set $\Z$. The function $f$ is linear combinations monotonicity-preserving if and only if there exist $\z' \in \Z$ and $\lambda_1,\lambda_2: \Z \to \Reals$ such that (i) $f(\z',\cdot)$ is monotonic, and (ii) $(\forall \z) \ f(\z, \cdot) = \lambda_1(\z) f(\z',\cdot) + \lambda_2(\z)$.
\end{proposition}

\subsection{Proof of \autoref{char_mdstar}}
\label{Proof_char_med_f}

\paragraph{$(\impliedby)$}
We omit the proof as it is similar to \hyperref[Proof_char_scd_f]{the proof} of \autoref{char_scd_f} in \appendixref{Proof_char_scd_f}.

\paragraph{$(\implies)$} 
The proof is trivial if $(\forall a, \theta)$ $u(a, \theta)=0$, so assume there exists $a_0$ such that $u(a_0, \cdot): \Theta \to \Reals$ is not a zero function. Define $f: A \times \Theta \to \Reals$ by $f(a, \theta) \equiv u(a, \theta) - u(a_0, \theta)$. Note that $(\forall a)$ $f(a, \theta)$ is a monotonic function of $\theta$.

Let $A' \equiv A \backslash \{a_0\}$. As in \hyperref[Proof_char_scd_f]{the proof} of \autoref{char_scd_f} in \appendixref{Proof_char_scd_f}, for every $\mu':A'\to \Reals$ with finite support, 
there exist convex utility combinations $\int_a u(a, \theta) \mathrm{d}P$ and $\int_a u(a, \theta) \mathrm{d}Q$, where $P$ and $Q$ have finite support, such that $\int_{a \in A'} f(a, \theta) \mathrm d\mu'$ is monotonic if and only if 
$\int_{a \in A} u(a, \theta) \mathrm{d} P- \int_{a \in A} u(a, \theta) \mathrm{d} Q$ is monotonic. Since the environment is convex, the latter utility difference is indeed monotonic, and so 
$\int_{a \in A'} f(a, \theta) \mathrm d\mu'$ is monotonic. By \autoref{equiv_cond_mon_general}, there exist $a' \in A \backslash a_0$ and $\lambda_1, \lambda_2: A \backslash \{a_0\} \to \Reals$ such that $(\forall a, \theta)$ $f(a, \theta) = \lambda_1(a)f(a', \theta) + \lambda_2(a)$.	Hence, there exist functions $g_1, g_2: A \to \mathbb{R}$ with $g_1(a_0) = g_2(a_0)=0$ such that $f(a, \theta) = g_1(a) f(a', \theta) + g_2(a)$, or equivalently, $u(a, \theta) = g_1(a) f(a', \theta) + g_2(a) + u(a_0, \theta).$

\vspace{.25in}
\singlespacing 
\addtocontents{toc}{\protect\setcounter{tocdepth}{1}} 
\bibliographystyle{ecta}
\bibliography{KLR}

\newpage
\onehalfspacing 

\renewcommand{\appendixtocname}{Supplementary Appendices}
\addappheadtotoc 
\section*{Supplementary Appendices (For Online Publication Only)}
\label{supp_app}


\addtocontents{toc}{\protect\setcounter{tocdepth}{-1}} 

\renewcommand*\thesection{SA.\arabic{section}}
\setcounter{section}{0}
\renewcommand*{\theHsection}{chX.\the\value{section}}

\renewcommand*\theequation{SA.\arabic{equation}}
\setcounter{equation}{0}
\renewcommand{\theHequation}{Supplement.\theequation}

\renewcommand*\thefigure{SA.\arabic{figure}}
\setcounter{figure}{0}
\renewcommand{\theHfigure}{Supplement.\thefigure}

\renewcommand{\thelemma}{SA.\arabic{lemma}}
\renewcommand{\theHlemma}{Supplement.\thelemma}
\setcounter{lemma}{0}

\renewcommand{\theproposition}{SA.\arabic{proposition}}
\setcounter{proposition}{0}
\renewcommand{\theHproposition}{Supplement.\theproposition}

\renewcommand{\theconjecture}{SA.\arabic{conjecture}}
\setcounter{conjecture}{0}
\renewcommand{\theHconjecture}{Supplement.\theconjecture}

\renewcommand{\theclaim}{SA.\arabic{claim}}
\setcounter{claim}{0}
\renewcommand{\theHclaim}{Supplement.\theclaim}

\appendixref{sec:monotone_selection} provides a monotone selection theorem (\autoref{prop_ms}); \appendixref{proof_str_sc} proves \autoref{char_scd_f}'s characterization of S\SCDstar; 
\appendixref{sec:proof_implications} proves the implications of \autoref{char_scd_f} (\autoref{cor:sced_char}, \autoref{lossfunctions}, \autoref{cor:multidim}, and \autoref{p:SCED-X}); and \appendixref{sec:MDproofs} contains proofs of the intermediate results \autoref{equiv_cond_mon} and \autoref{equiv_cond_mon_general} towards our \MDstar \ characterization (\autoref{char_mdstar}). Finally, \appendixref{sec:qs_karlin} elaborates on the connection between ratio ordering and \citepos{quah2012aggregating} signed-ratio monotonicity.

\section{SSCD and Monotone Selection}
\label{sec:monotone_selection}

\autoref{prop_mcs} established the connection between SCD and monotone comparative statics. Here we establish an analogous connection between SSCD and monotone selection.

\begin{definition}
$u: A \times \Theta \to \Reals$ has \textit{monotone selection (MS)} on $(A, \succeq)$ if for any $S\subseteq A$, every selection $s^*(\theta)$ from $\argmax_{a \in S} u(a, \theta)$ is increasing in $\theta$.\footnote{$s^*(\theta) \equiv \emptyset$ if $\argmax_{a \in S} u(a, \theta) = \emptyset$, and we extend $\succeq$ to $A \cup \{\emptyset\}$ by stipulating $a \succeq \emptyset \succeq a$ for every $a \in A$.}
\end{definition}

Define binary relations $\succ_{SSCD}$ and $\succeq_{SSCD}$ on $A$ as follows: $a \succ_{SSCD} a'$ if $D_{a,a'}$ is strictly single crossing only from below; $a \succeq_{SSCD} a'$ if either $a \succ_{SSCD} a'$ or $a=a'$. As before, if $u: A \times \Theta \to \mathbb{R}$ has SSCD, then $\succeq_{SSCD}$ is an order.

\begin{theorem}
\label{prop_ms}
$u:A \times \Theta \to \Reals$ has monotone selection on $(A, \succeq)$, where $A$ is minimal, if and only if $u$ has SSCD and $\succeq$ is a refinement of $\succeq_{SSCD}$.
\end{theorem}

Note that SSCD (and a refinement of $\succeq_{SSCD}$) is not only sufficient but also necessary in \autoref{prop_ms}. One can verify that, while not stated in their Theorem 4', \citepos{milgrom1994monotone} strict single-crossing property is also necessary for monotone selection in their sense.

\begin{proof}[Proof of \autoref{prop_ms}]
The proof is similar to the \hyperref[Proof_prop_mcs]{proof} of \autoref{prop_mcs} in \appendixref{Proof_prop_mcs}.

\paragraph{$(\implies)$}
Suppose $u:A \times \Theta \to \Reals$ has MS on $(A, \succeq)$.

To show that $u$ has SSCD on $A$, suppose not, towards contradiction. As we have shown in \hyperref[interval_choice_proofs]{the proof} of \autoref{lem:scd_interval_choice},
$$(\exists a', a'' \text{ with } a' \neq a'') (\exists \theta_l < \theta_m < \theta_h) \quad D_{a', a''}(\theta_l) \geq 0, D_{a', a''}(\theta_m) \leq 0, \text{ and } D_{a', a''}(\theta_h) \geq 0.$$
Let $S \equiv \{a', a''\}$ and consider a selection $s^*(\theta)$ from $\argmax_{a \in S} u(a, \theta)$ such that $s^*(\theta_l) = s^*(\theta_h) = a'$ and $s^*(\theta_m) = a''$. Since $u$ has MS on $(A, \succeq)$, we must have $a' \succeq a''$ and $a'' \succeq a'$, a contradiction to anti-symmetry of $\succeq$. 

To show that $\succeq$ is a refinement of $\succeq_{SSCD}$, it suffices to show that
$$
(\forall a', a'' \in A) \quad a' \succ_{SSCD} a'' \implies a' \succ a'',
$$
because both $\succeq$ and $\succeq_{SSCD}$ are anti-symmetric. Take any $a', a'' \in A$ such that $a' \succ_{SSCD} a''$. As $D_{a', a''}$ is strictly single crossing only from below, $\exists \theta_l<\theta_h$ such that $\sign [D_{a',a''}(\theta_l)]<\sign [D_{a',a''}(\theta_h)]$, which implies that $\sign [D_{a',a''}(\theta_l)] \leq 0$ and $\sign [D_{a',a''}(\theta_h)] \geq 0$. Consider a selection $a'' \in  \argmax_{a \in \{a', a''\}} u(a, \theta_l)$ and $a' \in \argmax_{a \in \{a', a''\}} u(a, \theta_h)$. Since $u$ has MS on $(A, \succeq)$, we have $a' \succeq a''$. Since $a' \neq a''$ and $\succeq$ is anti-symmetric, it must be that $a' \succ a''$.

\paragraph{$(\impliedby)$}
For any $S \subseteq A$, define $C_u(S) \equiv \bigcup_{\theta \in \Theta} \argmax_{a \in S} u(a, \theta)$. First, we claim that $C_u(S)$ is completely ordered by $\succeq_{SSCD}$. To see this, take any pair $a', a'' \in C_u(S)$ with $a' \neq a''$. As $u$ has SSCD, $D_{a', a''}$ is strictly single crossing in $\theta$. As $a', a'' \in C_u(S)$, $\sign[D_{a', a''}]$ is not a constant function with value either 1 or -1. Thus, $D_{a', a''}$ is strictly single crossing either only from below or only from above. It follows that $a' \succ_{SSCD} a''$ or $a'' \succ_{SSCD} a'$. Next, since $\succeq$ is a refinement of $\succeq_{SSCD}$, $\succeq$ coincides with $\succeq_{SSCD}$ on $C_u(S)$. By definition of $\succeq_{SSCD}$, when restricted to $C_u(S)\times \Theta$, $u$ satisfies \citeauthor{milgrom1994monotone}'s strict single-crossing property in $(a, \theta)$ with respect to $\succeq_{SSCD}$ and $\leq$.\footnote{\citeauthorpos{milgrom1994monotone} strict single-crossing property of $u$ on $C_u(S)\times \Theta$ 
is equivalent to $$(\forall a' \succ_{SSCD} a'') (\forall \theta_l < \theta_h) \quad u(a', \theta_l) \geq u(a'', \theta_l) \implies u(a', \theta_h) > u(a'', \theta_h).$$} As $C_u(S)$ is completely ordered by $\succeq_{SSCD}$, it follows from \citet[Theorem 4']{milgrom1994monotone} that any selection $s^*(\theta)$ from $\argmax_{a \in C_u(S)} u(a, \theta) (=\argmax_{a \in S} u(a, \theta))$ is increasing in $\theta$. 
\end{proof}

\section{Proof of \autoref{char_scd_f}'s S\SCDstar \ Characterization}
\label{proof_str_sc}

Similar to \hyperref[Proof_char_scd_f]{the proof} of \autoref{char_scd_f} for \SCDstar, our proof for S\SCDstar \ requires conditions ensuring that arbitrary linear combinations of functions are strictly single crossing. We state and discuss the analogs of \autoref{equiv_cond} and \autoref{equiv_cond_general} below in \appendixref{sec:aggreg_SSC}; their proofs are in \appendixref{proof:equiv_cond_str} and \appendixref{proof:equiv_cond_general_str} respectively. The proof of \autoref{char_scd_f} for S\SCDstar\ then follows in \appendixref{proof:char_scd_f_str}.

\subsection{Aggregating Strictly Single-Crossing Functions}
\label{sec:aggreg_SSC}

\begin{lemma}
\label{equiv_cond_str}
Let $f_1, f_2: \Theta \to \mathbb{R}$. The linear combination \mbox{$\alpha_1 f_1(\theta) + \alpha_2 f_2(\theta)$} 
is strictly single crossing $\forall \alpha \in \mathbb{R}^2 \backslash \{0\}$ if and only if $f_1$ and $f_2$ are strictly ratio ordered.
\end{lemma}

Besides the change to strict single crossing and, correspondingly, strict ratio ordering, \autoref{equiv_cond_str} has two other differences from \autoref{equiv_cond}.  First, we rule out $(\alpha_1,\alpha_2) = 0$; this is unavoidable because a zero function is not strictly single crossing.  Second, and more importantly, there is no explicit mention in \autoref{equiv_cond_str} that $f_1$ and $ f_2$ are each strictly single crossing. It turns out---as elaborated in the \hyperref[proof:equiv_cond_str]{Lemma's proof}---that when two functions are strictly ratio ordered, each of them must be strictly single crossing.

To extend \autoref{equiv_cond_str} to more than two functions, we say that $f:\Z\times \Theta \to \Reals$ is \textit{linear combinations SSC-preserving} if $\int_z f(\z,\theta) \mathrm{d} \mu$ is either a zero function or strictly single crossing in $\theta$ for every function $\mu:\Z \to \Reals$ with finite support. Parallel to \autoref{equiv_cond_general}:
\begin{proposition}
\label{equiv_cond_general_str}
Let $f:\Z \times \Theta \to \Reals$ for some set $\Z$, and assume there exist $\z_1,\z_2\in \Z$ such that $f(\z_1,\cdot):\Theta\to \Reals$ and $f(\z_2,\cdot):\Theta\to \Reals$ are linearly independent. The function $f$ is linear combinations SSC-preserving if and only if there exist $\lambda_1,\lambda_2:\Z\to \Reals$ such that
    \begin{enumerate}
    \item  $f(\z_1,\cdot):\Theta \to \Reals$ and $f(\z_2,\cdot):\Theta \to \Reals$ are strictly ratio ordered, and 
    \item \label{equiv_cond_general_2_str} $(\forall \z) \ f(\z,\cdot)= \lambda_{1}(\z) f(\z_1,\cdot) + \lambda_2(\z) f(\z_2,\cdot)$.
    \end{enumerate}
\end{proposition}
For the ``if'' direction of \autoref{equiv_cond_general_str}, the existence of a pair of linearly independent functions need not be assumed, because strict ratio ordering implies linear independence.  However, without that hypothesis, the ``only if'' direction would fail: given $\Z=\{\z_1,\z_2\}$, and $f(\z_1,\cdot)=2f(\z_2,\cdot)$ with $f(\z_1,\cdot)$ strictly single crossing, the function $f$ is linear combinations SSC-preserving even though $f(\z_1,\cdot)$ and $f(\z_2,\cdot)$ are not strictly ratio ordered.  

\subsection{Proof of \autoref{equiv_cond_str}}
\label{proof:equiv_cond_str}
\subsubsection*{When $|\Theta| \leq 2$.}
If $|\Theta|=1$, the proof is trivial as all functions are strictly single crossing and every pair of $f_1, f_2$ satisfy strict ratio ordering. So assume $|\Theta| =2$ and denote $\Theta = \{\theta_l, \theta_h\}$; without loss, we may assume $\theta_h>\theta_l$ because of our maintained assumption that upper and lower bounds exist for all pairs.

\paragraph{$(\implies)$}
Either $(f_1(\theta_l), f_2(\theta_l)) \neq0$ or $(f_1(\theta_h), f_2(\theta_h)) \neq0$: otherwise, for every $\alpha \in \Reals^2 \backslash \{0\}$,
$(\alpha\cdot f) (\theta_l) = (\alpha \cdot f) (\theta_h) = 0,$
and hence $\alpha \cdot f$ is a zero function, which is not strictly single crossing. 
Assume $(f_1(\theta_l), f_2(\theta_l)) \neq 0$; the proof for the other case is analogous. Let $\alpha_{l} \equiv (f_2(\theta_l), -f_1(\theta_l))$ and consider $(\alpha_l \cdot f) (\theta) = f_2(\theta_l)f_1(\theta) - f_1(\theta_l) f_2(\theta)$. We have
\mbox{$(\alpha_l \cdot f)(\theta_l) = 0$} and, by strict single crossing of $\alpha_l\cdot f$, $(\alpha_l\cdot f)(\theta_h) \neq 0$. That is, $f_2(\theta_l)f_1(\theta_h) \neq f_1(\theta_l) f_2(\theta_h)$, which means that $f_1$ and $f_2$ are strictly ratio ordered.

\paragraph{$(\impliedby)$}
For any $\alpha \in \mathbb{R}^2 \backslash \{0\}$, $\alpha \cdot f$ is not strictly single crossing if and only if $(\alpha\cdot f) (\theta_l) = (\alpha\cdot f) (\theta_h) = 0.$ This implies $\alpha_1f_1(\theta_l) = -\alpha_2f_2(\theta_l)$ and $\alpha_1f_1(\theta_h) = -\alpha_2f_2(\theta_h)$, and hence
\[\begin{split}
&\alpha_1f_1(\theta_l)f_2(\theta_h) = -\alpha_2f_2(\theta_l)f_2(\theta_h) =\alpha_1f_1(\theta_h)f_2(\theta_l) \quad\text{and}\\
&\alpha_2f_1(\theta_l)f_2(\theta_h) = -\alpha_1f_1(\theta_l)f_1(\theta_h) =\alpha_2f_1(\theta_h)f_2(\theta_l).
\end{split}\]
As $(\alpha_1,\alpha_2) \neq 0$, $f_1(\theta_l)f_2(\theta_h) = f_1(\theta_h)f_2(\theta_l)$, contradicting strict ratio ordering of $f_1$ and $f_2$.

\subsubsection*{When $|\Theta| \geq 3$.}
\paragraph{$(\implies)$}
Suppose, towards contradiction, that $f_1$ and $f_2$ are not strictly ratio ordered:
\begin{equation}\begin{split}
& (\exists \theta_l < \theta_h) \quad f_1(\theta_l)f_2(\theta_h) \leq f_1(\theta_h)f_2(\theta_l) \quad \text{ and }\\
& (\exists \theta' < \theta'') \quad f_1(\theta')f_2(\theta'') \geq f_1(\theta'')f_2(\theta').
\end{split}
\label{e:proof_prop1_str}
\end{equation}

Take any upper bound $\overline{\theta}$ of $\{\theta_l, \theta_h, \theta', \theta''\}$. Letting $\alpha_l \equiv (f_2(\theta_l), -f_1(\theta_l))$, it holds that $\alpha_l \cdot f$ is strictly single crossing only from below, as
$(\alpha_l \cdot f)(\theta_l) = (f_2(\theta_l), -f_1(\theta_l)) \cdot (f_1(\theta_l), f_2(\theta_l)) = 0$ and by \eqref{e:proof_prop1_str}, $(\alpha_l \cdot f) (\theta_h) \geq 0$. Hence $(\alpha_l \cdot f) (\overline \theta) \geq 0$. Analogously, letting $\alpha' \equiv (f_2(\theta'), -f_1(\theta'))$, we conclude that $(\alpha' \cdot f) (\overline \theta) \leq 0$. 
Now let $\overline{\alpha} \equiv (f_2(\overline{\theta}), -f_1(\overline{\theta}))$. It follows that
\[\begin{split}
& (\overline{\alpha}\cdot f)(\theta_l) = (f_2(\overline{\theta}), -f_1(\overline{\theta}))\cdot (f_1(\theta_l), f_2(\theta_l)) = - (\alpha_l \cdot f) (\overline{\theta}) \leq 0,\\
& (\overline{\alpha}\cdot f)(\theta') = (f_2({\theta'}), -f_1({\theta'}))\cdot (f_1(\theta'), f_2(\theta'))=- (\alpha' \cdot f) (\overline{\theta}) \geq 0, \text{ and }\\
& (\overline{\alpha}\cdot f)(\overline{\theta}) =0.\\
\end{split}\]
Therefore, $\overline{\alpha} \cdot f$ is not strictly single crossing. 

\paragraph{$(\impliedby)$}
We provide a proof for the case in which $f_1$ strictly ratio dominates $f_2$, and omit the other case's analogous proof. For any $\alpha \in \Reals^2\backslash \{0\}$, we prove that $\alpha \cdot f$ is single crossing. The argument is very similar to that used in proving \autoref{equiv_cond}, but note that here we do not assume that $f_1$ and $f_2$ are each strictly single crossing.

As $f_1$ strictly ratio dominates $f_2$,
\begin{align}
&(\forall \theta_l < \theta_h) \quad f_1(\theta_l) f_2(\theta_h) < f_1(\theta_h) f_2(\theta_l).
\label{e:RD_str}
\end{align}

Suppose, towards contradiction, that $\alpha \cdot f$ is not strictly single crossing.

\underline{Claim}: There exist $\theta_l, \theta_m, \theta_h$ with $\theta_l < \theta_m < \theta_h$ such that 
\begin{align}
& (\alpha \cdot f)(\theta_l) \leq 0, (\alpha \cdot f)(\theta_m) \geq 0, \text{ and } (\alpha \cdot f)(\theta_h) \leq 0, \quad \text{or}
\label{assumption1_in_claim2_str}\\
& (\alpha \cdot f)(\theta_l) \geq 0, (\alpha \cdot f)(\theta_m) \leq 0, \text{ and } (\alpha \cdot f)(\theta_h) \geq 0.
\label{assumption3_in_claim2_str}
\end{align}

\underline{Proof of claim}: Since $\alpha \cdot f$ is not strictly single crossing either from below or from above,
\[\begin{split}
&(\exists \theta_1 < \theta_2 ) \quad (\alpha \cdot f) (\theta_1) \geq 0 \geq (\alpha \cdot f)(\theta_2), \text{ and }\\
&(\exists \theta_3 < \theta_4 ) \quad (\alpha \cdot f) (\theta_3) \leq 0 \leq (\alpha \cdot f)(\theta_4).
\end{split}\]

Let $\Theta_0 \equiv \{\theta_1, \theta_2, \theta_3, \theta_4\}$ and let $\bar{\theta}$ and $\underline{\theta}$ be an upper and lower bound of $\Theta_0$, respectively. Either \mbox{$(\alpha \cdot f)(\underline{\theta}) \neq 0$} or $(\alpha \cdot f)(\overline{\theta}) \neq 0$, as otherwise $f_1(\underline{\theta}) f_2 (\overline{\theta}) = f_2 (\underline{\theta}) f_1(\overline{\theta})$, contradicting \eqref{e:RD_str}. Suppose $(\alpha \cdot f)(\bar{\theta}) \neq 0$. If $(\alpha \cdot f)(\bar{\theta}) < 0$, then we choose $(\theta_l, \theta_m, \theta_h) = (\theta_3, \theta_4, \bar{\theta})$, which satisfies \eqref{assumption1_in_claim2_str}. If $(\alpha \cdot f)(\bar{\theta}) > 0$, then we choose $(\theta_l, \theta_m, \theta_h) = (\theta_1, \theta_2, \bar{\theta})$, which satisfies \eqref{assumption3_in_claim2_str}. A similar argument applies when $(\alpha \cdot f)(\underline{\theta}) \neq 0$. $\parallel$

Condition \eqref{e:RD_str} implies that $f(\theta) \equiv (f_1(\theta), f_2(\theta)) \neq 0$ for all $\theta \in \{\theta_l, \theta_m, \theta_h\}$. Take any $\theta_1, \theta_2 \in \{\theta_l, \theta_m, \theta_h\}$ such that $\theta_1 < \theta_2$. By \eqref{e:RD_str}, $f(\theta_1)$ moves to $f(\theta_2)$ in a clockwise rotation with an angle $r_{12} \in (0, 180)$. Suppose \eqref{assumption1_in_claim2_str} holds;
the argument is analogous if \eqref{assumption3_in_claim2_str} holds.  It follows from
$0<r_{lh} <180$, $(\alpha \cdot f)(\theta_l) \leq 0$, and $(\alpha \cdot f)(\theta_h) \leq 0$ that $\{f(\theta_l), f(\theta_h)\} \subseteq \Reals^2_{\alpha, -} \cup \Reals^2_{\alpha, 0}$ with $\{f(\theta_l), f(\theta_h)\} \not \subseteq \Reals^2_{\alpha, 0}$.\footnote{Recall that $\Reals^2_{\alpha, +} \equiv \{ x \in \Reals^2 \,:\, \alpha \cdot x > 0\}$, $\Reals^2_{\alpha, 0} \equiv \{ x \in \Reals^2 \,:\, \alpha \cdot x = 0\}$, and $\Reals^2_{\alpha, -} \equiv \{ x \in \Reals^2 \,:\, \alpha \cdot x < 0\}$.}  This, together with $r_{lm}>0$ and $r_{mh}>0$, implies $f(\theta_m) \in \Reals^2_{\alpha, -}$, which contradicts \eqref{assumption1_in_claim2_str}.

\subsection{Proof of \autoref{equiv_cond_general_str}}
\label{proof:equiv_cond_general_str}

\appendixref{Proof_prop1} proved \autoref{equiv_cond_general} assuming certain functions are linearly independent. Essentially the same proof can be used for \autoref{equiv_cond_general_str}, replacing statements involving ``single crossing'' with ``either a zero function or strictly single crossing''. 

\subsection{Proof of the S\SCDstar \ Portion of \autoref{char_scd_f}}
\label{proof:char_scd_f_str}

The utility function $u: A \times \Theta \to \Reals$ has S\SCDstar \ if and only if $(\forall a, a' \in A)$ $D_{a, a'}$ is either a zero function or strictly single crossing. Most statements in \hyperref[Proof_char_scd_f]{the proof} of \autoref{char_scd_f} for \SCDstar\ go through for S\SCDstar\ 
when we replace ``single crossing'' with ``either a zero function or strictly single crossing''.

We need only to rewrite the proof of the ``only if'' part in the following two special cases:
    \begin{enumerate}
    \item $(\forall a', a'') (\forall \theta)$ $u(a', \theta) = u(a'', \theta)$, or
    \item $(\exists a', a'')$ such that (i) $u(a'', \theta) - u(a', \theta)$ is not a zero function of $\theta$, and (ii) $(\forall a)$ $u(a, \theta) - u(a', \theta)$ and $u(a'', \theta) - u(a', \theta)$ are linearly dependent functions of $\theta$. 
    \end{enumerate}

In the first case, we can write $u(a, \theta)$ in form of \eqref{e:funcform} 
where $g_1, g_2$ are zero functions, $h(\theta) \equiv u(a_0, \theta)$ for any $a_0$, $(\forall \theta) \ f_1(\theta)=1$, and $f_2(\theta)$ is any strictly decreasing function of $\theta$. Then,
$$(\forall \theta_l < \theta_h) \quad f_1(\theta_l)f_2(\theta_h) = f_2(\theta_h) < f_2(\theta_l) = f_1(\theta_h)f_2(\theta_l).$$

In the second case, for every $a$, there exists $\lambda \in \Reals^2 \backslash \{0\}$ such that $\lambda_1 \left( u(a, \cdot) - u(a', \cdot) \right) + \lambda_2 \left( u(a'', \cdot) - u(a', \cdot) \right)$ is a zero function. 
Note that $\lambda_1 \neq 0$, as otherwise $u(a'', \cdot) - u(a', \cdot)$ would be a zero function. It follows that there exists $\lambda: A \to \Reals$ such that 
$$(\forall a, \theta) \quad u(a, \theta) - u(a', \theta) = \lambda(a) \left(u(a'', \theta) - u(a', \theta)\right),$$
or equivalently,
$$(\forall a, \theta) \quad u(a, \theta) = \lambda(a) \left(u(a'', \theta) - u(a', \theta) \right) + u(a', \theta).$$
Note that $u(a'', \theta) - u(a', \theta)$ is a strictly single-crossing function of $\theta$: consider the expectational difference with distributions that put probability one on $a''$ and $a'$ respectively. If the difference is strictly single crossing from below, we can write $u(a, \theta)$ in the form of \eqref{e:funcform} where $g_1(a) = \lambda(a)$, $g_2(a)=0$, $f_1(\theta) = u(a'', \theta) - u(a', \theta)$, and $h(\theta) = u(a', \theta)$.
If the difference is strictly single crossing only from above, we let $g_1(a) = -\lambda(a)$ and $f_1(\theta) = u(a', \theta) - u(a'', \theta)$. 
Now take any strictly increasing function $h: \Theta \to \Reals$ and define
$$\hat{h}(\theta) \equiv 
\left\{\begin{array}{ll}
-e^{h(\theta)} & \text{if $f_1(\theta) \leq 0$}\\
e^{-h(\theta)} & \text{otherwise}
\end{array}\right.
\quad \text{ and } \quad
f_2(\theta) \equiv 
\left\{\begin{array}{ll}
\hat{h}(\theta) f_1(\theta) & \text{if $f_1(\theta) \neq 0$}\\
1 & \text{otherwise}.
\end{array}\right.
$$
To verify that $f_1$ and $f_2$ are strictly ratio ordered, take any $\theta_l < \theta_h$.  There are three possibilities to consider:
    \begin{enumerate}
    \item If $f_1(\theta_l)f_1(\theta_h) > 0$, then
    $$f_1(\theta_l) f_2(\theta_h) = f_1(\theta_l)f_1(\theta_h) \hat{h}(\theta_h) < f_1(\theta_l) f_1(\theta_h) \hat{h}(\theta_l) = f_1(\theta_h)f_2(\theta_l),$$
    as $\hat{h}(\theta)$ is strictly decreasing over $\{ \theta \,|\, f_1(\theta) < 0\}$ and $\{ \theta \,|\, f_1(\theta) > 0\}$.
    \item If $f_1(\theta_l)f_1(\theta_h) < 0$, then as $f_1(\theta)$ is strictly single crossing from below, we have $f_1(\theta_l) < 0 < f_1(\theta_h)$. Hence, 
    $$f_1(\theta_l) f_2(\theta_h) = f_1(\theta_l)f_1(\theta_h)\hat{h}(\theta_h) < 0 < f_1(\theta_l)f_1(\theta_h)\hat{h}(\theta_l) = f_1(\theta_h)f_2(\theta_l).$$
    \item If $f_1(\theta_l)f_1(\theta_h) = 0$, then because $f_1$ is strictly single crossing from below, we have either (i) $f_1(\theta_l) < 0 = f_1(\theta_h)$, which results in
    $f_1(\theta_l)f_2(\theta_h) = f_1(\theta_l) < 0 = f_1(\theta_h)f_2(\theta_l),$
    or (ii) $f_1(\theta_l) = 0 < f_1(\theta_h)$, which results in
    $f_1(\theta_l)f_2(\theta_h) = 0 < f_1(\theta_h) = f_1(\theta_h)f_2(\theta_l).$ \qedhere
    \end{enumerate}

\section{Proofs for the Implications of \autoref{char_scd_f}}
\label{sec:proof_implications}

\subsection{Proof of \autoref{lossfunctions}}
\label{proof_lossfunctions}

It is clear from \autoref{cor:sced_char} that $v(x, \theta) = - |x-\theta|^2 = -x^2 + 2x\theta - \theta^2$ has SCED, as \mbox{$f_1(\theta)=-1$} and $f_2(\theta)=2\theta$ are each single crossing and ratio ordered, and we take $g_1(x)=x^2$, $g_2(x)=x$, and $h(\theta)=-\theta^2$. 

For the converse, it is sufficient to prove the following claim.
\begin{claim}
\label{claim:powerloss}
If there exist $g_1, g_2: \Reals \to \Reals$ and $f_1, f_2, h: \Theta \to \Reals$ such that 
\begin{equation*}
v(x, \theta) \equiv - |x-\theta|^z = g_1(x)f_1(\theta) + g_2(x)f_2(\theta) + h(\theta),
\end{equation*}
then $z=2$.
\end{claim}

\begin{proof}[Proof of \autoref{claim:powerloss}]
Fix $x_0 \in \Reals$ and define $\tilde{v}(x, \theta) \equiv v(x, \theta) - v(x_0, \theta) =  \tilde{g}_1(x) f_1(\theta) + \tilde{g}_2 f_2(\theta),$
where $\tilde{g}_1(x) \equiv g_1(x) - g_1(x_0)$ and $\tilde{g}_2 \equiv g_2(x) - g_2(x_0)$. Fix any $\theta_l < \theta_m < \theta_h$. There exists $(\lambda_l, \lambda_m, \lambda_h) \in \Reals^3 \backslash \{0\}$ such that 
$$
\begin{bmatrix}
f_1(\theta_l) & f_1(\theta_m) & f_1(\theta_h)\\
f_2(\theta_l) & f_2(\theta_m) & f_2(\theta_h)
\end{bmatrix}
\begin{bmatrix}
\lambda_l \\ \lambda_m \\ \lambda_h
\end{bmatrix}
=
\begin{bmatrix}
0 \\ 0
\end{bmatrix}
.$$
Hence, for every $x \in \Reals$,
\begin{align*}
h(x) & \equiv  \lambda_l \tilde{v}( x,\theta_l) + \lambda_m \tilde{v}( x,\theta_m) + \lambda_h \tilde{v}( x,\theta_h)\\
& = 
\begin{bmatrix}
\tilde{g}_1(x) & \tilde{g}_2(x)\\
\end{bmatrix}
\begin{bmatrix}
f_1(\theta_l) & f_1(\theta_m) & f_1(\theta_h)\\
f_2(\theta_l) & f_2(\theta_m) & f_2(\theta_h)
\end{bmatrix}
\begin{bmatrix}
\lambda_l \\ \lambda_m \\ \lambda_h
\end{bmatrix}
= 0.\notag 
\end{align*}

We hereafter consider $\lambda_l \neq 0$ (and omit the proofs for the other two cases, $\lambda_m\neq 0$ and $\lambda_h\neq 0$, which are analogous).  The previous equation implies that for any $x \in \Reals$,
\begin{equation}
\tilde{v}(x, \theta_l) = -\frac{\lambda_m}{\lambda_l} \tilde{v}(x, \theta_m) - \frac{\lambda_h}{\lambda_l } \tilde{v}(x, \theta_h). \label{e:lossfunctionsproof}
\end{equation}
At any $x<\theta$, $\tilde v(x,\theta)=-(\theta-x)^z-v(x_0,\theta)$ is differentiable in $x$, and hence \eqref{e:lossfunctionsproof} implies that the partial derivative $\tilde{v}_x(x, \theta_l)$ exists at $x=\theta_l$. Thus, the right partial derivative $$\lim_{\epsilon \downarrow 0} \frac{\tilde{v}(\theta_l + \epsilon, \theta_l) - \tilde{v}(\theta_l, \theta_l)}{\epsilon} = - \lim_{\epsilon \downarrow 0} \epsilon^{z-1}$$ must equal the left partial derivative $$\lim_{\epsilon \downarrow 0} \frac{\tilde{u}(\theta_l - \epsilon, \theta_l) - \tilde{v}(\theta_l, \theta_l)}{-\epsilon} = \lim_{\epsilon \downarrow 0} \epsilon^{z-1},$$ which implies $\lim_{\epsilon \downarrow 0} \epsilon^{z-1}=0$, and thus $z>1$.

Now suppose to contradiction that $z\neq 2$.  At any $x >\theta_h$, \eqref{e:lossfunctionsproof} and $\tilde v(x,\theta)=-(x-\theta)^z-v(x_0,\theta)$ imply
$$-\lambda_l (x-\theta_l)^{z}=\lambda_m (x-\theta_m)^{z}+\lambda_h (x-\theta_h)^{z}+(\lambda_m+\lambda_h-\lambda_l)v(x_0,\theta),$$
and hence, differentiating with respect to $x$ and simplifying using $z>1$ and $z\neq 2$:
	\begin{align}
	-\lambda_l (x-\theta_l)^{z-1} &= \lambda_m (x-\theta_m)^{z-1} + \lambda_h (x-\theta_h)^{z-1},\label{e:lossfunctionsproof1}\\
	-\lambda_l (x-\theta_l)^{z-2} &= \lambda_m (a-\theta_m)^{z-2} + \lambda_h (x-\theta_h)^{z-2},\label{e:lossfunctionsproof2}\\
	-\lambda_l (x-\theta_l)^{z-3} &= \lambda_m (x-\theta_m)^{z-3} + \lambda_h (x-\theta_h)^{z-3}.\label{e:lossfunctionsproof3}
	\end{align}
It follows that $\lambda_m \lambda_h \neq 0$: if, for example, $\lambda_m=0$, then \eqref{e:lossfunctionsproof1} implies $\lambda_h \neq 0$ (as $\lambda_l\neq 0$), and then \eqref{e:lossfunctionsproof1} and \eqref{e:lossfunctionsproof2} imply $x -\theta_l = x - \theta_h$ for all $x > \theta_h$, contradicting $\theta_l <\theta_h$. Since $\left((x-\theta_l)^{z-2}\right)^2 = (x-\theta_l)^{z-1}(x-\theta_l)^{z-3}$, we manipulate the right-hand sides of \eqref{e:lossfunctionsproof1}--\eqref{e:lossfunctionsproof3} to obtain
$$	2\lambda_m\lambda_h(x-\theta_m)^{z-2}(x-\theta_h)^{z-2}=\lambda_m\lambda_h\left((x-\theta_m)^{z-1}(x-\theta_h)^{z-3} + (x-\theta_m)^{z-3}(x-\theta_h)^{z-1}\right),$$
	which simplifies, using $\lambda_m \lambda_h \neq 0$, to
	$$2=\frac{x-\theta_h}{x-\theta_m} + \frac{x-\theta_m}{x-\theta_h}.$$
	Therefore, $x-\theta_h = x-\theta_m$ for all $x  > \theta_h$, contradicting $\theta_m < \theta_h$. 
\end{proof}

\subsection{Proof of \autoref{cor:multidim}}

The ``if'' direction of the result follows directly from \autoref{char_scd_f}. For the ``only if'' direction, we apply \autoref{char_scd_f} and observe that, for any $a', a'' \in A$,
$$D_{a', a''}(\theta) =(g^I (a') - g^I(a'')) f^I (\theta) + (g^{II}(a')-g^{II}(a'')) f^{II}(\theta),$$
with some $f^I, f^{II}: \Theta \to \Reals$ each single-crossing and ratio ordered, and $g^I, g^{II}: A \to \Reals$.

For each dimension $i$ for which $g_i: A_i \to \Reals$ is non-constant, we take $a', a'' \in A$ with $g_i(a_i') \neq g_i(a_i'')$ and $a'_j = a''_j$ for $j \neq i$. It follows that $D_{a', a''}(\theta) = (g_i(a'_i) - g_i(a''_i) )f_i(\theta)$, and letting $\lambda^I_i \equiv \frac{g^I(a' )-g^{I}(a'')}{g_i(a'_i)-g_i(a''_i)}$ and $\lambda^{II}_i \equiv \frac{g^{II}(a')-g^{II}(a'')}{g_i(a'_i)-g_i(a''_i)}$, that
$$(\forall \theta) \quad f_i(\theta) = \lambda^I_i f^I(\theta) + \lambda^{II}_i f^{II}(\theta).$$
For each dimension $i$ for which $g_i:A_i \to \Reals$ is constant, we set $\lambda^I_i \equiv 0$ and $\lambda^{II}_i \equiv 0$.

We have
$$
u(a, \theta) = \sum_{i=1}^n g_i(a)f_i(\theta) = \left(\sum_{i=1}^n \lambda^{I}_i g_i(a_i) \right) f^I(\theta)+\left(\sum_{i=1}^n \lambda^{II}_i g_i(a_i) \right) f^{II}(\theta)+h(\theta),
$$
where $h(\theta) \equiv \sum_{i \in \{j: \text{$g_j$ is constant}\}} g_i f_i(\theta)$, with each $g_i$ a constant in the summation.

\subsection{Proof of \autoref{p:SCED-X}}
\label{sec:proof_info_design}

\paragraph{$(\impliedby)$}
For any $Q \in A\equiv \left\{ P \in \Delta\Delta \Omega : \int_{p\in \Delta \Omega} p \mathrm d P= p^*\right\}$, 
\begin{align*}
u(Q, \theta) & = \left(\int_{\Delta \Omega} g_1(p) \mathrm{d} Q \right)f_1(\theta) + \left(\int_{\Delta \Omega} g_2(p) \mathrm{d} Q \right)f_2(\theta)  + \int_{\Delta \Omega} \left(\sum_{\omega \in \Omega} v(\delta_{\omega}, \theta)p(\omega)\right) \mathrm{d} Q.
\end{align*}
The last term on the right-hand side is equal to
$$
\sum_{\omega} v(\delta_{\omega}, \theta) \left(\int_{\Delta \Omega} p(w) \mathrm{d} Q \right) = \sum_{\omega} v(\delta_{\omega}, \theta) p^*(\omega).
$$

Thus, for any $Q, R \in A$,
$$
D_{Q, R}(\theta)=
 \left(\int_{\Delta \Omega} g_1(p) \mathrm{d} Q - \int_{\Delta \Omega} g_1(p) \mathrm{d} R\right)f_1(\theta)+\left(\int_{\Delta \Omega} g_2(p) \mathrm{d} Q - \int_{\Delta \Omega} g_2(p) \mathrm{d} R\right)f_2(\theta),
$$
which is single crossing in $\theta$ by \autoref{equiv_cond}.

\paragraph{$(\implies)$}

Suppose that $v$ has SCED-X with a full-support prior $p^*$. By \autoref{char_scd_f}, for any experiment $Q \in A$,
$$
u(Q, \theta) = \int_{p \in \Delta \Omega} v(p,\theta) \mathrm{d}  Q = g_1(Q)f_1(\theta) + g_2(Q)f_2(\theta) + h(\theta),
$$
where $g_1, g_2: A \to \Reals$, $h: \Theta \to \Reals$, and $f_1, f_2:\Theta \to \Reals$ are each single crossing and ratio ordered.

Take any posterior $p \in \Delta \Omega$, and find $\alpha \in (0,1]$ and $q \in \Delta \Omega$ such that
$p^* = \alpha p + (1-\alpha) q$.
We consider two experiments: $Q_p$ yields posteriors $p$ and $q$ with probability $\alpha$ and $1-\alpha$, respectively, and $R_p$ yields each degenerate posterior $\delta_\omega \in \Delta \Omega$ with probability 
$\alpha p(\omega)$, and posterior $q$ with probability $1-\alpha$. Observe that $Q_p, R_p \in A$. Thus, 
\begin{align*}
& u(Q_p, \theta) = \alpha v(p, \theta) + (1-\alpha) v(q, \theta) = g_1(Q_p)f_1(\theta) + g_2(Q_p) f_2(\theta) + h(\theta), \quad \text{and}\\
& u(R_p, \theta) = \alpha \left(\sum_{\omega} v(\delta_\omega, \theta) p(\omega) \right) + (1-\alpha) v(q, \theta) = g_1(R_p)f_1(\theta) + g_2(R_p)f_2(\theta) + h(\theta).
\end{align*}
Hence, 
$u(Q_p,\theta) - u(R_p, \theta) = \alpha \left(v(p, \theta) - \sum_{\omega} v(\delta_{\omega}, \theta) p(\omega)\right)$,
which implies that
$$
v(p, \theta) - \sum_{\omega} v(\delta_{\omega}, \theta) p(\omega) =  \tilde{g}_1 (p) f_1(\theta) + \tilde{g}_2(p) f_2(\theta),
$$
where $\tilde{g}_i(p) = \frac{g_i(Q_p) - g_i(R_p)}{\alpha}$ for $i=1,2$.

\section{Proofs for Aggregating Monotonic Functions}
\label{sec:MDproofs}

\subsection{Proof of \autoref{equiv_cond_mon}}
\label{proof:equiv_cond_mon}

\paragraph{$(\impliedby)$} Suppose there exist $\lambda \in \Reals^2$ such that $f_2 = \lambda_1 f_1 + \lambda_2$. Then, for any $\alpha \in \Reals^2$, 
\begin{align*}
(\alpha \cdot f) (\theta) 
= \alpha_1 f_1(\theta) + \alpha_2 (\lambda_1 f_1(\theta) + \lambda_2)
= (\alpha_1 + \alpha_2 \lambda_1) f_1(\theta) + \lambda_2,
\end{align*}
which is monotonic.

\paragraph{$(\implies)$}
The proof is trivial if both $f_1$ and $f_2$ are constant functions. So we suppose that at least one function, say $f_1$, is not constant: 
\begin{equation}
(\exists \theta', \theta'') \quad f_1(\theta') \neq f_1(\theta'').
\label{e:selection_theta}
\end{equation}
This implies that $\rank[M_{\theta', \theta''}] = 2$, where
$$
M_{\theta', \theta''} \equiv 
\begin{bmatrix}
f_{1}(\theta') & 1 \\
f_{1}(\theta'') & 1 \\
\end{bmatrix}
.
$$
Hence, the system
\begin{equation}
\begin{bmatrix}
f_2(\theta')\\
f_2(\theta'')
\end{bmatrix}
=
\begin{bmatrix}
f_1(\theta') & 1 \\
f_1(\theta'') & 1 \\
\end{bmatrix}
\begin{bmatrix}
\lambda_1\\
\lambda_2
\end{bmatrix}
\label{e:extending_lin_depen_monotone}
\end{equation}
has a unique solution $\lambda^* \in \mathbb{R}^2$. We will show that $f_2 = \lambda^*_1f_1 + \lambda^*_2$. 

Suppose, towards contradiction, there exists $\theta^*$ such that 
\begin{equation}
\label{e:extending_lin_depend_contra_monotone}
f_2(\theta^*) \neq \lambda^*_1f_1(\theta^*) + \lambda^*_2. 
\end{equation}
Let $\underline{\theta}$ and $\overline{\theta}$ be a lower and upper bound of $\{\theta', \theta'', \theta^*\}$. If $\rank[M_{\underline{\theta}, \overline{\theta}}] < 2$, then $f_1(\overline{\theta}) = f_1(\underline{\theta})$. As $\theta'$ and $\theta''$ are in between $\underline{\theta}$ and $\overline{\theta}$ and $f_1$ is monotonic, we have $f_1(\theta') = f_1(\theta'')$, which contradicts \eqref{e:selection_theta}. 
If, on the other hand, $\rank[M_{\underline{\theta}, \overline{\theta}}]=2$, then the system
$$
\begin{bmatrix}
f_2(\underline{\theta})\\
f_2(\overline{\theta})
\end{bmatrix}
=
\begin{bmatrix}
f_1(\underline{\theta}) & 1 \\
f_1(\overline{\theta}) & 1 \\
\end{bmatrix}
\begin{bmatrix}
\lambda'_1\\
\lambda'_2
\end{bmatrix}
$$
has a unique solution $\lambda' \in \mathbb{R}^2$. 
As $\theta', \theta''$, and $\theta^*$ are in between $\underline{\theta}$ and $\overline{\theta}$, and $f_2 - \lambda'_1f_1$ is monotonic, we have
\begin{align}
\begin{bmatrix}
f_2(\theta')\\
f_2(\theta'')
\end{bmatrix}
&=
\begin{bmatrix}
f_1(\theta') & 1 \\
f_1(\theta'') & 1 \\
\end{bmatrix}
\begin{bmatrix}
\lambda'_1\\
\lambda'_2
\end{bmatrix}
\text{ and } \label{e:extending_lin_depen1_monotone}\\
f_2(\theta^*)& = \lambda'_1 f_1(\theta^*) + \lambda'_2.\label{e:extending_lin_depen2_monotone}
\end{align}
\autoref{e:extending_lin_depen1_monotone} implies that $\lambda'$ solves \eqref{e:extending_lin_depen_monotone}. As the unique solution to \eqref{e:extending_lin_depen_monotone} was $\lambda^*$, it follows that $\lambda'=\lambda^*$.  But then \eqref{e:extending_lin_depend_contra_monotone} and \eqref{e:extending_lin_depen2_monotone} are in contradiction.

\subsection{Proof of \autoref{equiv_cond_mon_general}}
\label{proof:equiv_cond_mon_general}

\paragraph{$(\impliedby)$} We omit the proof as it is similar to \hyperref[Proof_prop1]{the proof} of \autoref{equiv_cond_general} in \appendixref{Proof_prop1}.

\vspace{-6pt}
\paragraph{$(\implies)$} For the proof of necessity, if $(\forall \z)$ $f(\z, \theta)$ is a constant function of $\theta$, then we let $\lambda_1 (\z) = 0$ and $\lambda_2(\z) = f(\z, \theta)$. If there exists $\z' \in \Z$ such that $f(\z', \theta)$ is not a constant function of $\theta$, then \autoref{equiv_cond_mon} implies
$(\forall \z, \theta) \ f(\z, \theta) = \lambda_1(\z) f(\z', \theta) + \lambda_2(\z),$
with \mbox{$\lambda_1, \lambda_2: \Z \to \Reals$}.

\section{Relationship to Signed-Ratio Monotonicity} 
\label{sec:qs_karlin}

\citet{quah2012aggregating} establish that for any two functions $f_1:\Theta \to \Reals$ and $f_2:\Theta \to \Reals$ that are each single crossing from below, $\alpha_1f_1 + \alpha_2f_2$ is single crossing from below for all $\alpha \in \mathbb{R}^2_+$ if and only if $f_1$ and $f_2$ satisfy \textit{signed-ratio monotonicity}:
for all $i,j\in\{1,2\}$,
\begin{align}(\forall \theta_l  < \theta_h)\quad f_j(\theta_l)<0<f_i(\theta_l) &\implies f_i(\theta_h)f_j(\theta_l)\leq f_i(\theta_l)f_j(\theta_h).
\label{e:SRM}
\end{align}

\begin{figure}
\begin{center}
            \includegraphics[width=3.2in]{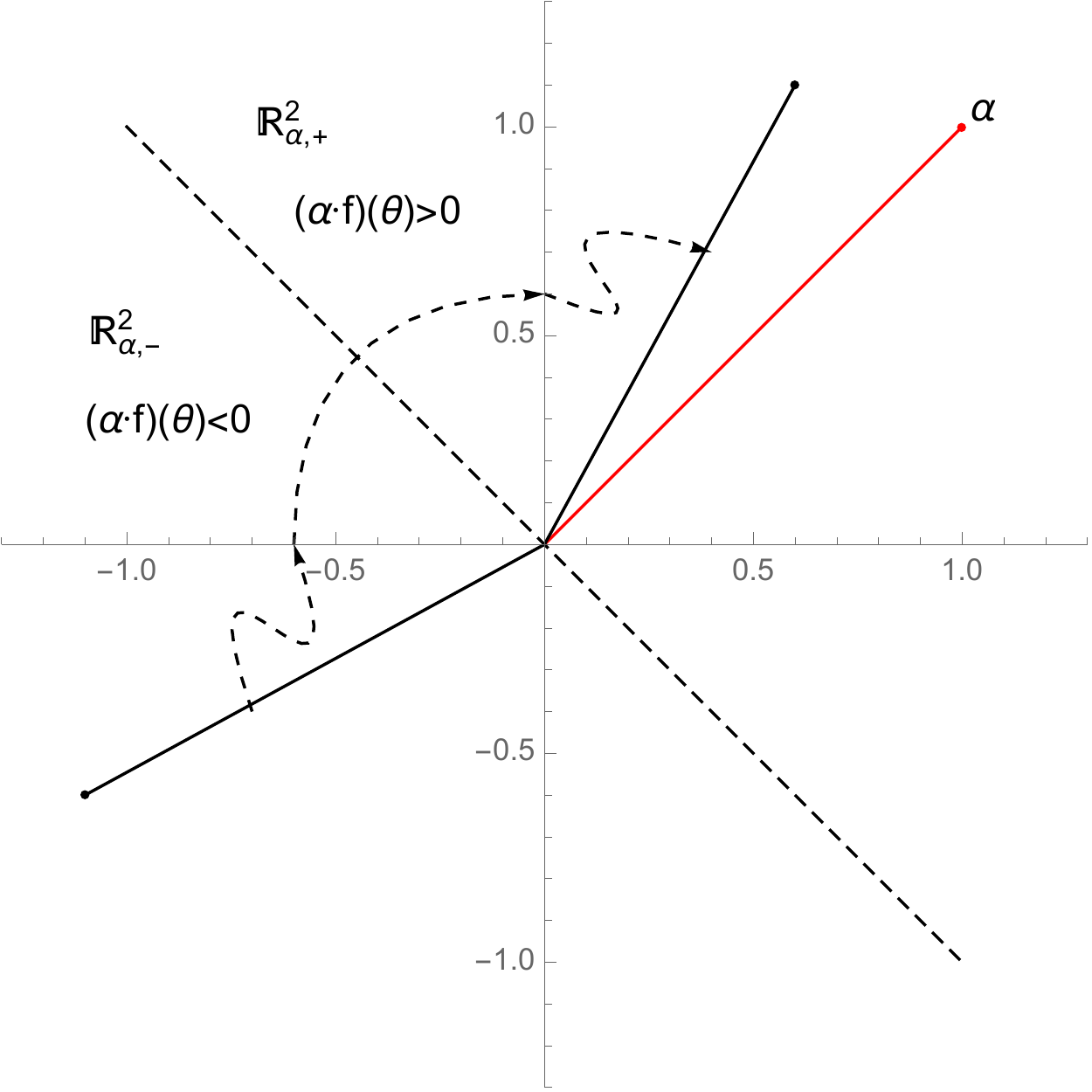}
    \caption{Signed-ratio monotonicity and single crossing of a convex combination.}
            \label{fig:signed_ratio_mon}
\end{center}
\end{figure}

Given our discussion in \autoref{sec:char} of a graphical interpretation of ratio ordering, one can see that Condition \eqref{e:SRM} implies that the vector $f(\theta) \equiv (f_1(\theta), f_2(\theta))$ rotates clockwise as $\theta$ increases within the upper-left quadrant (i.e., when $f_1(\cdot)<0<f_2(\cdot)$), while it rotates counterclockwise within the lower-right quadrant (i.e., when $f_1(\cdot)>0>f_2(\cdot)$); there are no restrictions in the other two quadrants.\footnote{To be precise: by ``quadrant'' we mean the interiors, i.e., excluding the axes.} The dashed curve with arrowheads in \autoref{fig:signed_ratio_mon} provides a depiction. Note that if $f_1$ and $f_2$ are both single crossing from below (or both from above), then there cannot exist $\theta_l<\theta_h$ such that one of $f(\theta_l)$ and $f(\theta_h)$ is in the upper-left quadrant and the other in the lower-right quadrant. It follows that if $f_1$ and $f_2$ are both single crossing from below, then ratio ordering implies signed-ratio monotonicity; more generally, however, the implication is not valid.

\medskip

\autoref{fig:signed_ratio_mon} also illustrates \citepos{quah2012aggregating} result, analogous to \autoref{fig:ordering_suff_nec} for \autoref{equiv_cond}. Any linear combination $\alpha \in \mathbb{R}^2_+ \backslash \{0\}$ defines two open half spaces, $\Reals^2_{\alpha, -} \equiv$ $\{ x \in \Reals^2 : \alpha \cdot x < 0\}$ and $\Reals^2_{\alpha, +} \equiv \{ x \in \Reals^2 : \alpha \cdot x > 0\}$, as indicated in \autoref{fig:signed_ratio_mon}. If the vector $f(\theta)$ rotates monotonically as $\theta$ increases from $\Reals^2_{\alpha, -}$ to $\Reals^2_{\alpha, +}$, or either half space contains the vector $f(\theta)$ for all $\theta$, then $\alpha \cdot f \equiv \alpha_1f_1 + \alpha_2 f_2$ is single crossing from below. Conversely, if $f(\theta)$ does not rotate monotonically in the upper-left or lower-right quadrant, then there exists $\alpha \in \mathbb{R}^2_+ \backslash \{0\}$ such that $\alpha \cdot f$ is not single crossing from below.

\end{document}